\documentclass[traditabstract]{aa} 
\usepackage{graphicx}
\graphicspath{{graphics/}{../graphics/}}   
\usepackage{txfonts, color}
\usepackage{soul}
\usepackage{tabularx}
\usepackage[draft=False,colorlinks=true,linkcolor=black,citecolor=blue,urlcolor=blue,linktoc=all]{hyperref}

\usepackage{calrsfs}
\DeclareMathAlphabet{\pazocal}{OMS}{zplm}{m}{n}
\newcommand{\significance}[1]{\pazocal{S}_{\mathrm{#1}}}
\newcommand{\flag}[2]{\pazocal{F}_{\mathrm{#1}}^{\mathrm{#2}}}

\usepackage{longtable,pdflscape}

\usepackage{multirow}

\usepackage[switch]{lineno}

\usepackage{enumerate}

\makeatletter
\def\LongtableFooter{%
  \multicolumn{\LT@cols}{r}{\framebox[1.1\width]{\textbf{continued on the following page}}}\\}
\makeatother

\usepackage{natbib}
\bibpunct{(}{)}{;}{a}{}{,} 

\usepackage{footmisc}

\newcommand{\chired}{\chi^{2}_{\mathrm{red}}}

\newcommand{\beamefficiency}{\eta_{\mathrm{mb}}}

\newcommand{\Fit}{\mathrm{fit}}

\newcommand{\Max}{\mathrm{max}}
\newcommand{\Min}{\mathrm{min}}
\newcommand{\Ncomp}{N_{\mathrm{comp}}}
\newcommand{\Nchan}{N_{\mathrm{chan}}}

\newcommand{\Nmin}{N_{\mathrm{min}}}
\newcommand{\Npad}{N_{\mathrm{pad}}}
\newcommand{\order}{\xi}
\newcommand{\plimit}{P_{\mathrm{Limit}}}
\newcommand{\rms}{\sigma_{\mathrm{rms}}}
\newcommand{\rmsavg}{\langle\sigma_{\mathrm{rms}}\rangle}
\newcommand{\rmsTa}{\sigma(\text{T}_{\mathrm{A}}^{*})}
\newcommand{\rmstrue}{\sigma_{\mathrm{rms,\,true}}}
\newcommand{\snmin}{\text{S/N}_{\mathrm{min}}}
\newcommand{\snminfit}{\text{S/N}_{\Min,\,\Fit}}
\newcommand{\snminneg}{\text{S/N}_{\Min,\,\mathrm{neg}}}
\newcommand{\snspike}{\text{S/N}_{\mathrm{spike}}}

\newcommand{\weight}[1]{\pazocal{W}_{\mathrm{#1}}}

\newcommand{\kms}{km$\,$s$^{-1}$}

\newcommand\gausspy{\textsc{GaussPy}}
\newcommand\gausspyplus{\textsc{GaussPy+}}
\newcommand\scousepy{\textsc{ScousePy}}

\usepackage{xfrac}

\usepackage{titlesec}

\titleclass{\subsubsubsection}{straight}[\subsection]

\newcounter{subsubsubsection}[subsubsection]
\renewcommand\thesubsubsubsection{\thesubsubsection.\arabic{subsubsubsection}}

\titleformat{\subsubsubsection}
  {\normalfont\normalsize}{\thesubsubsubsection}{1em}{}
\titlespacing*{\subsubsubsection}
{0pt}{3.25ex plus 1ex minus .2ex}{1.5ex plus .2ex}

\makeatletter
\renewcommand\paragraph{\@startsection{paragraph}{5}{\z@}%
  {3.25ex \@plus1ex \@minus.2ex}%
  {-1em}%
  {\normalfont\normalsize\bfseries}}
\renewcommand\subparagraph{\@startsection{subparagraph}{6}{\parindent}%
  {3.25ex \@plus1ex \@minus .2ex}%
  {-1em}%
  {\normalfont\normalsize\bfseries}}
\def\toclevel@subsubsubsection{4}
\def\toclevel@paragraph{5}
\def\toclevel@paragraph{6}
\def\l@subsubsubsection{\@dottedtocline{4}{7em}{4em}}
\def\l@paragraph{\@dottedtocline{5}{10em}{5em}}
\def\l@subparagraph{\@dottedtocline{6}{14em}{6em}}
\makeatother

\begin{document}
   \title{\gausspyplus\thanks{\url{https://github.com/mriener/gausspyplus}}: A fully automated Gaussian decomposition package for emission line spectra}

   \author{M. Riener
          \inst{1, }{\thanks{Member of the International Max-Planck Research School for Astronomy and Cosmic Physics at the University of Heidelberg (IMPRS-HD), Germany}}
          \and
          J. Kainulainen
          \inst{2, 1}
          \and
          J. D. Henshaw
          \inst{1}
          \and
          J. H. Orkisz
          \inst{2}
          \and
          C. E. Murray
          \inst{3, }{\thanks{NSF Astronomy \& Astrophysics Postdoctoral Fellow}}
          \and
          H. Beuther
          \inst{1}
          }

    \institute{Max-Planck Institute for Astronomy, K\"onigstuhl 17, 69117 Heidelberg, Germany
    \and
    Chalmers University of Technology, Department of Space, Earth and Environment, SE-412 93 Gothenburg, Sweden
    \and
    Department of Physics \& Astronomy, Johns Hopkins University, 3400 N. Charles Street, Baltimore, MD 21218
    }

   \date{Received ..., 2019; accepted ..., 2019}

 
  \abstract
  {Our understanding of the dynamics of the interstellar medium is informed by the study of the detailed velocity structure of emission line observations.
  One approach to study the velocity structure is to decompose the spectra into individual velocity components; this leads to a description of the dataset that is significantly reduced in complexity.
  However, this decomposition requires full automation lest it becomes prohibitive for large datasets, such as Galactic plane surveys.
  We developed \gausspyplus, a fully automated Gaussian decomposition package that can be applied to emission line datasets, especially large surveys of HI and isotopologues of CO.
  We built our package upon the existing \gausspy\ algorithm and significantly improved its performance for noisy data. 
  New functionalities of \gausspyplus\ include: i) automated preparatory steps, such as an accurate noise estimation, which can also be used as standalone applications; ii) an improved fitting routine; iii) an automated spatial refitting routine that can add spatial coherence to the decomposition results by refitting spectra based on neighbouring fit solutions.
  We thoroughly tested the performance of \gausspyplus\ on synthetic spectra and a test field from the Galactic Ring Survey.
  We found that \gausspyplus\ can deal with cases of complex emission and even low to moderate signal-to-noise values.}


   \maketitle
%


\section{Introduction}
Observations of emission lines are of fundamental importance in radio astronomy. 
Starting with the first detections of neutral hydrogen (HI) via the $21$~cm line at $1420.4$~MHz by \citet{Ewen1951} and the first detection of interstellar carbon monoxide (CO) in the Orion nebula by \citet{Wilson1970}, the study of emission lines at radio wavelengths has led to groundbreaking astrophysical insights.
Our knowledge about the interstellar medium (ISM) is to a large part shaped by observations of the emission of its gas molecules. 
In particular, we can use the radial velocity---corresponding to Doppler shifts of the emission line with respect to its rest frequency---to gain information about the kinematics and dynamics of the gas.

In our Milky Way, large Galactic surveys of HI \citep[e.g.,][]{Stil2006, Murray2015, Beuther2016} and isotopologues of CO \citep[e.g.,][]{Dame2001, Jackson2006, Dempsey2013, Barnes2015, Rigby2016, Umemoto2017, Schuller2017, Su2019} have been used to, e.g., study Galactic structure \citep[e.g.,][]{Dame2001, Nakanishi2006} and construct catalogues of molecular clouds and clumps \citep[e.g.,][]{Rathborne2009, Miville-Deschenes2017, Colombo2019}.
Such studies are usually more focused on the average properties of the gas on Galactic scales or on the scales of molecular clouds or clumps. 
However, there is a tremendous wealth of physically interesting information that can be gleaned from studying the detailed velocity structure of the gas, among them fundamental insights about turbulence properties in the ISM and molecular clouds (e.g., \citealt{Larson1981}, \citealt{Ossenkopf2002}, \citealt{Heyer2004}, \citealt{burkhart2010}, \citealt{Orkisz2017}; for reviews, see \citealt{Elmegreen2004} and \citealt{hennebelle2012review}) and dense cores \citep[e.g.,][]{Falgarone2009, Pineda2010, keto2015, chen2018droplets}, inference about imprints of shear in molecular clouds \citep[e.g.,][]{Hily-Blant2009}, and the internal velocity structure of filaments \citep[e.g.,][]{Arzoumanian2013, Arzoumanian2018, Hacar2013, Henshaw2014, Orkisz2019}.

While the gas dynamics on smaller scales has been already well studied, the detailed velocity structure of the gas on Galactic scales remains as yet unexplored.
We currently do not know whether the velocity structure across large scales shows properties that could serve as diagnostics of phenomena such as molecular cloud formation and evolution or the impact of the Galactic structure on the ISM.
To facilitate such analyses, we would ideally like to apply the methods and techniques of the small-scale studies to the large surveys of the Galactic plane.

One approach that has substantial potential is quantifying and analysing the complex spectra taken through the Galactic plane by decomposing them into velocity components and then analysing the properties and statistics of these components. In such analyses, the components are usually assumed to have Gaussian shapes, as random thermal and non-thermal motions in the gas lead to Doppler motions with a Gaussian distribution of gas velocities. Moreover, adopting the Gaussian shape is mathematically simple and leads to a significant reduction in complexity and enables easier post-analysis steps through a rich set of available Gaussian statistics tools.

Recently, several semi-automatic \citep[e.g.,][]{Ginsburg2011, Hacar2013, Henshaw2016, Henshaw2019} and fully automated \citep[e.g.,][]{Haud2000, Lindner2015, Miville-Deschenes2017, Clarke2018, Marchal2019-rohsa} spectral fitting techniques have been introduced.
The semi-automated techniques require user interaction, usually in deciding how many velocity components to fit. 
This can be achieved, for instance, using spatially smoothed spectra to inform the fit. 
However, the user-dependent decisions introduce subjectivity to the fitting procedure that reduces reproducibility of the results.
The required interactivity with the user can also make it difficult to distribute the analysis to multiple processors.
Therefore, while semi-automated approaches are well-suited for small data sets (individual molecular clouds or nearby galaxies at high or low spatial resolution, respectively), they can become prohibitively time-consuming for the analysis of big surveys with millions of spectra and components.

The automated methods overcome these drawbacks by removing the user interaction. The initial number of components can either be a guess \citep{Miville-Deschenes2017, Marchal2019-rohsa} or based on the derivatives of the spectrum \citep{Lindner2015, Clarke2018}.
However currently, these automated routines either: 
fit the spectra independently from each other \citep{Lindner2015, Clarke2018}, which might introduce unphysical differences between the fit results of neighbouring spectra; use a fixed number of velocity components as initial guesses \citep{Miville-Deschenes2017, Marchal2019-rohsa}, which can be computationally expensive; or are not freely available to the community.
Also, the current versions of the automated methods listed above are of the "first generation"; there is still potential to improve the decomposition techniques and their applicability to different datasets.

In this work, we present \gausspyplus, an automated decomposition package that is based on the existing \gausspy\ algorithm \citep{Lindner2015}, but with physically-motivated developments specifically designed for analysing the dynamics of the ISM.
We developed \gausspyplus\ with the specific aim of analysing CO surveys of the Galactic plane, such as the Galactic Ring Survey \citep[GRS;][]{Jackson2006} and SEDIGISM \citep{Schuller2017}. 
However, \gausspyplus\ should be easily adaptable to other emission line surveys for which Gaussian shapes provide a good approximation of the line shapes.
Some of the line-analysis tasks of \gausspyplus, such as the estimation of noise and the identification of signal peaks, can also be used as independent standalone modules to serve more specific purposes.

In this paper, we present the algorithm and test it thoroughly on synthetic spectra and a GRS test field.
A full application of \gausspyplus\ on the entire GRS dataset is in preparation and will be presented in a subsequent paper.


\section{Archival data and methods}
\label{cha:data}

\subsection{The \gausspy\ algorithm}
\label{sec:gausspy}
In this work we extend and modify the \gausspy\ algorithm \citep{Lindner2015}, which is an autonomous Gaussian decomposition technique for automatically decomposing spectra into Gaussian components.
While \gausspy\ was developed for the decomposition of HI spectra \citep[e.g.][]{Murray2018, Denes2018} it can in principle be used for the decomposition of any spectra that can be approximated well by Gaussian functions (e.g., CO).

One of the strengths of the \gausspy\ algorithm is that it automatically determines the initial guesses for Gaussian fit components for each spectrum with a technique called derivative spectroscopy.
This technique is based on finding functional maxima and minima in the spectrum to gauge which of the features are real signal peaks.
Since the estimation of maxima and minima requires the calculation of higher derivatives (up to the fourth order), an essential preparatory step in \gausspy\ is to smooth the spectra in such a way as to get rid of the noise peaks without smoothing over signal peaks \citep[cf. Fig.$\,2$ in][]{Lindner2015}.
If the dataset contains signal peaks that show a limited range in widths, smoothing with a single parameter $\alpha_{1}$ may already lead to good results in the fitting.
In the original \gausspy\ algorithm users can choose between two different versions of denoising the spectrum before derivatives of the data are calculated: a total variation regularization algorithm and filtering with a Gaussian kernel.
We use exclusively the latter approach, in which the parameter $\alpha_{1}$ refers to the size of the Gaussian kernel that is used to Gaussian-filter the spectrum.
The decomposition of datasets that show a mix of both narrow and broad linewidths likely requires an additional smoothing parameter $\alpha_{2}$ to yield good fitting outcomes.
The fitting procedure using a single or two smoothing parameters is referred to as one-phase or two-phase decomposition, respectively.

It is essential for the best performance of the derivate spectroscopy technique to find the optimal smoothing parameters for the original spectra. 
The \gausspy\ algorithm achieves this via an incorporated supervised machine learning technique, for which the user has to supply the algorithm with a couple of hundred well-fit spectra, from which the algorithm then deduces the best smoothing parameters.

More specifically, \gausspy\ uses the gradient descent technique -- a first-order iterative optimization algorithm -- to find values for $\alpha_{1}$ and $\alpha_{2}$ that yield the most accurate decomposition of the training set. 
This accuracy is measured via the F$_{1}$ score, which is defined as:

\begin{equation}
	F_{1} = 2\cdot\dfrac{\text{precision}\cdot\text{recall}}{\text{precision} + \text{recall}},
\end{equation}

where precision refers to the fraction of fit components that are correct and recall refers to the fraction of true components that were found in the decomposition of the training set with guesses for $\alpha_{1}$ and $\alpha_{2}$.
See \citet{Lindner2015} for more details on how the training set is evaluated.

\subsection{$^{13}$CO data}
We test \gausspyplus\ on data from the Boston University–Five College Radio Astronomy Observatory Galactic Ring Survey \citep[GRS;][]{Jackson2006} that we downloaded from the online repository of the Boston University Astronomy Department\footnote{\url{https://www.bu.edu/galacticring/new_data.html}\label{foot:grs}}. 
This survey covered the lowest rotational transition of the $^{13}$CO isotopologue with an angular resolution of $46^{\prime\prime}$, a pixel sampling of $22^{\prime\prime}$, and a spectral resolution of $0.21$~\kms.
The values in the GRS dataset are given in antenna temperatures, which we converted to main beam temperatures by dividing them with the main beam efficiency of $\beamefficiency = 0.48$\footref{foot:grs}.

The lowest rotational transition of $^{12}$CO can show strong self-absorption that can severely affect the lineshape \citep[e.g.][]{Hacar2016-opacity}. 
A decomposition of the spectrum can therefore lead to incorrect results, as strong self-absorption features can be erroneously fit with multiple components.
We do not expect such strong opacity effects for $^{13}$CO observations, but it can still become optically thick in very bright regions \citep[e.g.][]{Hacar2016-opacity}. 
Optical depth effects are also expected for $^{13}$CO observations of nearby regions or observations with high spatial resolution, for which the opacity effects are not smoothed out as in a larger physical beam.
For the moderate spatial resolution of the GRS survey one would thus not expect severe optical depth effects, even though the analysis by \citet{Roman-Duval2010} suggests that opacity effects do indeed have to be taken into account for the GRS dataset.
In this work we will not address the potential problems of optical depth effects or self-absorption on the decomposition results, but we caution that fitting $^{13}$CO peaks with Gaussian components might lead to incorrect fits of multiple components for a single self-absorbed emission line in case regions of optically thick $^{13}$CO are expected to be present in the dataset. 

Note that even though in this paper we demonstrate the functionality of \gausspyplus\ only for a small GRS test field, we used the entire dataset in testing and developing the algorithm.
A forthcoming paper will present and discuss the decomposition results of \gausspyplus\ for the whole GRS dataset and will also discuss the effects and implications of possible optical depth effects for the $^{13}$CO emission and the fitting results.

\section{New decomposition package: \gausspyplus}

\begin{figure}
\centering
\includegraphics[width=\columnwidth]{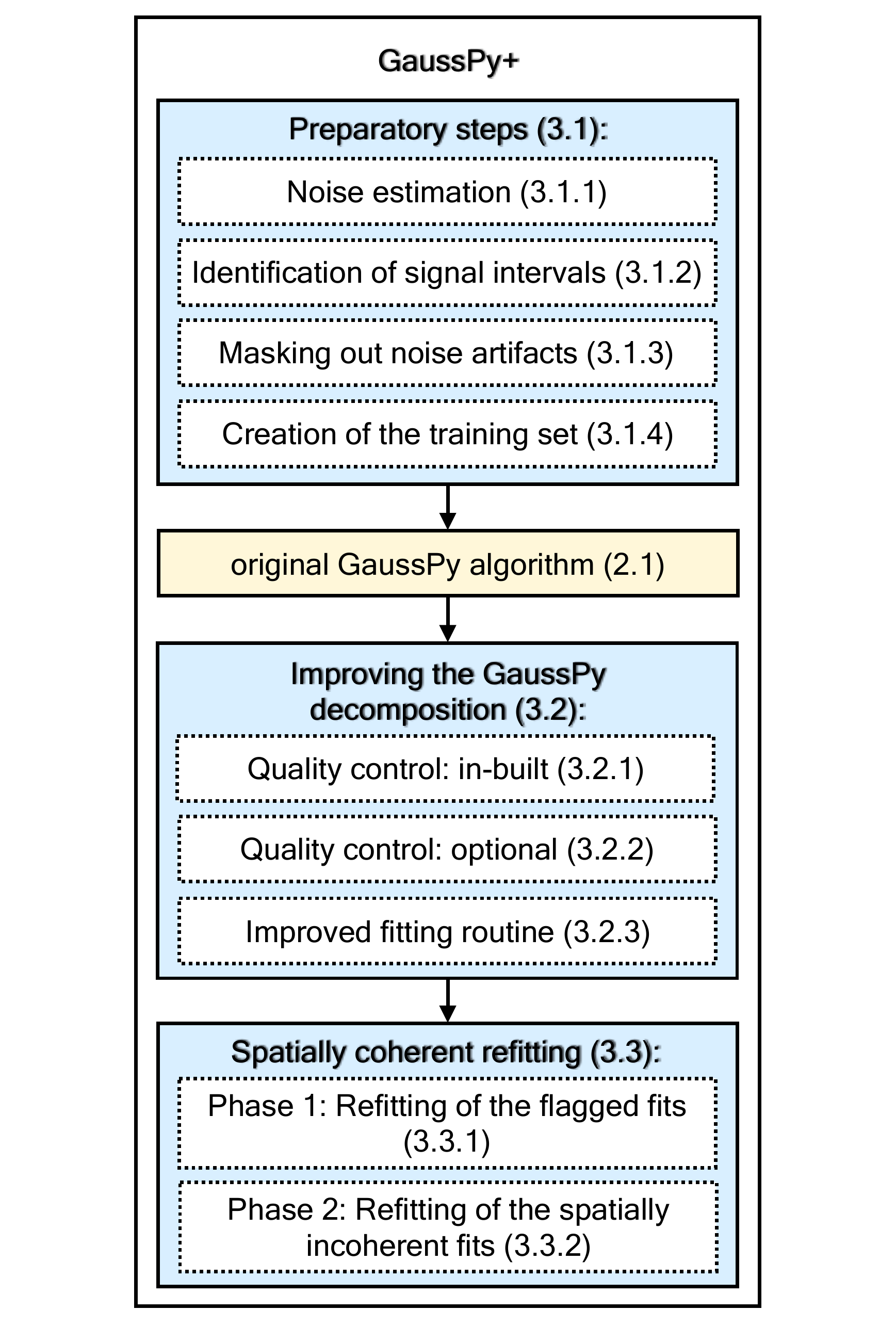}
\caption{Schematic outline describing the new automated methods and procedures included in \gausspyplus, along with the corresponding sections in this paper.
}
\label{fig:schematic_added_procedures}
\end{figure}

The methods and procedures described in this section are all either new preparatory steps for, or extensions to, the original \gausspy\ algorithm. They aim at either improving the performance of \gausspy\ or automating required preparatory steps. Figure \ref{fig:schematic_added_procedures} presents a schematic outline of the \gausspyplus\ algorithm.

The main shortcomings of the original \gausspy\ algorithm that we aim at improving are: i) the noise values are calculated from a fixed fraction of channels in the spectrum, which is not ideal in cases where signal peaks might occur at all spectral channels; ii) the user has to supply the training set; iii) there is no in-built quality control of the fit results; iv) the fit of each spectrum is treated independently of its neighbours.
The last point might lead to drastic jumps between the number of Gaussian components between neighbouring spectra.
From a physical point of view we would not expect such component jumps for resolved extended objects with sizes larger than the beam.
Moreover, observations are often Nyquist sampled, in which case the beam size or resolution element is larger than the pixel size. 
Therefore neighbouring pixels will contain part of the same emission, which also introduces coherence between the number of components between neighbouring spectra.

To develop a fitting algorithm that improves on the above points, we have included in \gausspyplus: i) automated preparatory steps for the noise calculation and creation of the training set (see Sect.$\,\ref{sec:preparatory-steps}$); ii) automated quality checks for the decomposition, some of which can be customized by the user and are used to flag and refit unphysical or unwanted fit solutions (see Sect.$\,\ref{sec:improving-gausspy}$); iii) automated routines that check the spatial coherence of the decomposition and in case of conflicting results try to refit the spectrum based on neighbouring fits (see Sect.$\,\ref{sec:spatial-refitting}$).

In the following, the \gausspyplus\ algorithm is described in detail, following the outline presented in Fig. \ref{fig:schematic_added_procedures}.

A description of \gausspyplus\ keywords including their default values and other symbols used throughout the paper can be found in the Appendix~\ref{tbl:gausspyplus-keywords}.

\subsection{Preparatory steps}
\label{sec:preparatory-steps}

\subsubsection{Noise estimation}
\label{sec:noise-estimation}

The original \gausspy\ algorithm either requires the user to supply noise estimates or uses a certain fraction of the spectral channels, assumed to contain no signal, for the noise estimation. 
However, the latter approach only leads to correct noise 
estimates if one can exclude the presence of signal peaks in the spectral channels used to calculate the noise.

A reliable noise estimation is of fundamental importance for the decomposition---key steps of \gausspy\ depend on the noise value, and also the new procedures in \gausspyplus\ rely on accurate noise estimation: the signal-to-noise (S/N) threshold is used for the initial guesses for the number of components in \gausspy\ and the noise estimate is needed for the quality assessments of the fit components in \gausspyplus.
Because of the key role of the noise, we developed a new, automated noise estimation routine as a preparatory step for the decomposition.

The fundamental, underlying assumptions in our noise estimation process are: i) the noise statistics are Gaussian, i.e. "white noise"; ii) the spectral channels are uncorrelated; and iii) the noise is fluctuating around a baseline of zero. These assumptions enable us to make use of the number statistics of negative/positive channels in the noise estimation process (elaborated further in item 1 below).

\begin{figure*}
    \centering
    \includegraphics[width=2\columnwidth]{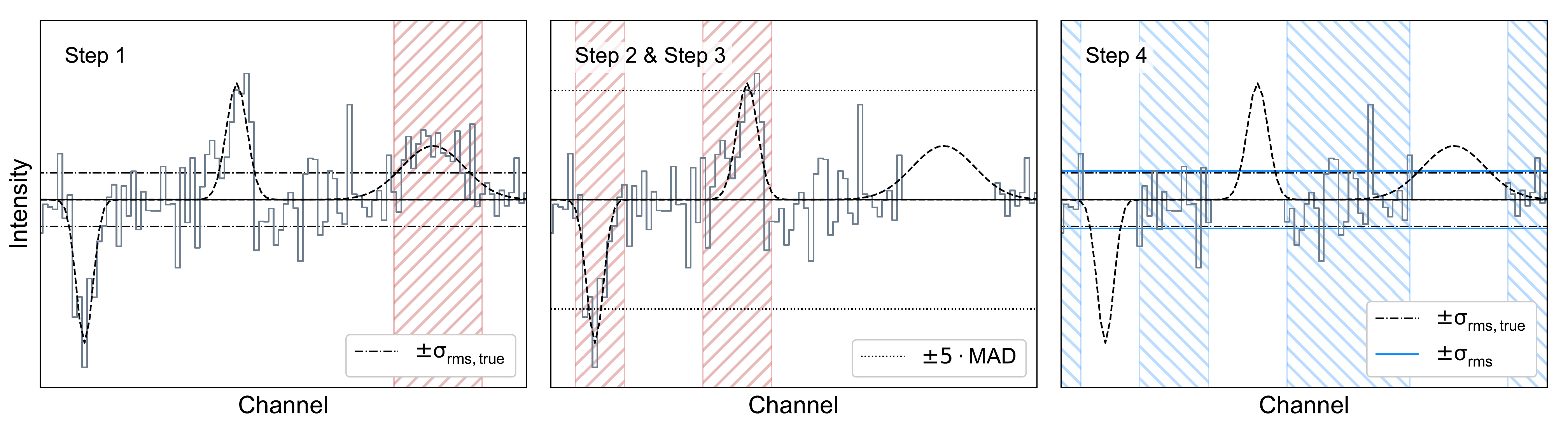}
    \caption{Illustration of our automated noise estimation routine for a mock spectrum containing two signal peaks and a negative noise spike. 
    Hatched red areas indicate spectral channels that are masked out and hatched blue areas indicate all remaining spectral channels used in the noise calculation.
    The lower right panel compares the true noise value ($\rmstrue$; black dash-dotted lines) with the noise value estimated by our automated routine ($\rms$, blue solid lines).
    See Sect.$\,\ref{sec:noise-estimation}$ for more details.
    }
    \label{fig:noise_scheme}
\end{figure*}

In the following, we describe how our automated noise estimation proceeds. The overall idea is to identify the spectral channels that can be used for noise estimation and maximize their number. To do so, the routine has to identify as many channels as possible that are free from signal and instrumental effects. We demonstrate the steps of the process for a mock spectrum in Fig.~\ref{fig:noise_scheme}. The spectrum has 100 channels and contains two challenging features for the noise estimation: a negative noise spike in the first few channels and a broad signal feature with a maximum amplitude of two times the root-mean-square noise $\rms$.

The steps to estimate the noise are the following:
\begin{enumerate}
    \item 
    Mask out broad features in the spectrum; such features are likely to be either positive signal or instrumental artifacts due to, e.g., insufficient baseline corrections. 
    Given our basic assumptions (see above), spectra containing (only) noise have the same number of positive and negative spectral channels on average. We can use this fact to determine the probability of having a number of consecutive positive or negative channels in the spectrum, i.e., the probability that a given feature is noise (or signal/artefact). This provides a mean to mask out features that are likely not noise.
    We estimate the probability that a consecutive number of positive or negative channels is due to noise with a Markov chain (see Appendix~\ref{app:markov} for more details).
    We then mask out all features whose probability to be caused by noise is below a user-defined threshold $\plimit$. 
    For the example spectrum in Fig.~\ref{fig:noise_scheme} we used the default value of $\plimit = 2\%$\footnote{$\plimit = 2\%$ yielded good results in our tests and represents a good compromise between excluding signal peaks with low amplitude values from the noise estimation without masking out too many noise features.}.
    From the Markov chain calculations for a spectrum with $100$ spectral channels we get that all features with more than twelve consecutive positive or negative channels have a probability less than $\plimit = 2\%$ to be the result of random noise fluctuations and are thus masked out (one feature; see left panel in Fig.~\ref{fig:noise_scheme}).
    In many cases, peaks will still continue on both sides of the identified consecutive channels.
    To take this into account, the user can specify how many additional channels $\Npad$ will be masked out on both sides of the identified feature.
    In the example spectrum (Fig.~\ref{fig:noise_scheme}) we set $\Npad = 2$, so two additional channels on both sides of the identified features got masked out. 
    \item
    Use the unmasked negative channels to calculate their median absolute deviation (MAD).
    We use the MAD statistic because it is very robust against outliers in the dataset, such as noise spikes.
    The relationship of MAD to the standard deviation $\sigma$ is $\text{MAD} \approx 0.67 \sigma$.
    We restrict the calculation of the MAD to spectral channels with negative values, since the positive channels can still contain multiple narrow high signal peaks that were not identified in the previous step.
    Note that narrow negative spikes will still be included in this calculation but we assume that their presence is sufficiently uncommon so that they will not significantly affect the estimation of the MAD.
    \item
    Identify intensity values with absolute value higher than $5\,\times\,$MAD.
    We then mask out all consecutively negative or positive channels of all features that contain an intensity value higher than $\pm\,5\,\times\,$MAD.\footnote{
    We choose $\pm\,5\,\times\,$MAD as our threshold because it is a good trade-off: lower thresholds would remove too many valid noise peaks and higher thresholds could miss too many narrow signal peaks with low amplitude values.
    }
    The mask is extended again on both sides by the user-defined number of channels $\Npad$. 
    In the example spectrum, two regions are masked out in this step (middle panel in Fig.~\ref{fig:noise_scheme}). Note how this step is able to identify the second positive signal and the negative noise spike in the spectrum.
    \item
    Use all remaining unmasked channels to calculate the rms noise. 
    The example spectrum is left with 51 unmasked channels (blue hatched areas in the right panel of Fig.~\ref{fig:noise_scheme}) from which the noise is estimated.
\end{enumerate}

The right panel of Fig.~\ref{fig:noise_scheme} shows the determined $\rms$ value (blue solid line), which is very close to the true value $\rmstrue$ (black dash-dotted line) that was used to generate the noise.
This example represents a case in which estimating the noise from a fixed fraction of channels in the beginning or the end of the spectrum would obviously not work well.
Had we estimated the noise with the first or last $20\%$ of spectral channels, we would have overestimated the noise by factors of 2.3 and 1.3, respectively.

In case of residual continuum in the spectrum or signal peaks covering almost all of the spectral channels, the noise estimation can be skewed and biased towards low values.
To circumvent this problem, the user can supply an average noise value $\rmsavg$ or calculate $\rmsavg$ directly from the datacube by randomly sampling a specified number of spectra throughout the cube.
This $\rmsavg$ value is adopted instead of the value resulting from steps $1\text{--}4$ above, if 1) the fraction of spectral channels available for noise calculation from steps $1\text{--}4$ is less than a user-defined value (default: $10\%$), and 2) the noise value resulting from steps $1\text{--}4$ is less than a user-defined fraction of $\rmsavg$ (default: $10\%$)\footnote{The default values are deliberately set to low values to target only spectra with anomalies such as severe baseline effects.}.
If no $\rmsavg$ value is supplied or calculated, the spectra that do not reach the required minimum fraction of spectral channels for the noise calculation are masked out.

We performed thorough testing of the effects of random noise fluctuations on our noise estimation routine. 
A detailed description of the tests is given in Appendix~\ref{app:test_noise}. 
The tests showed that the routine is robust in typical situations (pure white noise, white noise with signal, white noise with signal and negative noise spikes, white noise with weak signal and negative noise spikes).

\subsubsection{Identification of signal intervals}
\label{sec:signal-interval}

If a spectrum contains a high fraction of signal-free spectral channels, goodness of fit calculations can be completely dominated by noise and their value thus may decrease to acceptable numbers even in cases for which the fit did not work out. 
Therefore, we added a routine to \gausspyplus\ that automatically identifies intervals of spectral channels that contain signal; goodness of fit calculations are subsequently restricted to these channels\footnote{With the exception of one normality test that we perform over the whole channel range. See Sect.$\,\ref{sec:goodness-of-fit}$ and App~\ref{app:normality-tests}.}.
Note that the fitting itself is still performed on all spectral channels.

As part of our automated noise estimation routine (outlined in Sect.$\,\ref{sec:noise-estimation}$) we already identify consecutive positive spectral channels that can potentially contain signal (see Fig.~\ref{fig:noise_scheme}). 
We identify these features as signal intervals using a criterion that takes both the S/N ratio and the extent of the feature into account (this criterion is described in more detail in Sect.$\,\ref{sec:significance}$).
For spectra that contain a single narrow peak, only a small fraction of the spectrum might be identified as signal interval. 
To ensure that for such cases the goodness of fit values are not artificially increased by a too small number of spectral channels, the user can require that a minimum number of spectral channels be adopted as signal intervals ($\Nmin$; default value: $100$).
If the signal intervals identified in the spectrum contain fewer channels than required by $\Nmin$, the size of all individual signal intervals identified in the spectrum is incrementally increased on both sides by $\Npad$, until $\Nmin$ is reached.
This incremental padding will not include regions masked out as negative noise spikes (see next section). 
If no signal intervals could be identified in the spectrum, all channels are used for goodness of fit calculations, even though it is unlikely in this case that there are peaks in the spectrum that will be fit. 
We tested the performance of the signal interval identification on synthetic spectra and found that it is able to reliably determine weak and strong signal peaks without being sensitive to smaller peaks caused by random noise fluctuations (see Appendix~\ref{app:test_signal_interval}).

\subsubsection{Masking out noise artifacts}
\label{sec:noise-spikes}

Spectra can sometimes contain negative noise spikes, which can bias the goodness of fit calculations.
In principle, candidate regions with negative noise spikes are already identified in the automated noise estimation routine (Sect.$\,\ref{sec:noise-estimation}$).
However, since the MAD-based threshold is set to a conservative value to exclude most of the narrow signal peaks from the noise estimation, it will also incorrectly remove an increased fraction of regular noise peaks or false positives (see the distribution for sample A of our synthetic spectra in Fig.~\ref{app:test_noise}).
To avoid such contamination of identified noise artifacts by regular noise peaks, the user can decide below which negative value features get masked out by supplying the value in terms of the S/N-ratio ($\snspike$; default value: $5$).
Setting $\snspike=5$ means that any region of consecutive negative channels that contains at least one channel with a value lower than $-5\times\rms$ will get masked out. 
We tested the performance of the identification of noise spikes on synthetic spectra and found that we are able to reliably mask such features out (see Appendix~\ref{app:test_noise_spike}).

\subsubsection{Creation of the training set}
\label{sec:training-set}

As described in Sect.$\,\ref{sec:gausspy}$, \gausspy\ needs a sample of already decomposed spectra to determine the smoothing parameters used in the decomposition. 
In principle, this training set can be composed of synthetic spectra whose noise and emission properties are similar to the dataset the user wants to analyze. 
Another approach is to use actual spectra from the dataset for which the user can supply a reliable decomposition. 
We added a routine to \gausspyplus\ that adopts the latter approach and automatically decomposes a user-defined number of spectra from the dataset.
These decomposition results are then supplied to \gausspy, which uses its machine learning functionality to infer the most appropriate smoothing parameters for the dataset.

In principle, we could use \gausspy\ itself to construct decompositions for this training sample by first guessing the smoothing parameters and correcting them accordingly to get good fitting results.
However, since it can be tricky and time-consuming to guess the correct smoothing parameters for a dataset we added a routine to \gausspyplus\ that decomposes spectra for a training set.

Our key requirement for this decomposition routine was that it should be able to produce high quality fits for a small subset of the dataset.
We recommend to use training set sizes of about $200 \text{--} 500$ decomposed spectra, as these should already give very good values for the smoothing parameter.
In principle also larger training sets can be created, but users should be aware that in this case it can become time-consuming to train \gausspy, as it might be necessary to use different starting values for the smoothing parameters $\alpha_{1}$ and $\alpha_{2}$ to make sure that the search for optimal smoothing parameters explored the parameter space properly and did not get stuck in a local minimum (see Fig.~3 in \citealt{Lindner2015}).
Training sets containing $< 200$ spectra bear the risk of higher uncertainties for the resulting smoothing parameter values, as incorrectly fitted features in the training set may have a large negative impact on the F$_{1}$ score.
While deviations of the smoothing parameters from the optimal values will impact the decomposition with \gausspy, the improved fitting (Sect.$\,\ref{sec:improved-fitting}$) and spatially coherent refitting (Sect.$\,\ref{sec:spatial-refitting}$) routines in \gausspyplus\ should be able to mitigate such incorrect or insufficient decomposition results.
Thus the decomposition of \gausspyplus\ also has a bigger margin for deviations of the smoothing parameters from their optimal values than the decomposition with \gausspy, which allows the use of smaller training set sizes.

For the decomposition of the spectra for the training set we use the SLSQP optimization algorithm and least squares statistic (\textsc{SLSQPLSQFitter}) of the \textsc{astropy.modeling} package, which produced good fits to the spectra in our tests of the routine.
We have to supply the \textsc{SLSQPLSQFitter} routine with initial guesses for possible Gaussian fit components. 
We determine the number of Gaussian fit component candidates and their initial guesses by estimating how many local positive extreme values or maxima are present in the spectrum.
To find these local extreme values, we first set all values to zero that are below a user defined S/N threshold ($\snmin$; default value: 3).
The remaining positive values are then searched for local maxima.
We define a local maximum as a peak that exceeds all values for a minimum number of neighbouring spectral channels on either side of the peak.
This required minimum number of spectral channels on either side can be defined by the user with the $\order$ parameter (default value: 6).

To infer a good value for $\order$, users are advised to check the shape of the components present in the spectra or make a test run for a small training set size and check the decomposition results (routines for plotting the spectra, decomposition results, and residuals are contained in our method).

Our routine then tries to fit a number of Gaussian components according to the inferred peaks of local positive maxima present in the spectrum.
We therefore likely start out with the maximum possible number of Gaussian fit components for the spectrum.
The individual fit parameters of each Gaussian parameter (amplitude $a_{i}$, mean position $\mu_{i}$, standard deviation $\sigma_{i}$) are then checked for the following criteria:
\begin{itemize}
    \item
    amplitude $a_{i} \geq$ $\snmin\,\times\,\rms$
    \item 
    significance $\significance{fit} \geq \significance{\Min}$. 
    See Sect.$\,\ref{sec:significance}$ for more information about this criterion.
    \item
    the standard deviation $\sigma_{i}$ is between user defined limits: $\sigma_{\Min} \leq \sigma_{i} \leq \sigma_{\Max}$, where the limits for the standard deviation can be specified in terms of the full width at half maximum (FWHM) given as fraction of channels ($\Theta_{\Min}$ and $\Theta_{\Max}$; default values: 1. and \texttt{None}, respectively).
\end{itemize}

We do not check if components are blended in the creation of the training set.
If any of the individual Gaussian components do not satisfy all these requirements, its values are removed from the list of initial guess values and a new fit is performed.
These checks and the subsequent refitting is performed as long as some of the individual Gaussians are not satisfying all the criteria or there are no more Gaussian parameters remaining.
In the process of refitting a spectrum we do not add any new fit component candidates.

We thoroughly tested the routine outlined in this section on samples of synthetic spectra and found that it is able to create reliable training sets that allow inferring optimal smoothing parameters with \gausspy\ (see Appendix$\,$\ref{app:test_trainingset}).

Note that we did not optimize the \textsc{SLSQPLSQFitter} decomposition routine for speed, which is why we recommend to only use this fitting technique for the creation of training sets.
See Appendix~\ref{app:time_training_set} for a quantitative comparison between the \textsc{SLSQPLSQFitter} fitting routine and the improved fitting routine of \gausspyplus\ (Sect.$\,\ref{sec:improved-fitting}$) in terms of execution time and performance of the decomposition.

\subsection{Improving the \gausspy\ decomposition}
\label{sec:improving-gausspy}

\subsubsection{Quality control: in-built}
\label{sec:checks}

In this section, we describe the automated quality checks for the decomposition results we implemented in \gausspyplus.
If individual Gaussian components do not satisfy one of the criteria outlined in Sect.$\,\ref{sec:fwhm} - \text{Sect.}\,\ref{sec:channel-range}$ they get discarded.

\begin{figure}
    \centering
    \includegraphics[width=\columnwidth]{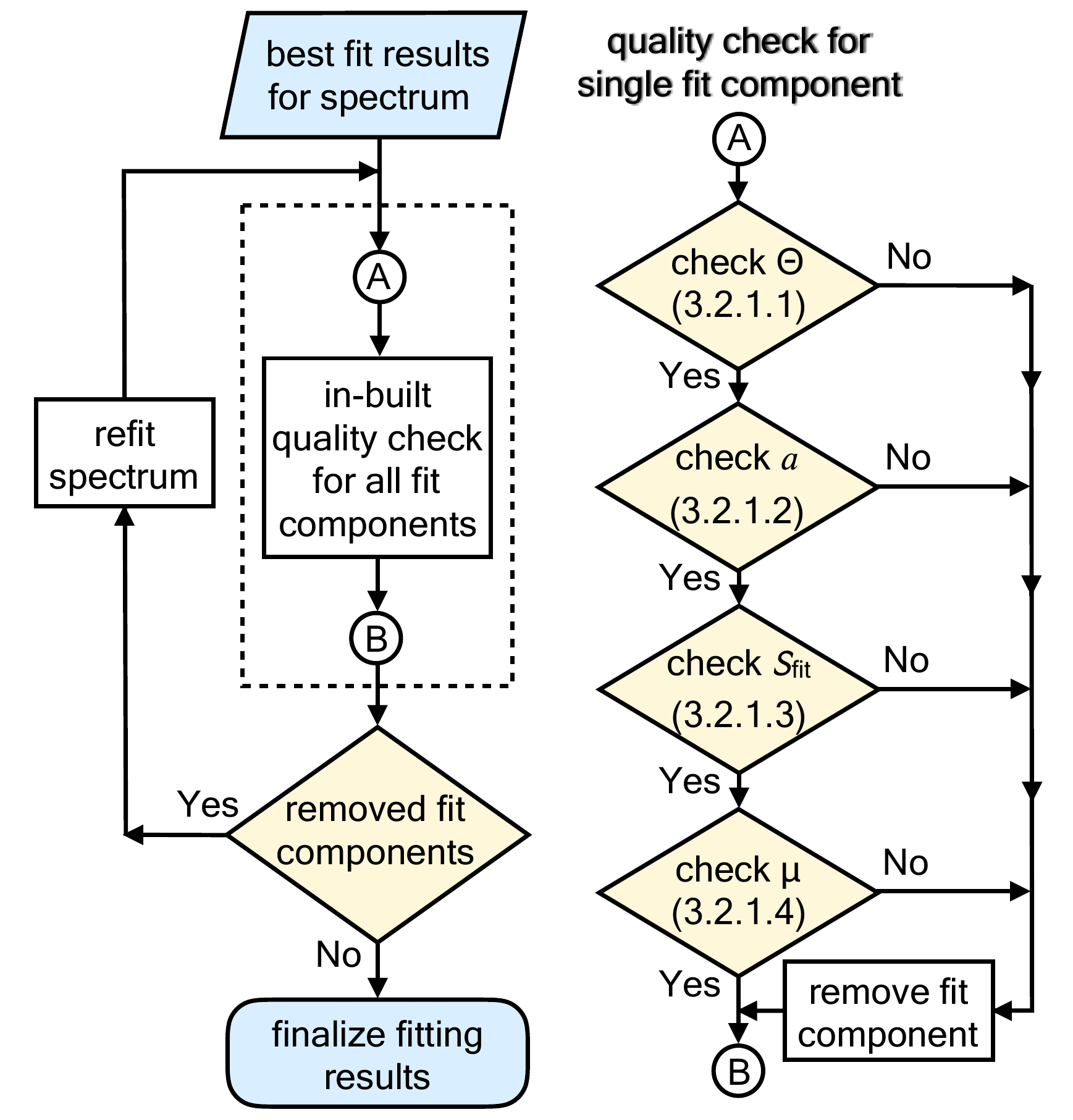}
    \caption{Flowchart outlining how the in-built quality controls from Sect.$\,\ref{sec:fwhm} - \text{Sect.}\,\ref{sec:channel-range}$ are applied to the fit results of a spectrum.
    }
    \label{fig:flowchart_in-built_quality_check}
\end{figure}

Figure$\,\ref{fig:flowchart_in-built_quality_check}$ illustrates how the in-built quality controls explained in Sect.$\,\ref{sec:fwhm} - \text{Sect.}\,\ref{sec:channel-range}$ are used to improve the fit results for a spectrum.
This refitting procedure using the in-built quality controls is applied to all fit solutions obtained in the decomposition steps of \gausspyplus\ (Sect.$\,\ref{sec:improved-fitting} \text{--} \ref{sec:phase_2}$).

The corrected Akaike information criterion and normality tests for the normalised residual are used to decide between different fit solutions of a spectrum and to assess whether a spectrum needs to be refitted, respectively; both methods are described in Sect.$\,\ref{sec:goodness-of-fit}$.

See App.~\ref{app:performance_quality_control} for a discussion about the performance of the in-built quality controls on the decomposition results of the synthetic spectra (Sect.$\,\ref{sec:synthetic-spectra}$) and the GRS test field (Sect.$\,\ref{sec:test-field}$).

\subsubsubsection{FWHM value}
\label{sec:fwhm}

If users supply limits for the lower and upper values of the FWHM ($\Theta_{\Min}$ and $\Theta_{\Max}$, respectively) all fitted components with FWHM values outside this defined range are removed.
In the \gausspyplus\ default settings $\Theta_{\Min} = 1$, which means that the FWHM value of a fit component has to be at least one spectral channel.
By default, \gausspyplus\ does not set any value for $\Theta_{\Max}$.
Users are advised to use the $\Theta_{\Max}$ parameter with caution, as it can produce artefacts in the decomposition, such as an increase of the number of fit components whose widths are close to or exactly at this predefined upper limit.

\subsubsubsection{Signal-to-noise ratio}
\label{sec:snratio}
The user-defined minimum signal S/N ratio $\snmin$ (default value: 3) is in the default settings used as the S/N threshold for: i) the original spectrum and the second derivative of the smoothed spectrum in the \gausspy\ decomposition (i.e. $\text{SNR}_{1} = \snmin$ and $\text{SNR}_{2} = \snmin$); ii) the search for new peaks in the residual (Sect.$\,\ref{sec:improved-fitting}$); iii) the search for negative residual peaks (i.e. $\snminneg = \snmin$, Sect.$\,\ref{sec:negative-residual}$); iv) the decomposition of the training set (Sect.$\,\ref{sec:training-set}$).
These parameters can all be set to different values from each other to improve the fitting results but we advise to keep them at the same value for consistency.

The minimum required amplitude values of Gaussian fit components are determined by the $\snminfit$ parameter, whose default value is half the value of $\snmin$.
All Gaussian components with $a_{i} < \snminfit\times\sigma_{\mathrm{rms}}$ will be removed from the fit.
We recommend setting $\snminfit < \snmin$ to allow fit components to also converge to an amplitude value that is below $\snmin$, as such smaller unfit peaks might otherwise negatively influence the fitting results of higher signal peaks that are close by (cf. panel b in Fig.$\,\ref{fig:schematic_flags}$). 
A smaller value for $\snminfit$ can also be beneficial if it cannot be excluded that some of the spectra might be affected by insufficient baseline subtraction effects, in which case the spectra would show a very broad but low-amplitude feature that can stretch over all spectral channels.
However, the $\snminfit$ can also be supplied by the user directly in case the default settings do not yield good results.

\subsubsubsection{Significance}
\label{sec:significance}

To further check the validity of fitted Gaussian components, we use the integrated area of the Gaussian as a proxy for the significance of the component.
Assuming that the noise properties are Gaussian (white noise), random noise fluctuations are more likely to cause narrow features with a higher amplitude than broader, extended features with a lower amplitude.
With this significance criterion we basically require that the fit components, or data peaks, have either very high intensity or are extended over a wide channel range. 

The integrated area $W_{i}$ of a Gaussian component can be calculated from its amplitude and FWHM value $\Theta$ in terms of spectral channels:

\begin{equation}
	W_{i} = a_{i}\cdot c\cdot \sqrt{2 \pi}
\end{equation}

with the parameter $c$ defined as

\begin{equation}
	c = \frac{\Theta_{i}}{2 \sqrt{2\ln 2}}.
\end{equation}

For the calculation of the significance value, we compare the area of the Gaussian component to the integrated $\rms$ interval of the channels from the interval $\mu_{i}\pm \Theta_{i}$, which gives a good approximation for the total width of the emission line:

\begin{equation}
	\significance{fit} = \frac{W_{i}}{\sqrt{2\cdot \Theta_{i}}\cdot \sigma_{\mathrm{rms}}}.
\end{equation}

The $\significance{fit}$ value is then compared to a user-defined minimum $\significance{min}$ (default value: 5) and the Gaussian component is discarded if $\significance{fit} < \significance{min}$.

This check helps to remove noise peaks that might have been fit and were not discarded in the checks for the S/N ratio.

We can use the significance parameter also as a threshold to decide whether peaks in the data are valid signal peaks.
For this estimate of the significance ($\significance{data}$), we first search for peaks in the data above the user-defined S/N threshold and then compare the integrated intensity of all positive consecutive channels belonging to this feature to the integrated $\rms$ interval of the channels spanned by this feature.
We discard the peak as a valid signal feature if $\significance{data} < \significance{min}$.

\begin{figure}
    \centering
    \includegraphics[width=\columnwidth]{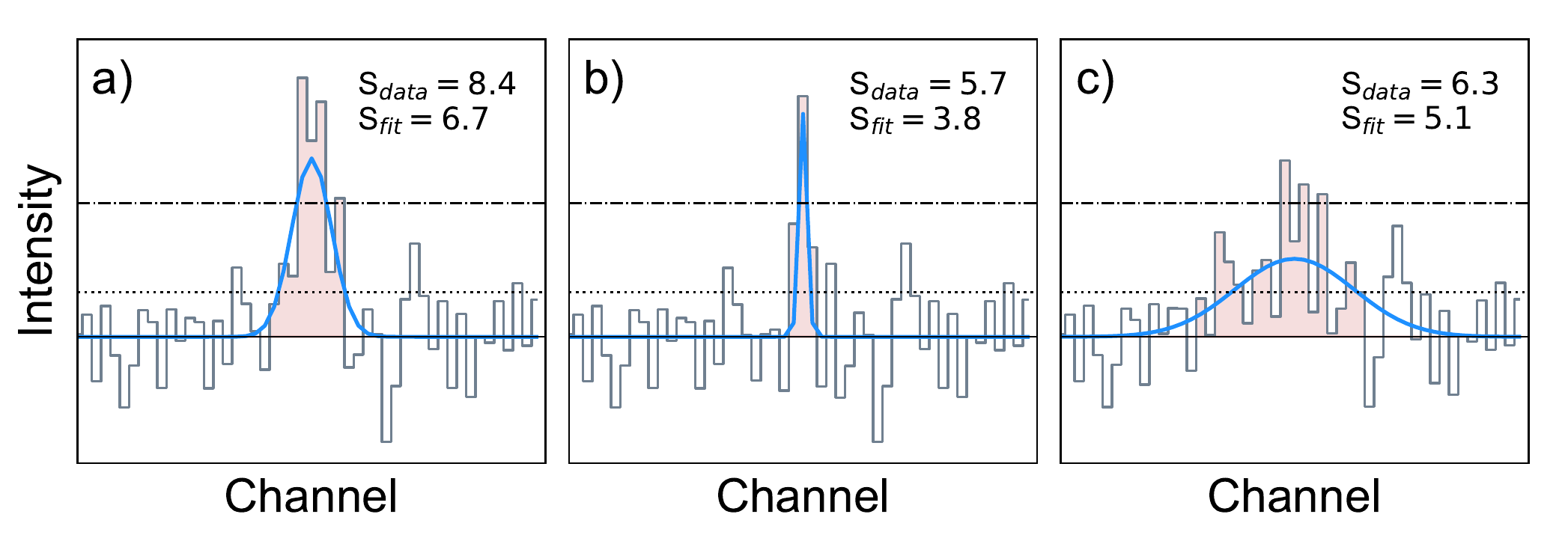}
    \caption{Calculation of the significance for Gaussian fit components ($\significance{fit}$; blue dashed lines) or peaks in the data ($\significance{data}$; red-shaded areas).
    The dotted and dash-dotted lines indicate the $\rms$ value and S/N thresholds of $3$, respectively.
    }
    \label{fig:schematic_significance}
\end{figure}

Figure~\ref{fig:schematic_significance} illustrates this significance measure for three different cases. 
Panel~(a) shows a signal peak and fit component that is very likely corresponding to a true signal, with the significance measures for the data peak and the fit both above the critical default value of $5$.
Panel~(b) shows a data peak with narrow linewidth that might be caused by random fluctuations of the noise. 
The $\significance{data}$ value of this feature passes the threshold value $\significance{\Min} = 5$, but the depicted Gaussian fit component for this data feature only has a $\significance{fit}$ value of $3.8$.
This low $\significance{fit}$ value would cause the algorithm to reject this fit component even though its peak has a high $\text{S/N}$ ratio of about $5$.
Panel~(c) shows a broader feature, which has only low S/N values.
However, since this feature is spread over more spectral channels than the feature shown in panel~(b), we would accept it based on its $\significance{data}$ value. 
With the default settings of \gausspyplus\ we would also keep the depicted fit component.
As already mentioned in Sect.$\,\ref{sec:snratio}$, it can be beneficial to keep Gaussian components with such low S/N ratios in the decomposition results, as to not negatively influence the fitting of nearby data peaks (cf. panel b in Fig.$\,\ref{fig:schematic_flags}$).

For a fitted feature or signal peak containing $N_\mathrm{feat}$ spectral channels, the $\significance{min}$ parameter implies an average S/N ratio $\langle \text{S/N} \rangle$ of

\begin{equation}
    \langle \text{S/N} \rangle = \dfrac{\significance{min}}{\sqrt{N_\mathrm{feat}}}.
\end{equation}

Users can apply this relation to judge which value for $\significance{min}$ is most suitable for their dataset.
For the default value of $\significance{min} = 5$, Gaussian fits or signal peaks spanning $4$ or $9$ spectral channels would require $\langle \text{S/N} \rangle$ values across the feature of $2.5$ and $\sim 1.7$, respectively.
See App.~\ref{app:snratio_vs_significance} for a discussion about the effects a variation of the $\snmin$ and $\significance{}$ parameters has on the decomposition results.

\subsubsubsection{Mean position outside channel range or signal intervals}
\label{sec:channel-range}
All Gaussian components whose mean positions $\mu_{i}$ are outside the channel range $\left[0, N_{\mathrm{chan}}\right]$ are automatically discarded from the fit.

If the mean position of a fit component is located outside the estimated signal intervals (Sect.$\,\ref{sec:signal-interval}$), we check the significance value of the fitted data peak $\significance{data}$ (Sect.$\,\ref{sec:significance}$).
We discard the corresponding fit component, if $\significance{data}$ is smaller than the user-defined threshold for the significance $\significance{min}$.

\subsubsubsection{Estimation of the goodness of fit}
\label{sec:goodness-of-fit}

When we fit a model to data whose errors are Gaussian distributed and homoscedastic, we can arrive at a good fit solution by minimizing the chi-squared ($\chi^{2}$), which is defined as the weighted sum of the squared residuals:

\begin{equation}
	\chi^{2} = \sum^{N}_{i=1} \frac{\left(y_{i} - Y_{i}\right)^{2}}{\sigma_{\mathrm{rms}}^{2}},
\end{equation}
with $y_{i}$ and $Y_{i}$ denoting the data and fit value at channel position $i$, respectively.

The reduced chi-square ($\chired$) value is often used as an estimate for the goodness of fit, since it also takes the sample size (in our case the number of spectral channels) and number of fit parameters into account.
$\chired$ is defined as the chi-squared per degrees of freedom:

\begin{equation}
	\chi^{2}_{\mathrm{red}} = \frac{\chi^{2}}{N - k},
\end{equation}
with $N$ being the sample size (in our case this corresponds to the number of considered spectral channels) and $k$ denoting the degrees of freedom, which in the case of a Gaussian decomposition would be three times the number of fitted Gaussian components.
It thus may seem straightforward to use the $\chired$ value to judge whether all signal peaks in a spectrum were fit, as one would expect $\chired \sim 1$ in this case.
However, as \citet{Andrae2010} pointed out, in case of non-linear models such as a combination of Gaussian functions, the exact value for $k$ cannot be reliably determined and can vary between $0$ and $N - 1$ and need not even stay constant during the fit.

The $\chired$ estimate is thus not the best metric to decide between different fit solutions for a spectrum\footnote{We thus use maps of the determined $\chired$ values only for qualitative comparisons in Sect.$\,\ref{sec:testfield-decomp}$.}.
A more suited criterion for model selection is the Akaike information criterion \citep[AIC;][]{Akaike1973}, which aims for a compromise between the goodness of fit of a model and its simplicity, by penalizing the use of a large number of fit components that do not contribute to a significant increase in the fit quality. 

The AIC is defined as

\begin{equation}
	\text{AIC} = 2k - 2\,\text{ln}(\hat{L}),
\end{equation}

with $\hat{L}$ being the maximum value of the likelihood function for the model. 
If the parameters of a model are estimated using the least squares statistic--as in our case--the AIC is given as\footnote{For a derivation of Eq.$\,\ref{eq:aic-lsqr}$ see e.g. \citet{Banks2017-aic}.}:

\begin{equation}
	\text{AIC} = N \cdot \text{ln} \left( \frac{\sum^{N}_{i=1} \left(y_{i} - Y_{i}\right)^{2}}{N}\right) + 2k.
	\label{eq:aic-lsqr}
\end{equation}

For small sample sizes, the AIC tends to select models that have too many parameters, meaning that it will overfit the data. 
Therefore a correction to the AIC was introduced for small sample sizes\footnote{\citet{Burnham1998} recommend to use the corrected AIC instead of the AIC if $N/k < 40$.
If the sample size $N \rightarrow \infty$, the corrected AIC value converges to the AIC value.} -- the corrected Akaike information criterion \citep[AICc;][]{Hurvich1989} that is defined as:

\begin{equation}
	\text{AICc} = \text{AIC} + \frac{2k^{2} + 2k}{N - k - 1}.
\end{equation}

We employ the AICc as our model selection criterion to decide between different fit solutions.
The AICc value is meaningful only in relative terms, i.e. if the AICc values for two different fit solutions are compared with each other.
In such a comparison, the fit solution with the lower AICc value is preferred as it incorporates a better trade-off between the used number of components and the goodness of fit of the model.

As an alternative to goodness of fit determinations based on the $\chired$ value, \citet{Andrae2010} suggest to check whether the normalised residuals show a Gaussian distribution. 
We implement this additional goodness of fit criterion in \gausspyplus\ by subjecting the normalised residuals to two different normality tests: the \textsc{Scipy.Stats.Kstest}, which is a two-sided Kolmogorov-Smirnov test \citep{Kolmogorov1933, Smirnov1939}; and the \textsc{Scipy.Stats.Normaltest}, which is a based on \citet{Dagostino1971} and \citet{Dagostino1973} and analyses the skew and kurtosis of the data points.
Both of these normality tests examine the null hypothesis that the residual resembles a normal distribution, as would be expected if we are only left with Gaussian noise after we subtract the fit solution from the data.
If the $p$-value from one of these test is less than a user-defined threshold (default: $1\%$), we reject the null hypothesis and will try to refit the spectrum.
We found that the combined results of these two hypothesis tests allows a robust conclusion of whether the residual is consistent with Gaussian noise (see Appendix~\ref{app:normality-tests} for more details).

\subsubsection{Quality control: optional}
\label{sec:quality-optional}
The automated checks described in the previous section should already help to reject many fit components that are not satisfying our quality requirements.
However, depending on the dataset, the user might want to flag and refit the decomposition based on more criteria, which we outline in this section.\footnote{All quality checks or flags in this section can be selected or deselected by the user.}
The quality criteria discussed in this section are used to flag and refit spectra in the improved fitting and spatially coherent refitting routines discussed in Sect.$\,\ref{sec:improved-fitting}$ and Sect.$\,\ref{sec:spatial-refitting}$, respectively\footnote{The criterion comparing the number of fit components between neighbouring spectra (Sect.$\,\ref{sec:ncomps}$) is only used in the spatially coherent refitting routines.}.
See App.~\ref{app:performance_quality_control} for a discussion about the performance of the optional quality controls on the fitting results of the synthetic spectra (Sect.$\,\ref{sec:synthetic-spectra}$) and the GRS test field (Sect.$\,\ref{sec:test-field}$).

\begin{figure}
    \centering
    \includegraphics[width=\columnwidth]{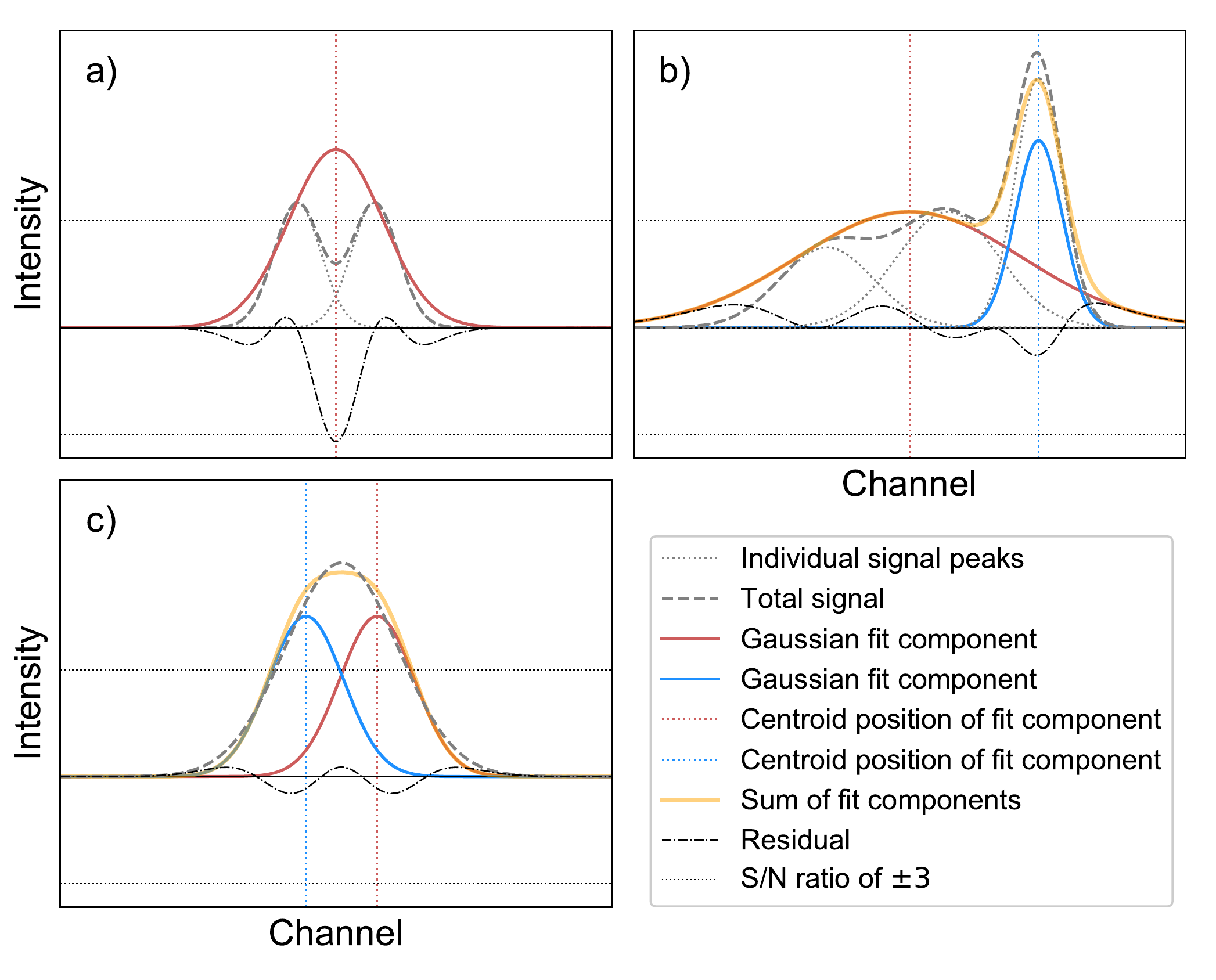}
    \caption{Optional criteria used to flag fits in the improved fitting routine and in the spatially coherent refitting stage: a) negative residual features introduced by the fit, b) broad components, c) blended components.
    }
    \label{fig:schematic_flags}
\end{figure}

\subsubsubsection{Negative peaks in the residual}
\label{sec:negative-residual}

The first quality check examines negative peaks in the residual, since these can indicate a poor fit.
Panel~(a) in Fig.~\ref{fig:schematic_flags} presents a scenario in which a double peaked profile (shown in dashed grey lines) is fit with a single Gaussian component (red line), leading to a significant negative peak in the residual (dash-dotted black line) at the position between the two data peaks.
The search for negative peaks in the residual can be controlled by the user with the $\snminneg$ parameter, which defines the minimum S/N ratio that the negative peak has to have (in the default settings $\snminneg = \snmin$).
To be flagged as a negative residual feature, a negative peak has to satisfy $\vert y_{i} - Y_{i} \vert \geq \snminneg \times \rms$, with $y_{i}$ and $Y_{i}$ denoting the data and corresponding fit value at channel position $i$.
This requirement takes into account that negative peaks could have already been present in the original spectrum and requires that a significant part of the negative peak was introduced by the fit.

\subsubsubsection{Gaussian components with a broad FWHM}
\label{sec:broad-components}
It can occur that a single, broad Gaussian component is fit over multiple peaks in the spectrum, which can be an undesired property. 
A broad feature can be caused by peaks being close to the noise limit, multiple blended components, or issues in the data reduction, e.g., insufficient baseline corrections or unsubtracted continuum emission.
Panel~(b) in Fig.~\ref{fig:schematic_flags} shows an example of a broad component that was incorrectly fit over multiple data peaks without introducing significant residual features as in panel~(a).
This would lead to wrong estimates of the total number of components present in this spectrum, a severe overestimate of the linewidth for the two smaller peaks incorrectly fit with one component, and an underestimate of the amplitude of the rightmost component.
The example presented in panel~(b) also highlights why it can be beneficial to set the required minimum S/N threshold for fitted component $\snminfit$ to lower values than the S/N threshold for data peaks $\snmin$ (see Sect.$\,\ref{sec:snratio}$).
If $\snminfit$ were set equal to $\snmin$, the fit component for the leftmost peak in panel~(b) will get discarded, forcing the fit of a broad component over the two leftmost peaks to minimize the residual.

Unfortunately, it can be difficult to set a maximum allowed FWHM value for the Gaussian components, as the range of expected values in the data may not be known.
Setting a strict limit for the maximum FWHM value might also lead to a large number of components which have their linewidth equal to the limiting value.
To prevent such an undesired effect, we flag a component as broad if it is broader by a user-defined factor $f_{\Theta,\,\Max}$ (default value: 2.) than the second broadest fit component.
This obviously does not work for spectra with only one Gaussian component fit, but this case is taken into account during the spatially coherent refitting (Sect. \ref{sec:phase_1}).

Another physical cause for the broadening of the lines could be opacity broadening, which is especially relevant for optically thick emission lines such as the $^{12}$CO(1-0) rotational transition \citep{Hacar2016-opacity}. 
In case the user expects opacity broadening for a significant number of spectra in the dataset, we recommend to not flag or refit broad fit components.

\subsubsubsection{Blended Gaussian components}
\label{sec:blended-comps}

We define a Gaussian component $i$ as blended with a neighbouring component $j$, if the distance between their mean positions $\mu_{i}$ and $\mu_{j}$ is less than the minimum required separation $\mu_{\mathrm{sep}}$.
This minimum required separation is determined by multiplying the lower FWHM value of the two components with a user-defined factor $f_{\mathrm{sep}}$:

\begin{equation}
	\mu_{\mathrm{sep}} = f_{\mathrm{sep}} \times \text{min}(\Theta_{i}, \Theta_{j}).
\end{equation}

The default value of $f_{\mathrm{sep}}$ is $1 / \sqrt{2\, \text{ln}\,2}$.
This value was chosen so that the required separation between two identical Gaussian components defaults to two times their standard deviation.
If two identical Gaussian fit components are separated by a distance larger than two times their standard deviation, their combined signal would have a local minimum between the two peak positions, which we define as a requirement for well resolved Gaussian fit components. 
Panel~(c) in Fig.~\ref{fig:schematic_flags} shows a case in which the minimum separation between the peak positions of the two identical Gaussian fit components is not reached.
The combined signal of the fit components (shown in orange) shows no local minimum between the peak positions and a single Gaussian component that corresponds to the sum of the two individual components would thus be evaluated as a better fit.

Without additional information from neighbouring spectra it can be very difficult to reliably conclude whether a two-component fit is a better choice than the fit of a single component.
If this quality criterion is selected by the user we will therefore always try to replace two blended components with a single bigger component in the improved fitting routine (Sect.~\ref{sec:improved-fitting}), where each spectrum is still treated independently.

\subsubsubsection{Residuals not normally distributed}
\label{sec:normaltest}

This flag checks whether the normalised residuals show a Gaussian distribution. 
We subject the normalised residual to two different tests for normality (see Sect.$\,\ref{sec:goodness-of-fit}$ for more details), with the null hypothesis that the residual values are normally distributed.
We reject this null hypothesis if the $p$-value of at least one of the normality tests is less than a user-defined threshold (default: $1\%$), in which case the spectrum gets flagged.

\subsubsubsection{Different number of components compared to neighbouring spectra}
\label{sec:ncomps}

\begin{figure}
    \centering
    \includegraphics[width=\columnwidth]{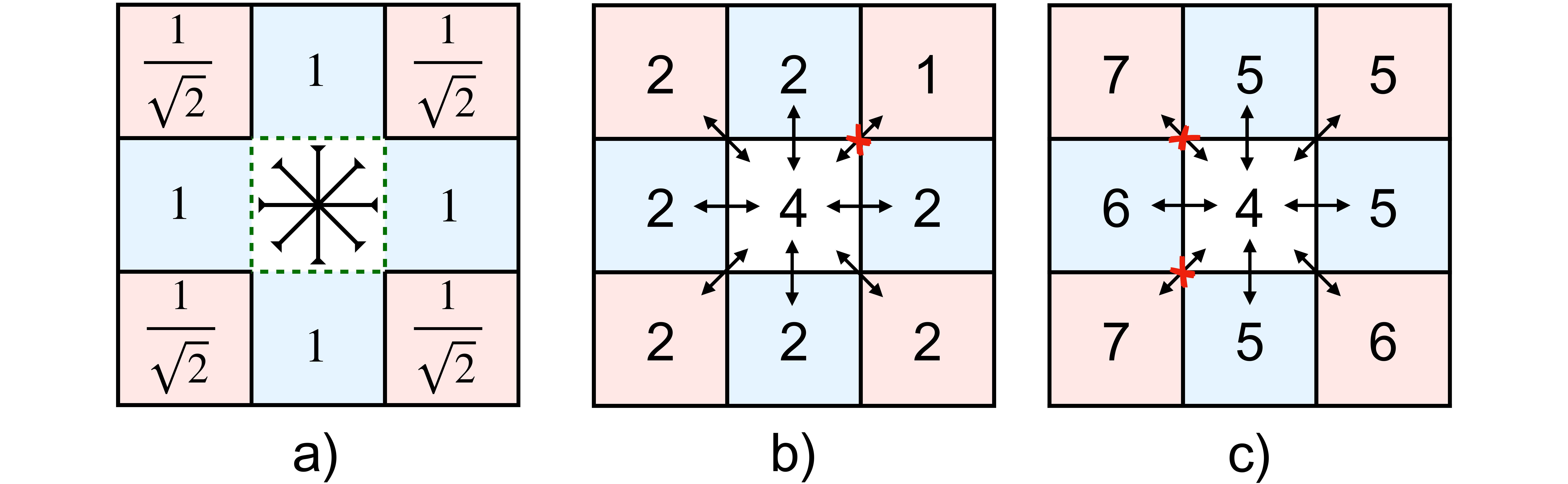}
    \caption{Illustration of the flagging of spectra based on their number of components with the default settings of our algorithm.
    Each $3\times3$ square shows the central spectrum (in white) and the surrounding immediate neighbours colored according to their weights.
    Panel~(a) shows the weights we apply to each neighbouring fit solution to calculate their weighted median.
    Panel~(b) and (c) show two cases where the fitted number of components of the central spectrum would be flagged as incompatible with the fitted number of components of their neighbours.
    See Sect.$\,\ref{sec:ncomps}$ for more details.
    }
    \label{fig:schematic_ncomps}
\end{figure}

This quality criterion compares the number of fitted Gaussian components of a spectrum with its immediate neighbouring spectra.
We include the fit solutions of all neighbouring spectra in this comparison, irrespective of whether they were already flagged by another optional quality criterion.

There are two conditions for which a spectrum can be flagged by this check: 
\begin{itemize}
    \item
    The number of components $\Ncomp$ in the spectrum is different by more than a user defined value $\Delta N_{\Max}$ (default value: 1) from the weighted median number of components determined from all its immediate neighbours.
    For a sequence of $n$ ordered elements $x_{1}, x_{2}, ..., x_{n}$ with corresponding positive weights $w_{1}, w_{2}, ..., w_{n}$ that sum up to $w_{\mathrm{tot}}$, the weighted median is defined as the element $x_{k}$ for which $\sum^{k - 1}_{i=0} w_{i}  < 0.5\times w_{\mathrm{tot}}$ and $\sum^{n}_{i=k + 1} w_{i}  < 0.5\times w_{\mathrm{tot}}$.
    Panel~(a) in Fig.~\ref{fig:schematic_ncomps} shows the weights we apply to the immediate neighbours, which are inversely proportional to their distance to the central spectrum.
    \item
    The spectrum shows differences in $\Ncomp$ towards individual neighbours that exceed a user defined value $\Delta N_{\mathrm{jump}}$ (default value: 2).
    We flag a spectrum if these differences occur towards more than $N_{\mathrm{jump}}$ (default value: 1) of its neighbouring spectra.
\end{itemize}

We illustrate this criterion in Fig.~\ref{fig:schematic_ncomps} for two cases and the default settings of \gausspyplus.
Panel~(b) shows an instance where the number of components of the central spectrum shows no component jumps $>2$ to any of its neighbours. 
However, we would still flag the central spectrum for its number of fitted components, since it differs by more than $\Delta N_{\Max}$ to the weighted median number of components as inferred from the neighbouring fit solutions (2 components).
Panel~(c) shows the opposite case, where the median number of components of 5 is still compatible with the actual number of components but the fit solution of the central spectrum would be flagged as inconsistent with its neighbours as it shows two component jumps $>2$ with two of its neighbours.

\subsubsection{Improved fitting routine}
\label{sec:improved-fitting}

The improved fitting routine in \gausspyplus\ aims to improve the fitting results of the original \gausspy\ algorithm via the use of the quality controls described in Sect.$\,\ref{sec:checks}$ and Sect.$\,\ref{sec:quality-optional}$.
The original version of \gausspy\ hands over its initial guesses to a least squares minimization routine without restricting the fitting parameters, apart from a requirement of positive amplitude values. 
This means that the individual Gaussian components are allowed to freely vary their FWHM and mean positions.
Moreover, the number of Gaussian components is set and fixed by the initial guesses, so if \gausspy\ determined that the fit should contain a certain number of Gaussian components, it will try to fit all those components even if one of them does not contribute to improving the fit or is making the fit worse. 
This unrestricted fitting can lead to unphysical results or conflicting fit solutions between neighbouring spectra (see the quality flags discussed in Sect.$\,\ref{sec:quality-optional}$).

\begin{figure}
    \centering
    \includegraphics[width=\columnwidth]{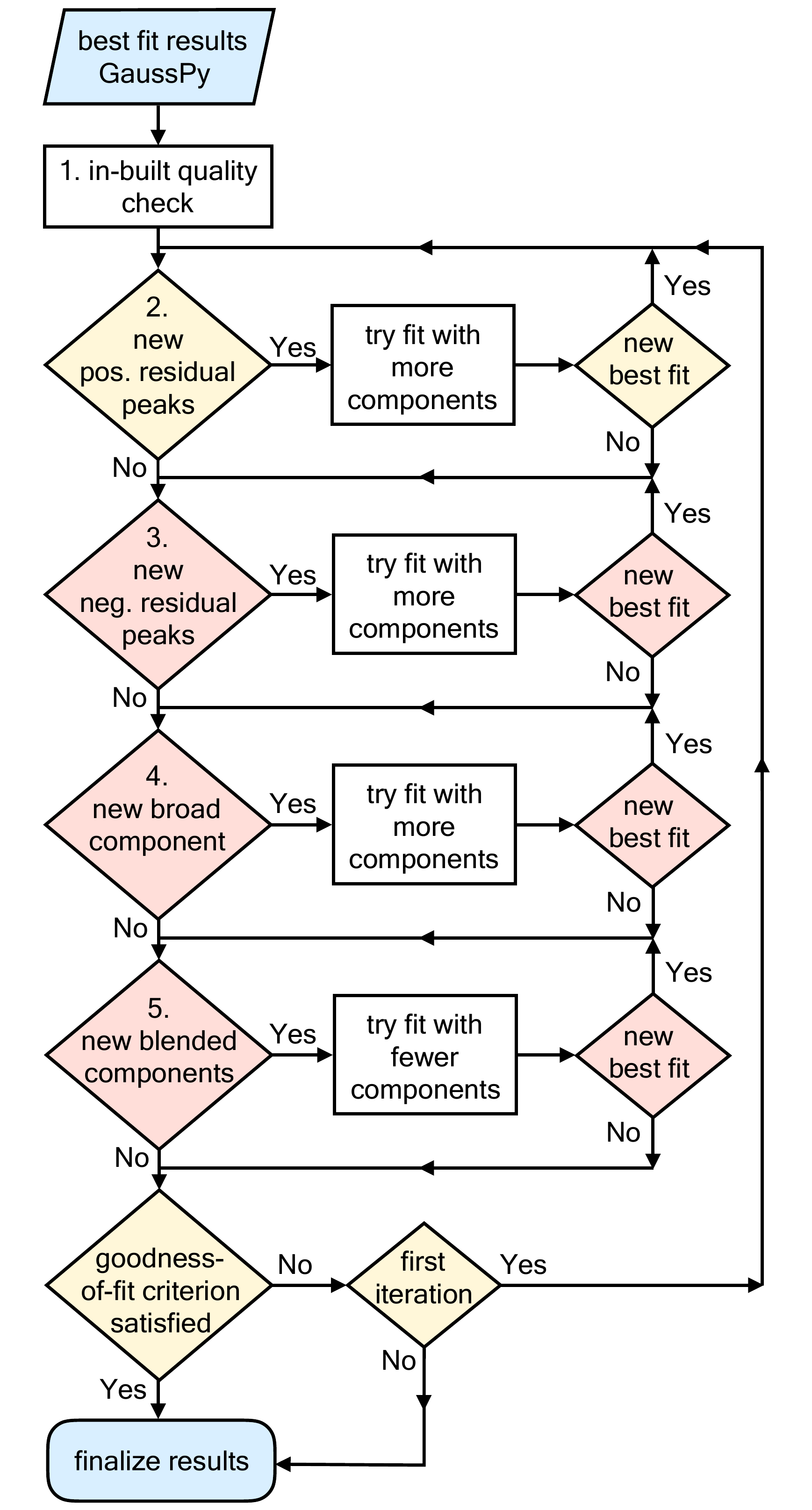}
    \caption{Flowchart outlining the basic steps of our improved fitting routine.
    The conditional stages in red correspond to optional stages that can be selected by the user.
    See Sect.$\,\ref{sec:improved-fitting}$ for more details.
    }
    \label{fig:flowchart_improved_fitting}
\end{figure}

The general idea of our routine is to try to improve the fit based on the residual and optional user-selected quality criteria (Sect.$\,\ref{sec:negative-residual} \text{--} \ref{sec:blended-comps}$). 
This improved fitting phase is applied to every spectrum.
The steps of this routine proceed as follows (see also Fig.~\ref{fig:flowchart_improved_fitting}):

\begin{enumerate}
    \item
    Check the best fit result of \gausspy\ with the quality criteria outlined in Sect.$\,\ref{sec:fwhm} \text{--} \ref{sec:channel-range}$ (see Fig.$\,\ref{fig:flowchart_in-built_quality_check}$).
    All Gaussian components not satisfying any of these criteria are removed from the best fit solution of \gausspy\ and the spectrum is refit with the remaining fit components; this procedure gets repeated until all of the leftover fit components satisfy all quality criteria.
    
    \item
    Try to iteratively improve the fit by adding new Gaussian components based on positive peaks in the residual of the best fit solution.
    Requirements for the acceptance of residual peaks as additional Gaussian component candidates are that: i) the maximum value of the residual peak is higher than $\snmin$; ii) the consecutive positive spectral channels of the residual peak satisfy the significance criterion $\significance{data} \geq \significance{min}$ outlined in Sect.$\,\ref{sec:significance}$. 
    If one or multiple peaks are found in the residual that satisfy these requirements for being new Gaussian component candidates, a refit of the spectrum is performed by adding all of these new candidates.
    For the refit, the initial Gaussian parameter guesses for the accepted residual peaks are set to: the maximum positive value of the residual peak for the amplitude; the spectral channel containing the maximum positive value of the residual peak for the mean position; the number of consecutive positive channels of the residual peak for the FWHM parameter.
    After a successful pass of all quality criteria, we adopt the new fit as the new best fit if its AICc value is lower than the AICc value of the previous best fit solution.
    If a new best fit was chosen, a new iteration with a search for peaks in the residual of the new best fit solution continues.
    We proceed to the next step if no new positive peaks are found in the residual or no new best fit could be assigned.
    
    \item
    Optional: Check whether a negative residual feature (Sect.$\,\ref{sec:negative-residual}$) was introduced by the fit components.
    This check is only performed if it is the first pass through the main loop or a new best fit was assigned.
    Negative residual features can be indicative of a poor fit with multiple signal peaks fit by a single broad component.
    In case such a feature is present, we try to replace the broadest Gaussian component at the place of the residual feature with two narrower components.
    The initial guesses for the two new narrow components are estimated from the residual obtained if the broad component is removed, which proceeds in a similar way as in the previous step.    
    If the new fit with the two narrow components passes all quality requirements and its AICc value is lower than the AICc value of the current best fit, we will assign it as the new best fit and repeat the search for negative residual peaks. 
    In case multiple negative residual features are present in a spectrum, we deal with the features in order of increasing negative residual values, i.e. we will first try to replace the Gaussian component causing the residual feature that contains the most negative value.
    We proceed to the next step if no new negative peaks are found in the residual or no new best fit could be assigned.
    
    \item
    Optional: Check for broad components (Sect.$\,\ref{sec:broad-components}$).
    If a broad Gaussian component is present we will try to replace it in this step with multiple narrower components.
    The number of narrow components and their initial parameter guesses are estimated from the residual we get if the broadest component is removed from the fit.
    If this results in a new best fit we will repeat this procedure with the resulting next broadest component.
    We proceed to the next step if no excessively broad component is identified anymore, or no new best fit could be assigned.
    
    \item
    Optional: Check for blended components (Sect.$\,\ref{sec:blended-comps}$).
    If this is the case we will try to refit the spectrum by in turn omitting one of the blended components and checking whether the AICc value of the resulting best fit is better than the AICc value of the currently best fit.
    Blended components are omitted in order of increasing amplitude value, i.e. we will first try to refit the spectrum by excluding the blended component with the lowest amplitude value.
    If no new best fit is assigned or no blended components are present in the spectrum we exit the improved fitting procedure and finalize the fitting results if the normalised residuals of the best fit solution show a normal distribution, which we verify with two different normality tests (Sect.$\,\ref{sec:goodness-of-fit}$).
    If this is not the case, we repeat the whole improved fitting procedure beginning with step 2, the search for positive peaks in the residual.
\end{enumerate}

We tested the performance of our improved fitting routine on synthetic spectra and found that it yields a significant improvement in the decomposition compared to the original \gausspy\ algorithm.
In Sect.$\,\ref{sec:synthetic-spectra}$ and App.~\ref{app:test_decomposition} we give a detailed discussion about the decomposition results for the synthetic spectra.

\subsection{Spatially coherent refitting}
\label{sec:spatial-refitting}

So far all steps of the fitting routine treated each spectrum separately and independently from its neighbours.
Here we describe a new routine that aims to also incorporate the information from neighbouring spectra and tries to refit spectra according to this information.
Our routine proceeds iteratively and starts from the fitting results obtained with the method outlined in the previous section (Sect.$\,\ref{sec:improved-fitting}$).
Note that this is different to algorithms such as \scousepy, which first start with an averaged spectrum and use its decomposition result to fit the individual spectra.
We proceed in a reverse manner: we first produce a sample of high quality fits for each spectrum without regarding their neighbours and then refit them, if it is deemed to be necessary, using the fit solutions of the immediate neighbouring spectra.\footnote{In the current implementation of \gausspyplus\ we only consider directly neighbouring spectra, whereas algorithms such as \scousepy\ allow the user to also include information from larger spatial areas.}

The spatial refitting proceeds in two phases.
In Phase 1, we try to improve the fit solutions based on a flagging system, for which the fitting results from the previous stage are checked and flagged according to user-selected criteria.
We subsequently try to refit each flagged spectrum with the fit solutions from its neighbours and thereby already introduce a limited form of local spatial coherence.
In Phase 2, we use a weighting system to try to enforce spatial coherence more globally. 
We check for the entire dataset if the Gaussian components of each spectrum are spatially consistent with the neighbouring spectra, by comparing the centroid positions of the Gaussian components.
We then try to refit spectra whose Gaussian components show centroid velocity values that are inconsistent with the fit solutions from neighbouring spectra.

\subsubsection{Phase 1: Refitting of the flagged fits}
\label{sec:phase_1}

\begin{figure}
    \centering
    \includegraphics[width=\columnwidth]{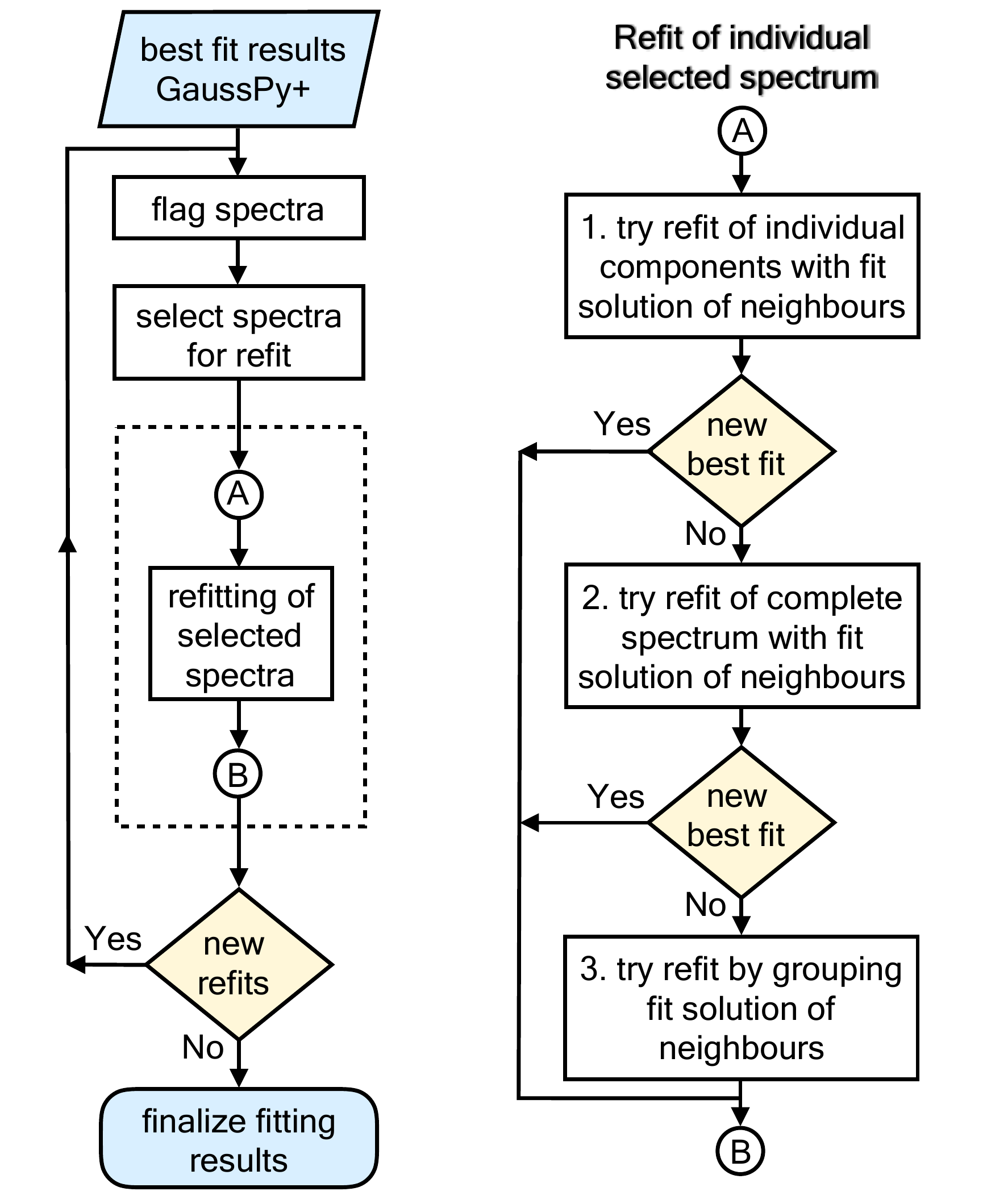}
    \caption{Flowchart outlining the steps of the first phase of spatially coherent refitting routine.
    See Sect.$\,\ref{sec:phase_1}$ for more details.}
    \label{fig:flowchart_spatial_refitting}
\end{figure}

The steps of the first phase of the spatially coherent refitting method are outlined in Fig.~\ref{fig:flowchart_spatial_refitting}.
The idea here is to determine which of the spectra need to be refit based on flags set by the user.
We try to refit all spectra that show features that do not satisfy the quality requirements imposed on the fits (these are also retained as flags indicating bad quality fits in case the spectrum cannot be successfully refit).
Depending on the dataset, the user might not always want to flag or refit spectra that show one or more of these features. Therefore, all of the following flags can be chosen as required by the user.
In the current version of \gausspyplus\, the following features can be flagged by the user:
\begin{enumerate}[(i)]
    \item
    $\flag{neg.\,res.\,peak}{}$: The presence of negative peaks in the residual (Sect.$\,\ref{sec:negative-residual}$).
    \item
    $\pazocal{F}_{\Theta}$: Gaussian components with a broad FWHM value (Sect.$\,\ref{sec:broad-components}$). 
    For the spatial refitting we additionally flag a component as broad if it is broader by a user-defined factor ($f_{\Theta,\,\Max}$) than the broadest component in more than half of its neighbours.
    \item
    $\flag{blended}{}$: The presence of blended Gaussian components in the fit (Sect.$\,\ref{sec:blended-comps}$).
    \item
    $\flag{residual}{}$: Fits whose normalised residual values do not pass the tests for normality (Sect.$\,\ref{sec:normaltest}$).
    \item
    $\pazocal{F}_{N_{\mathrm{comp}}}$: The number of components $\Ncomp$ differs significantly from its neighbours (see Sect.$\,\ref{sec:ncomps}$).
\end{enumerate}

Flags (i) -- (v) are recomputed in each new iteration.

We then try to refit each flagged spectrum with the help of one or all of the best fit solutions of its neighbouring unflagged spectra.
In the default settings of the algorithm we try to refit all flagged spectra by using fit solutions from unflagged neighbouring spectra.
At maximum, this provides eight new different fit solutions for the flagged spectrum (if all of its eight neighbouring spectra are unflagged).
If there are multiple unflagged neighbours, they get ranked according to their $\chired$ values, and the neighbouring fit solution with the lowest $\chired$ value is used first.

It is also possible to only flag fit solutions without refitting them, though this has to be selected by the user.
This might be useful, for example, if users want to exclude neighbouring fit solutions whose normalized residuals did not satisfy the normality tests as templates for the refit but do not want to refit these spectra themselves.

\begin{figure}
    \centering
    \includegraphics[width=\columnwidth]{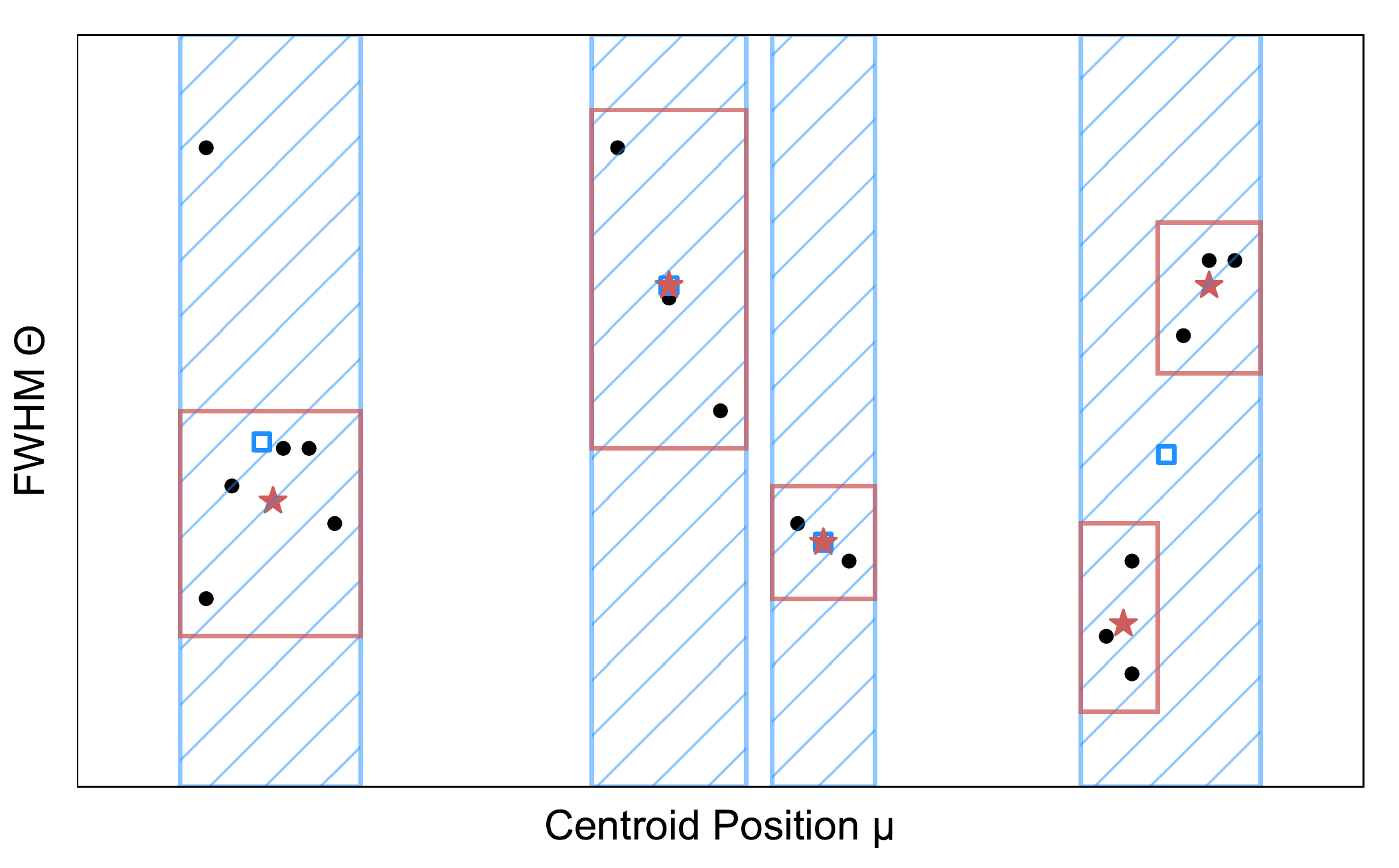}
    \caption{Illustration of the grouping routine.
    Black points indicate centroid ($\mu$) and FWHM  ($\Theta$) values of Gaussian components from the best fit solutions of unflagged neighbouring spectra.
    Blue hatched areas indicate the results of the first grouping, in which data points are only separated according to their $\mu$ values.
    Red rectangular areas mark the results of the second grouping in which data points are additionally separated according to their $\Theta$ values.
    Blue squares and red stars indicate the initial guesses for the refitting with the first and second grouping approach, respectively.}
    \label{fig:clustering_schematic}
\end{figure}

The refitting of an individual flagged spectrum proceeds in the following way (see right part of Fig.$\,\ref{fig:flowchart_spatial_refitting}$):
\begin{enumerate}
    \item
    Use the fit solutions of unflagged neighbouring spectra to refit individual components of the flagged spectrum.
    Spectra that are flagged as having negative residual features, broad, or blended components might show a good fit solution apart from the flagged features.
    Therefore we first try to replace the Gaussian components of such flagged features by using the Gaussian components of neighbouring unflagged fit solutions that cover the same region in the spectrum as new input guesses. 
    The refit attempt is then performed for the entire spectrum by combining these new initial guesses from a neighbouring fit solution with the remaining fit components of the old fit solution of the spectrum that were not affected by the flagged feature.
    If multiple regions of a spectrum are flagged with different flags we will try to refit the flagged features in the order of: negative residual feature, broad component, and blended components. 
    As soon as a flagged feature is successfully refit we stop the refitting iteration, even if other flagged features should still be present in the spectrum.
    We impose no selection criteria on the neighbouring Gaussian components, i.e. we will in turn use all unflagged neighbouring fit solutions as new initial guesses, starting with the fit solution that has the lowest $\chired$ value.
    If one of the input guesses of the unflagged neighbours leads to a new improved fit the refitting of the flagged spectrum is successfully terminated, otherwise we proceed with the next step.
    
	\item
	Use the fit solutions of unflagged neighbouring spectra to refit the complete flagged spectrum.
	In this step all fit components of a neighbouring spectrum are used as new input guesses for refitting the entire spectrum.
	We again loop through all unflagged neighbouring fit solutions, starting with the one that has the lowest $\chired$ value.
    The refitting of the flagged spectrum is successfully terminated as soon as one of the neighbouring fit solutions leads to a new improved fit, otherwise we continue with the next step.
    
    \item
    Obtain a new set of fit parameters from the fit solutions of all unflagged neighbouring spectra, by grouping and averaging the parameters of all their Gaussian components in a parameter space spanned by the fitted velocity centroid and FWHM values.
    Figure~\ref{fig:clustering_schematic} illustrates how the grouping proceeds. 
    First, the grouping is only performed for the $\mu$ values (blue hatched areas).
    The requirement for group membership is that data points are at maximum located at a distance of $\Delta\mu_{\Max}$ (default value: 2 channels) from any other point of this group.
	We require a minimum group membership of two points, which means that single points that do not belong to any group are treated as outliers.
    The blue points and hatched areas show the new fitting constraints used for the refitting.
    As initial guesses for the amplitude, FWHM value and centroid position we use the corresponding average values of all the data points belonging to a group.
    The fitting constraints for the centroid positions are based on the extent of the groups along the $\mu$ axis.
    For each amplitude value we require that it has a positive value and set its maximum limit to the maximum data point in the original spectrum that occurs in the range that encompasses all $\mu$ values of this group multiplied by a user-defined factor $f_{a}$.
    FWHM values are not allowed to be smaller than the user-defined parameter $\Theta_{\Min}$ but there is no upper constraint for their values.
    If this first grouping approach does not lead to a successful refit, we use a second grouping approach that additionally groups the data points according to their FWHM values (red boxes in Fig.~\ref{fig:clustering_schematic}).
    A group membership for a data point is established if its $\mu$ and $\Theta$ values are at maximum located at a distance of $\Delta\mu_{\Max}$ (default value: 2 channels) and $\Delta\Theta_{\Max}$ (default value: 4 channels), respectively, from any other point of this group.
    The points in each group are then averaged in a similar way as for the first grouping approach and supplied as new fit parameters for the refitting.
\end{enumerate}

Grouping only by the centroid values has the advantage that it will try to fit the spectrum with the least amount of components inferred from its neighbours.
A disadvantage is that outliers in the FWHM regime can negatively influence the initial fit values.
The second grouping approach should be able to deal better with the fidelity of the data even though some of the initial guesses for Gaussian fits could overlap heavily.

For the decision of whether to accept a refit as the new fit solution we define a total flag value $\flag{tot}{}$ that increases by one for each of the user-selected flags the fit solution does not satisfy.
For the proposed new fit solutions, the total flag value increases in addition by one for each flagged criterion that got worse than in the current best fit solution, i.e. for an increase in the number of blended components or negative residual features, broad components that got broader, smaller $p$-values for the null hypothesis testing for normally-distributed residuals, and a greater difference in the number of components compared to the neighbouring fit solutions.

In the stage where all spectra were treated independently (Sect.$\,\ref{sec:improved-fitting}$), the decision to accept a fit model was made via the AICc.
In the spatial refitting phase this decision is mainly guided by the comparison of the total flag value of the new fit solution ($\flag{tot}{new}$) with the old best fit solution ($\flag{tot}{old}$).
There are three possible scenarios:
\begin{itemize}
    \item
	$\flag{tot}{new} > \flag{tot}{old}$. 
	In this case the new fit solution is rejected.
	\item
	$\flag{tot}{new} = \flag{tot}{old}$.
	The new fit solution is accepted if its AICc value is smaller than the AICc value for the best fit solution we started out with.
	\item
	$\flag{tot}{new} < \flag{tot}{old}$.
	The new fit solution is accepted if the data points of the normalised residual pass the normality tests.
\end{itemize}

In the last case we have to test whether new fit solutions incorrectly decreased $\flag{tot}{new} $ by removing valid fit components.
For example, both $\flag{blended}{}$ and $\pazocal{F}_{\Theta}$ could be reduced by one if a broad component is deleted.
To prevent such incorrect fit solutions we require that the normalised residual resembles a Gaussian distribution, which we check with two different normality tests (see Sect.$\,\ref{sec:goodness-of-fit}$).
The null hypothesis of normally distributed residual values gets rejected if the $p$-value is less than a user-defined threshold (default: $1\%$), in which case we do not accept the new fit solution.

\subsubsection{Phase 2: Refitting of the spatially incoherent fits}
\label{sec:phase_2}

In the second phase of the spatially coherent refitting, we check for coherence of the centroid positions of the fitted Gaussian components for all spectra. 
The motivation for this step is that we would expect coherence in the centroid positions of the fitted Gaussian components for resolved extended objects, especially for oversampled observations where the size of a pixel is smaller than the beam size or resolution element.

\begin{figure}
    \centering
    \includegraphics[width=\columnwidth]{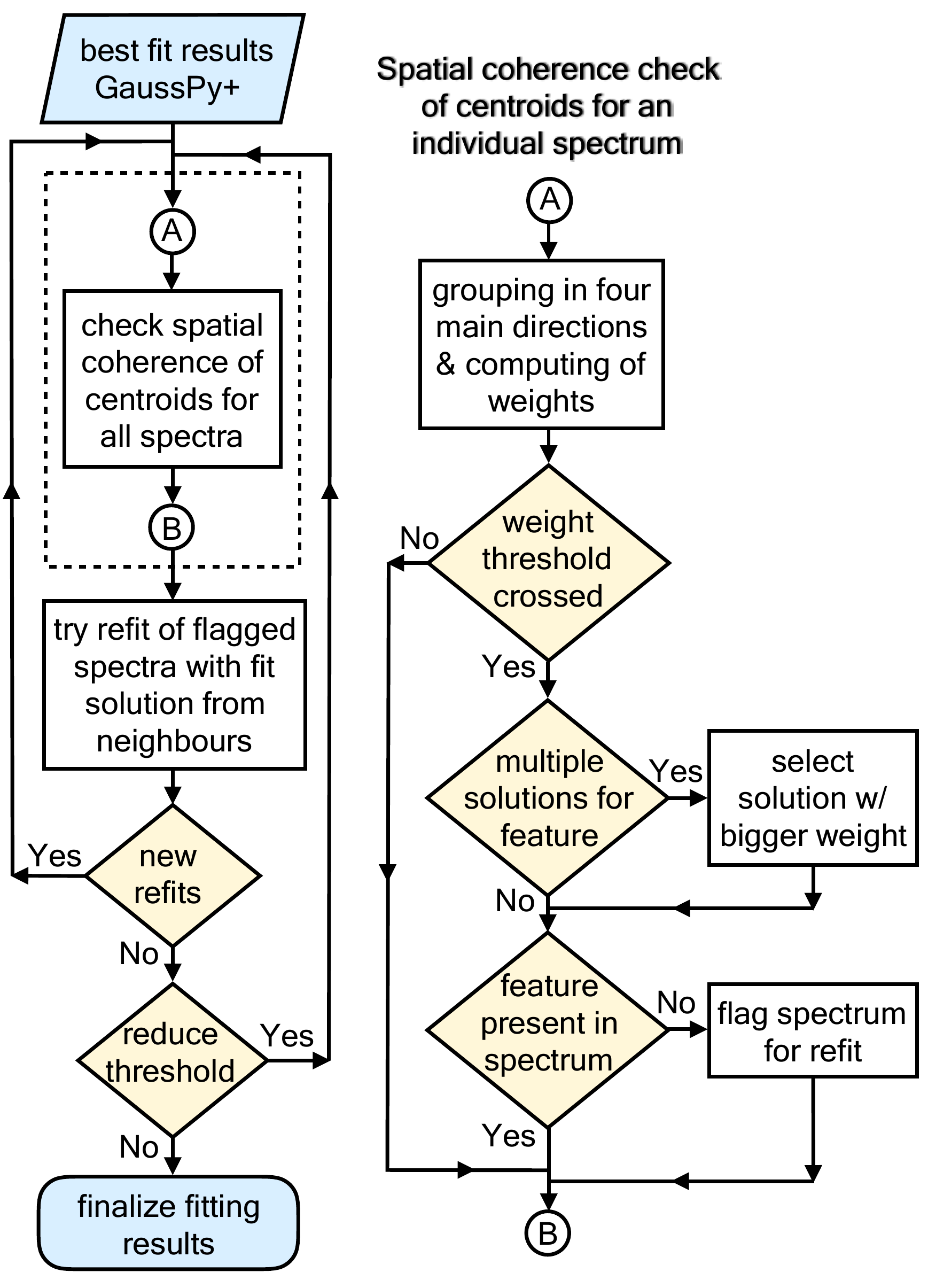}
    \caption{Flowchart outlining the basic steps of the second phase of the spatial refitting routine.
    See Sect.$\,\ref{sec:phase_2}$ for more details.}
    \label{fig:flowchart_spatial_refitting_p2}
\end{figure}

\begin{figure}
    \centering
    \includegraphics[width=\columnwidth]{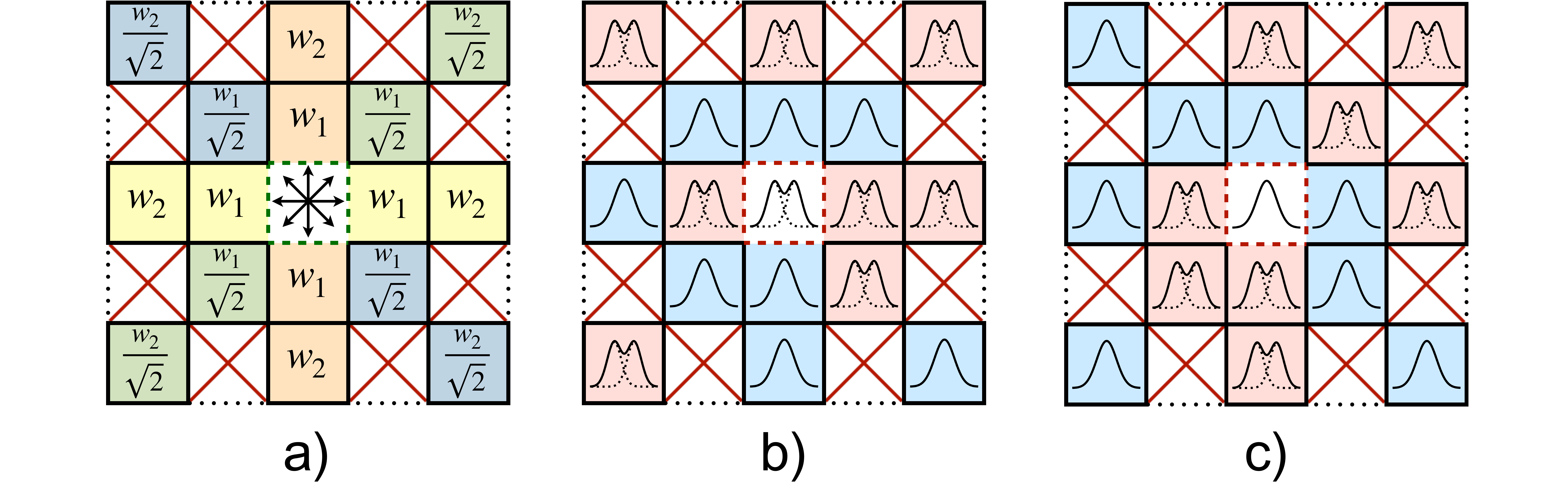}
    \caption{Illustration of phase two of the spatial refitting routine of \gausspyplus.
    Each $5\times5$ square shows a central spectrum (in white) and its surrounding neighbours.
    White squares that are crossed out are not considered.
    The left panel shows the principal directions for which we check for consistency of the centroid position and shows the applied weights $w_{1}$ and $w_{2}$ attached to the neighbouring spectra.
    The middle and right panels show two different example cases with simple fits of one and two Gaussian components shaded in blue and red, respectively.
    Based on the fits of the neighbouring spectra we would try to refit the central spectrum in panel~(b) with one Gaussian component, whereas the central spectrum in panel~(c) is already consistent with what we would expect from our spatial consistency check of the centroid positions.
    See Sect.$\,\ref{sec:phase_2}$ for more details.
    }
    \label{fig:schematic_phase_2}
\end{figure}

The spatial consistency check, in which we determine whether a spectrum should contain Gaussian components in specific spectral ranges based on the fitting results from neighbouring spectra, proceeds in an iterative way. 
For that, we use 16 neighbours along the 4 main directions (see panel a in Fig.$\,\ref{fig:schematic_phase_2}$)\footnote{This number is reduced accordingly in case neighbouring spectra are masked out or the central spectrum happens to be close to or at the border of the image.}.
For simplicity we do not consider the off-diagonal pixels.

Users can specify the ratio of the weight of the closest neighbour ($w_{1}$) to the weight of the neighbour located one pixel farther away ($w_{2}$) with the parameter $f_{w} = w_{1} / w_{2}$ (default value: 2).
In the default settings the contribution of the neighbours is inversely proportional to their distance to the central spectrum (see left panel of Fig.~\ref{fig:schematic_phase_2}).
The weights $w_{1}$ and $w_{2}$ are normalized so that $2 w_{1} + 2 w_{2} = 1$, which means that along the horizontal and vertical direction the weights sum up to a value of 1.
Setting the parameter $f_{w}$ to higher values than the default value has the effect of decreasing the contribution of neighbours that are located at a distance of two pixels and thus puts even more emphasis on the closest neighbours.

In case the central spectrum has Gaussian components whose centroid positions do not match with what would be expected from the fit results of its neighbouring spectra, we try to refit the spectrum with a better-matching fit solution from one of its neighbours.

In the following, we outline the spatial consistency check of the centroid positions in more detail (see also Fig.~\ref{fig:flowchart_spatial_refitting_p2}):

\begin{enumerate}
    \item
    Check for a consistent feature in the neighbouring spectra along any of the main directions indicated in the left panel of Fig.~\ref{fig:schematic_phase_2}.
    For each of the four directions, we group the centroid positions of the fitted Gaussian components as described in section Sect.$\,\ref{sec:phase_1}$ and shown schematically in Fig.~\ref{fig:clustering_schematic} (blue hatched areas). 
    We perform the grouping in each direction rather than globally to simplify the grouping, which might get too confused if all 16 neighbours are considered together.
    \item
    Compute the total weight $\weight{tot}$ for each group of centroid position data points by summing up the weights of the neighbouring spectra that contributed data points to the group and check if it exceeds a predefined weight threshold $\pazocal{W}$.
    \item
    Check whether the central spectrum has Gaussian components compatible with the required Gaussian components inferred from its neighbours (i.e. all centroid position groups that reached the required weight threshold $\pazocal{W}$).
    We try to refit the central spectrum with the fit solution from individual neighbours if its Gaussian components are incompatible with the inferred required components.
\end{enumerate}

In the default settings of \gausspyplus, the first set of iterations use a weight threshold of $\pazocal{W} = 1 - w_{2}$; this threshold can only be reached in the horizontal or vertical direction if two immediate spectra and an additional spectrum further out contributed data points to the group, i.e. show a common feature.
The threshold of $\pazocal{W} = 1 - w_{2}$ is used as long as it leads to new successful refits of spectra. In case no new refits were possible $\pazocal{W}$ is reduced again by a value of $w_{2}$ so that the new threshold is $\pazocal{W} = 1 - 2\cdot w_{2}$. This iterative procedure continues until $\pazocal{W}$ gets below a user defined minimum threshold $\weight{min}$ (default value: $0.5$).

We only start the refitting procedure after we looped through all spectra of the dataset and determined the spatial consistency of the centroid position values for all of them.
This means that the fit solutions are not dynamically updated or propagating outwards during an iteration.
New fit solutions are accepted based on the flagging system introduced in the previous section.
We add a new flag in this phase that increases the total flag value $\flag{tot}{}$ by a value of 2 if the fit solution is inconsistent with the required centroid positions inferred from the spatial consistency check. 

Panels (b) and (c) of Fig.~\ref{fig:schematic_phase_2} show example cases for the spatial consistency check of centroid values for the case of a simple emission line feature. 
Based on the fit solutions in the neighbouring spectra we want to establish whether a one- or two-component fit should be used for the central spectrum.
For this example we use the default settings of the algorithm, i.e. $\weight{min} = 0.5$ and $f_{w} = 2$, which sets $w_1 = \sfrac{1}{3}$ and $w_2 = \sfrac{1}{6}$. 

For the case depicted in panel~(b) the required weight threshold for the first set of iterations is $\pazocal{W} = 1 - w_1 = \sfrac{5}{6}$.
The $\weight{tot}$ value for the vertical and horizontal direction would reach this threshold, giving us two conflicting fit solutions for the central spectrum.
In such a case, we recompute $\weight{tot}$ for the fit solutions by grouping the eight immediate surrounding neighbouring spectra together and choose the fit solution with the higher $\weight{tot}$ value.
For the setup depicted in panel~(b) the fit solution with one Gaussian component would be selected, as the immediate surrounding neighbours with this fit solution have a bigger total weight of $\weight{tot} = 2 w_{1} + 3 w_{1} / \sqrt{2}$ (compared to $\weight{tot} = 2 w_{1} + w_{1} / \sqrt{2}$ for the two-component fit solution).\footnote{In case both fit solutions have the same total weight as calculated from its immediate surrounding neighbours and this way to decide on the fit solutions thus should be inconclusive, we repeat this total weight calculation for all 16 considered neighbours (colored squares in panel~(a) of Fig.~\ref{fig:schematic_phase_2}).
If this is also inconclusive we choose the fit solution that uses less Gaussian components.}
We would thus try to refit the central spectrum with a fit solution that uses only one Gaussian component. 
Note, however, that the fit solution for the central spectrum is only updated if the total flag value for the fit solution using one component is lower or equal than the total flag value for the fit solution using two components in addition to the requirements that the distribution of the residual data points resembles a normal distribution (see Sect.$\,\ref{sec:phase_1}$).

For the example case depicted in panel~(c) of Fig.~\ref{fig:schematic_phase_2} none of the four main directions would contain fit solutions that pass a weight threshold of $\pazocal{W} = \sfrac{5}{6}$.
However, both the vertical and the diagonal direction from upper left to lower right would reach a weight threshold of $\pazocal{W} = \sfrac{4}{6}$, which is used in the second round of iterations. 
The total weight for the single component fit solution in the diagonal direction ($\weight{tot} =  2\cdot w_{1} / \sqrt{2}\,+ 2\,\cdot w_{2} / \sqrt{2} \approx 0.7$) is bigger than the total weight for the two component fit solution in the vertical direction ($\weight{tot} =  w_{1}  + 2\cdot w_{2} = \sfrac{2}{3}$) and thus gets selected.
Since the central spectrum already has a single component fit we would not try to refit it.


\section{Performance of \gausspyplus\ on samples of synthetic spectra}
\label{sec:synthetic-spectra}

In this section, we compare the decomposition results of the improved fitting routine of \gausspyplus\ (Sect.$\,\ref{sec:improved-fitting}$) with the original \gausspy\ algorithm.
We applied both algorithms on samples of synthetic spectra containing: white noise (A); white noise and signal (B); white noise, signal, and negative noise spikes (C); white noise, weak signal, and negative noise spikes (D).
We then determine how well the two algorithms were able to recover the mean position, amplitudes and FWHM values of the Gaussian components used to create the synthetic spectra.
For more details about the synthetic spectra, see Appendix~\ref{app:test_sample}. 

To facilitate the comparison, we supplied the results from the noise calculation of \gausspyplus\ (Sect.$\,\ref{sec:noise-estimation}$) also to the decomposition with the original \gausspy\ algorithm. 
We also use the same S/N thresholds for the original spectrum ($\text{SNR}_{1} = 3$) and the second derivative of the smoothed spectrum ($\text{SNR}_{2} = 3$) for the decompositions with \gausspy\ and \gausspyplus. 
We use the smoothing parameters $\alpha_{1}$ and $\alpha_{2}$ we obtained from the training sets decomposed with the method outlined in Sect.$\,\ref{sec:training-set}$\footnote{For sample~A we use the same smoothing parameters as for sample~D.} (see Appendix~\ref{app:test_trainingset} for more details). 
We left all additional parameters of \gausspyplus\ at their default settings.

\begin{table}
    \caption{Percentage of correctly and incorrectly identified mean positions of Gaussian components for decomposition runs on samples of synthetic spectra.}
    \centering
    \small
    \renewcommand{\arraystretch}{1.2}
    \begin{tabular}{ccccc}
    \hline\hline
     & \multicolumn{2}{c}{\gausspy} & \multicolumn{2}{c}{\gausspyplus} \\
    Sample & correct\tablefootmark{a} & incorrect\tablefootmark{b} & correct\tablefootmark{a} & incorrect\tablefootmark{b}\\
    \hline
    A & -- & $2.8\%$ & -- & $0.0\%$ \\
    B & $78.0\%$ & $3.9\%$ & $93.7\%$ & $1.6\%$ \\
    C & $72.6\%$ & $3.7\%$ & $93.4\%$ & $1.8\%$ \\
    D & $29.4\%$ & $6.5\%$ & $81.7\%$ & $4.5\%$ \\
    \hline
    \end{tabular}
    \label{tbl:decomposition_correct_means}
    \tablefoot{ 
    \footnotesize 
    \tablefoottext{a}{We define the mean position of a Gaussian component as correctly identified if it is within $\pm 2$ spectral channels of the true value.\\}
    \tablefoottext{b}{We define the fraction of incorrect identifications for sample~A as all spectra for which noise features were fitted.
    The percentage of incorrect identifications for sample~B--D refers to the fraction of fitted Gaussian components whose mean position was located at a distance of more than $4$ spectral channels to the true value.}}
\end{table}

Table~\ref{tbl:decomposition_correct_means} presents quality metrics of the results of the decomposition runs with \gausspy\ and \gausspyplus\ of the four samples of synthetic spectra.
The percentage of correct detections refers to the number of Gaussian components that were fit within $\pm 2$ spectral channels of the true position.
Note that for a correct identification of a peak position we do not consider whether the amplitude and FWHM values of the Gaussian component were fit correctly.
The fraction of incorrect detections refers either to all spectra for which at least one noise feature was fitted (in case of sample~A) or the percentage of Gaussian fit components that were placed at a distance of more than 4 spectral channels away from the true position.

Table~\ref{tbl:decomposition_correct_means} demonstrates that \gausspyplus\ manages to fit significantly more Gaussian components at the correct positions in the spectrum than \gausspy\ while decreasing the fraction of incorrect identifications\footnote{Note that a limiting factor for the performance of \gausspyplus\ was that the synthetic spectra were not set up to show spatial coherence. 
Thus, the algorithm will have had difficulty in the decomposition of some spectra to correctly decide whether a structure might be blended and better fit by multiple peaks.}.
This improvement is especially striking for weak signal peaks (sample~D), where the number of correctly placed Gaussian fit components increased by more than a factor of $2.7$ in the \gausspyplus\ decomposition. 
The performance of the \gausspyplus\ decomposition is not affected by the presence of negative noise spikes in the spectrum (sample~C), whereas this has a more significant impact on the performance of \gausspy.
Note also that \gausspyplus\ did not incorrectly fit any Gaussian components in sample~A, whereas \gausspy\ mistook noise features as signal peaks for $2.8\%$ of the spectra.

\begin{figure}
    \centering
    \includegraphics[width=\columnwidth]{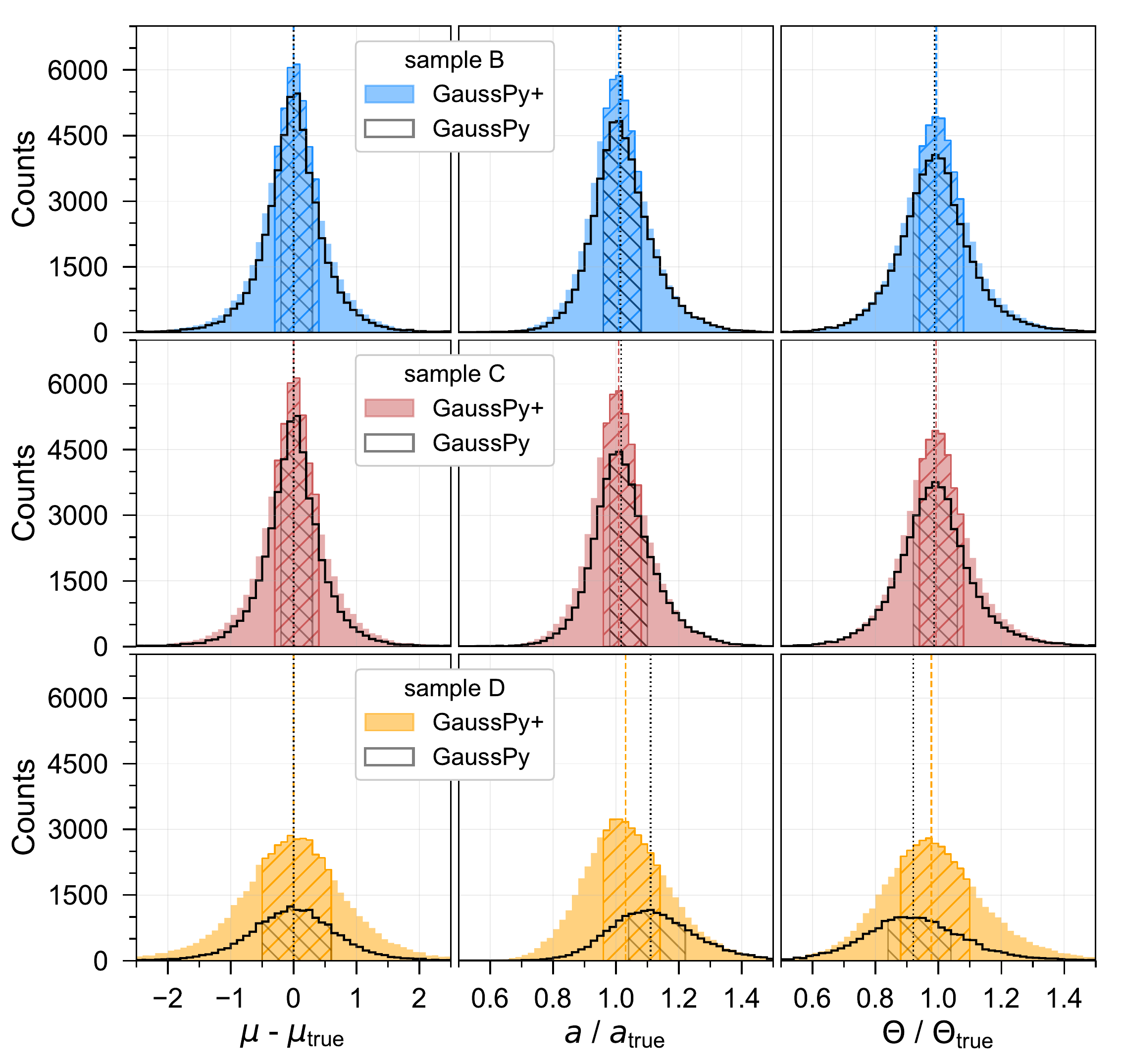}
    \caption{Comparison of the performance results of decompositions with \gausspyplus\ and \gausspy\ for different samples of synthetic spectra.
    The distribution shows how the fitted parameter values (mean position $\mu$, amplitude $a$, and FWHM $\Theta$ from left to right, respectively) compare to the true parameter values used to create the synthetic spectra.
    The unfilled and colored histograms show the distribution of fit components obtained with \gausspy\ and \gausspyplus, respectively.
    Hatched areas correspond to the interquartile ranges and the vertical lines indicate the median value of the distribution (colored and black for the \gausspyplus\ and \gausspy\ results, respectively.
    The improved fitting routine of \gausspyplus\ leads to a significant increase of correctly fitted parameters (see also Table~\ref{tbl:decomposition_correct_means} and Sect.$\,\ref{sec:synthetic-spectra}$ for more details).}
    \label{fig:improved_fitting_vs_gausspy}
\end{figure}

Figure~\ref{fig:improved_fitting_vs_gausspy} compares the fitted Gaussian parameters to the true values used to create the synthetic spectra.
The \gausspyplus\ decomposition results for sample~B, C, and D are shown in blue, red, and orange, respectively and the corresponding \gausspy\ results are indicated with the black line.
The left column of panels shows the distribution of fitted mean positions from which the true mean position was subtracted. 
As already demonstrated in Table~\ref{tbl:decomposition_correct_means}, the vast majority of components were fitted close to the true mean position.
There were fewer detected peaks in sample~D because the signal in these spectra was constructed to be close to or below the detection limit.

The middle and right column of panels in Figure~\ref{fig:improved_fitting_vs_gausspy} show the distribution of amplitude and FWHM values, respectively, both normalized by the corresponding true parameter values.
In these distributions we only included those fitted Gaussian components whose mean position was less than two channels away from the true mean position of the corresponding Gaussian component in the synthetic spectrum (corresponding to the percentages of correctly identified components in Table~\ref{tbl:decomposition_correct_means}). 
For all three samples of synthetic spectra the vast majority of fitted parameters are within $\pm 10\%$ of the true values for both decompositions, but due to the higher amount of correctly identified peak positions, \gausspyplus\ manages to fit many more components correctly.
Moreover, for sample~D the median values of the distribution are closer to the true values for the \gausspyplus\ decomposition results.
In contrast, \gausspy\ tends to fit the spectra of sample~D with components that have too large amplitude values and too narrow linewidths, as demonstrated from the shape of the distributions and their median values.

We found that the decomposition performance of \gausspyplus\ also shows much less dependence on the number of signal peaks, their S/N ratio, their linewidth, or their closest distance to a neighbouring signal peak than the decomposition with \gausspy\ (see Appendix~\ref{app:test_decomposition}).


\section{Performance of \gausspyplus\ on a GRS test field}
\label{sec:test-field}

In this section, we focus on a sub-region of the GRS survey and perform a detailed analysis and discussion of the decomposition results with \gausspyplus\ to showcase its performance.

The test field we chose is a $0.43^{\circ} \times 0.37^{\circ}$ region located towards the outer part of the GRS coverage at Galactic coordinates of $l = 55.48^{\circ}$ and $b = 0.19^{\circ}$.
This GRS region contains $4200$ spectra with $424$ spectral channels that cover $v_{\mathrm{LSR}}$ values of $-5 - 85$~\kms.
The chosen region contains three molecular clouds (G055.64+00.14, G055.39+00.14, G055.34+00.19) and 19 clumps as identified by \citet{Rathborne2009}.

In the following sections we will first describe the best way to compare flux between the original dataset and the decomposition and show the improvements we gain by using the noise estimation technique built into \gausspyplus.
We then make a detailed comparison between the decomposition results of \gausspy\ and \gausspyplus.

Details about the execution time for the entire decomposition and the performance of the spatially coherent refitting can be found in App.~\ref{app:execution_time_grs} and App.~\ref{sec:refit_iterations}, respectively.

\subsection{Optimal flux estimate for fair comparisons between the dataset and decomposition results}
\label{sec:testfield-0_mom-original}

\begin{figure}
  \centering
  \includegraphics[width=\columnwidth]{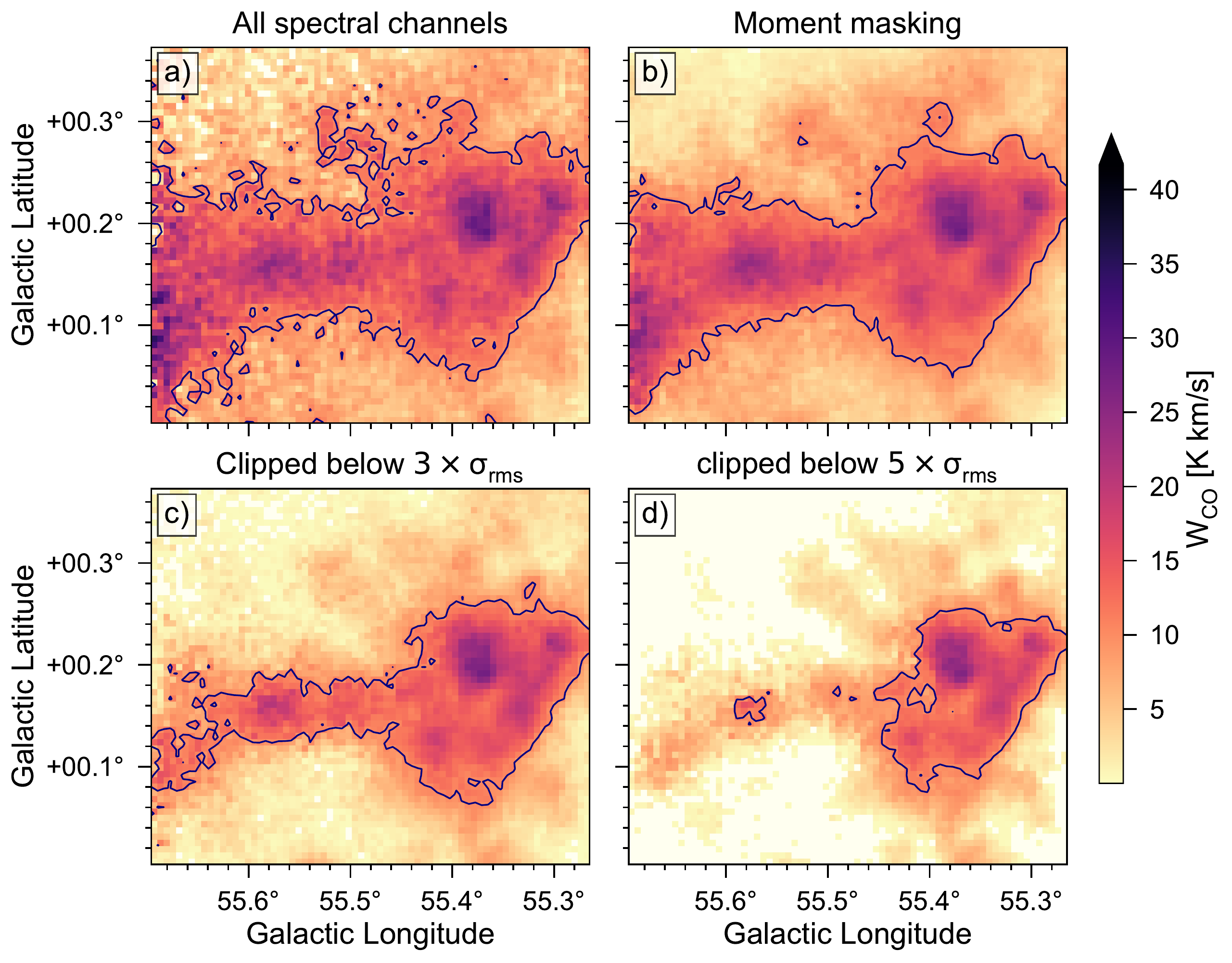}
  \caption{Zeroth moment maps for a region in the outer parts of the GRS.
  Panels~(a) -- (d) show the results obtained by summing up all spectral channels, applying a moment masking technique (see Sect.$\,\ref{sec:testfield-0_mom-original}$ for details), and clipping all spectral channels with values below $3\cdot\sigma_{\mathrm{rms}}$ and $5\cdot\sigma_{\mathrm{rms}}$, respectively.
  The contour indicates a $W_{\mathrm{CO}}$ level of $5$~K~\kms.
  }
  \label{fig:grs-01_0mom_vs_sigma}
\end{figure}

One measure of the quality of the decomposition results is the fraction of recovered flux from the comparison of zeroth moment maps; we aim at inspecting this fraction in Section \ref{sec:testfield-decomp}. 
However, imperfect baseline corrections and noise spikes can lead to wrong flux estimates if all spectral channels are integrated along the spectral axis.
It is therefore recommended to mask out all spectral channels that do not contain signal.

For our comparisons of the recovered flux in the decomposition (Sect.$\,\ref{sec:testfield-decomp}$) we opted to use the moment masking technique outlined in \citet{Dame2011}.
The basic idea of moment masking is to mask out spectral channels based on S/N cuts on a spatially and spectrally smoothed version of the original dataset.
For the smoothed data cube, \citet{Dame2011} suggests to degrade the spatial resolution by a factor of 2 and degrade the spectral resolution to the width of the narrowest spectral lines contained in the dataset. 
\citet{Dame2011} found that a threshold of $5\cdot\sigma_{\mathrm{rms,\,smoothed}}$ gives the best results, where $\sigma_{\mathrm{rms,\,smoothed}}$ refers to the rms-noise of the smoothed spectra.
If a spectral channel in the smoothed cube exceeds this S/N threshold, we unmask this channel and all channels that were within the spatial and spectral smoothing kernels in the original datacube.
Moment masking thus allows us to also include spectral channels whose value has low S/N levels and would be masked out if we based the clipping of spectral channels on a S/N threshold of the original dataset.
Moreover, the high S/N requirement for spectral channels of the smoothed dataset guarantees that most of the channels containing noise are masked out.

For the moment masking of the GRS test field, we created a smoothed version of the data cube by smoothing the original dataset spatially with a Gaussian kernel with a FWHM value of $92^{\prime\prime}$ (corresponding to twice the beam size); spectrally, we smoothed the dataset with a Gaussian kernel with a FWHM value of $0.42\,$\kms, which corresponds to twice the spectral resolution or 2 spectral channels.
We then masked out all spectral channels whose value in the smoothed data cube was below a S/N threshold of 5.

Figure~\ref{fig:grs-01_0mom_vs_sigma} shows zeroth moment maps of our test region obtained by: summing up all spectral channels (panel~a); using the moment masking technique described above (panel~b); masking out all spectral channels with S/N values below 3 and 5 (panel~c and d, respectively). 
The contour in the panels marks a value of $W_{\mathrm{CO}} = 5$~K~\kms, with $W_{\mathrm{CO}}$ being the integrated CO intensity along the spectral axis.
By summing up all intensity values along the spectral axis we also include a significant contribution from noise, which is clearly visible in the fraying of the contour line in panel~(a) of Fig.~\ref{fig:grs-01_0mom_vs_sigma}.
If we mask out all spectral channels with S/N values lower than 3 or 5 times the $\rms$ (panel~c and d), we also cut away a significant fraction of real signal, leading to a severe underestimate of the total flux contained in the region.
Conversely, the zeroth moment map constructed with the moment masking technique (panel~b in Fig.~\ref{fig:grs-01_0mom_vs_sigma}) replicates well the flux distribution of panel~(a), and excludes most of the noise contributions.

We quantify the recovered flux by summing up all intensity values above a value of $5$~K~\kms\ (contours in Fig.~\ref{fig:grs-01_0mom_vs_sigma}).
The summed up value inside the contour of the map obtained with the moment masking technique is only $8\%$ smaller than the corresponding value of the map in which we sum up all spectral channels.
This small difference is likely due to contributions of spectral channels containing only noise that are also included in the zeroth moment map shown in panel~(a) in Fig.~\ref{fig:grs-01_0mom_vs_sigma}.
In contrast, the summed up value inside the contours of panel~(c) and panel~(d) of Fig.~\ref{fig:grs-01_0mom_vs_sigma} is smaller by $33\%$ and $56\%$ respectively than the summed up value for the corresponding contour in panel~(a).

We conclude that by summing up all spectral channels or masking out spectral channels via a S/N threshold based on the original dataset we would either slightly overestimate or severely underestimate the flux contained in the dataset, respectively.
On the other hand, the moment masking technique gives a good estimate of the total flux contained in a dataset and we therefore use it in comparisons of the flux between the decomposition results and the original dataset.
We assume here implicitly that noise contributions in the remaining spectral channels average out when the intensity values are integrated.

\subsection{Noise map}
\label{sec:testfield-noise}

A good estimate of the noise is crucial for obtaining good fitting results if parameters of the decomposition technique are based on S/N thresholds (see Sect.$\,\ref{sec:noise-estimation}$).

\begin{figure}
  \centering
  \includegraphics[width=\columnwidth]{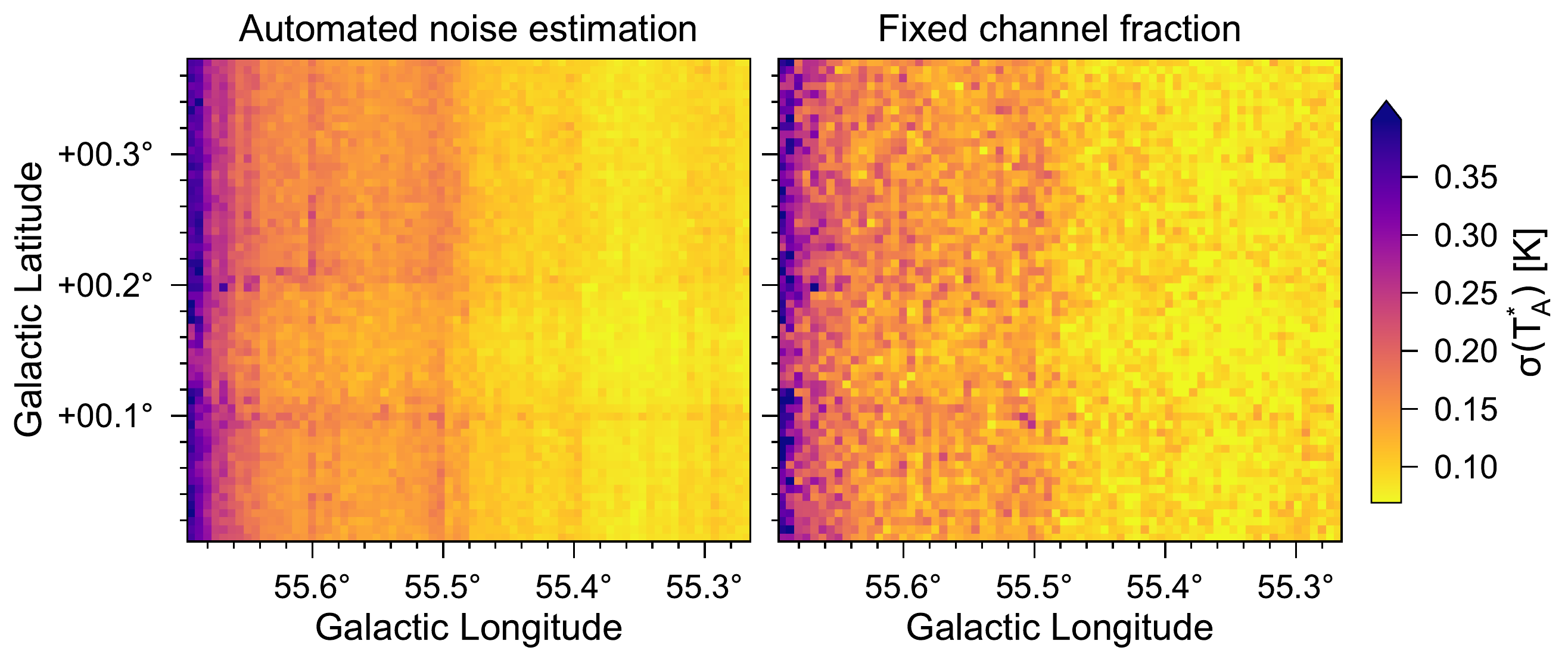}
  \caption{Noise maps for the region in Fig.~\ref{fig:grs-01_0mom_vs_sigma}.
  (\textit{left}): Results from the automated noise estimation technique discussed in Sect.$\,\ref{sec:noise-estimation}$.
  (\textit{right}): Results from using a fixed amount of spectral channels for the noise calculation.
  }
  \label{fig:grs-01_noise}
\end{figure}

Figure~\ref{fig:grs-01_noise} shows noise maps for the region depicted in Fig.~\ref{fig:grs-01_noise} that were obtained with the noise estimation routine of \gausspyplus\ (panel~a) and a much simpler approach that uses a fixed number of channels to calculate the $\rms$ value (panel~b).
For the latter approach we used 24 spectral channels from $80 - 85$~\kms (corresponding to $\sim 6\%$ of all available spectral channels), similar as it was done in \citet{Jackson2006} for this region.
For the GRS dataset, the fixed channel approach can be problematic, since there is not really a channel interval that is guaranteed to be emission-free over the entire survey region.
It can be clearly seen that the \gausspyplus\ noise estimation routine gives a much better estimate of the $\rms$ values, as artifacts from the map-making procedures become more pronounced. 
Note that there is also a clear gradient in the $\rmsTa$ values in this region, which makes the \gausspyplus\ decomposition challenging, since it uses the same decomposition parameters throughout the whole region.
We would thus expect to have more difficulty in the decomposition of spectra with high $\rmsTa$ values, leading to small S/N values of the signal peaks in these spectra.

\begin{figure*}
  \centering
  \includegraphics[width=1.5\columnwidth]{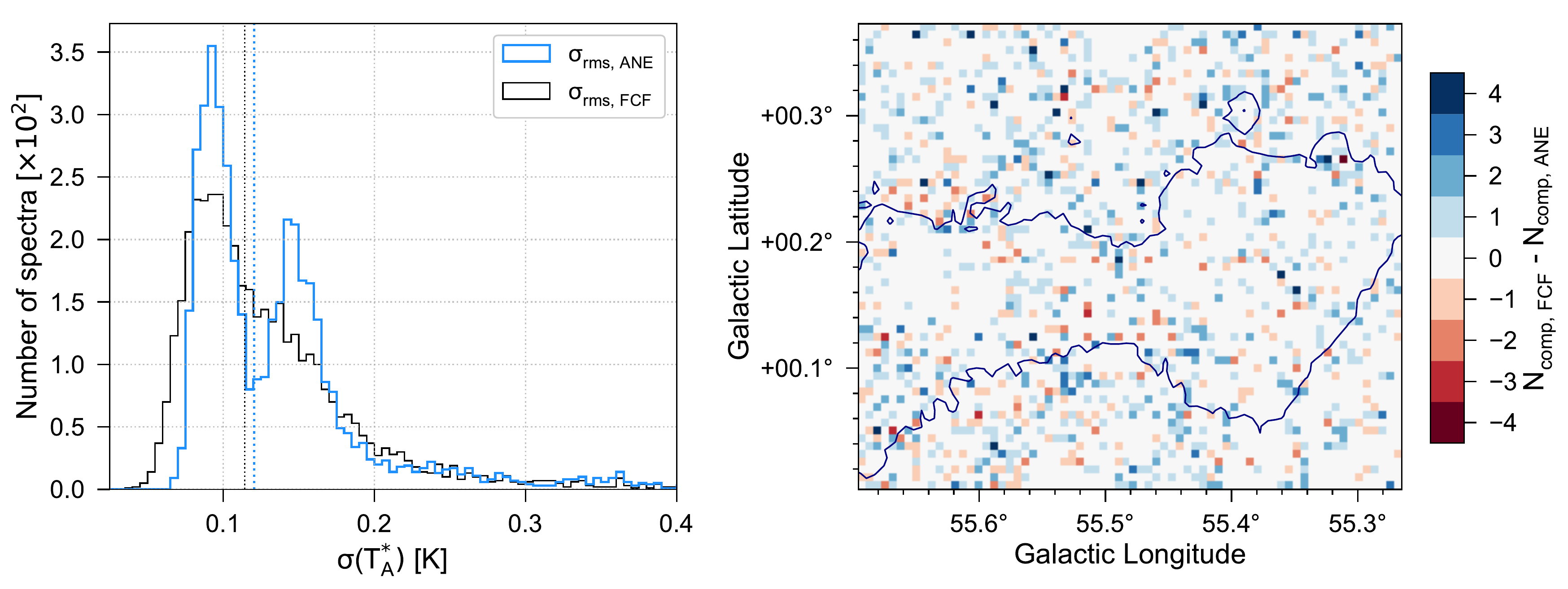}
  \caption{(\textit{left}): Histogram of the $\rms$ values shown in Fig.~\ref{fig:grs-01_noise} for the automated noise estimation of \gausspyplus\ (ANE, blue) and the fixed channel fraction approach (FCF, black).
  The dotted vertical lines show the corresponding median values of the two distributions. 
  (\textit{right}): Map showing the difference in the number of fitted components for the automated noise estimation ($N_{\mathrm{comp,\,ANE}}$) and the fixed channel fraction approach ($N_{\mathrm{comp,\,FCF}}$).
  }
  \label{fig:grs-01_noise_histogram}
\end{figure*}

Panel~(a) in Fig.~\ref{fig:grs-01_noise_histogram} displays histograms of the noise maps of Fig.~\ref{fig:grs-01_noise}; the automated noise estimate shows a clear bimodal distribution, whereas the fixed channel fraction approach is more influenced by random fluctuations of the noise in the limited fixed number of channels used for the noise calculation.
The median $\rmsTa$ value of our automated noise estimation is only $\sim 6\%$ higher by than the median value obtained via the fixed channel approach, so globally the two methods give similar results. 
However, Fig.$\,\ref{fig:grs-01_noise}$ shows that there are considerable differences on the individual line of sight scale, which will lead to large differences in the decomposition.

To quantify the impact of the estimated noise on the fitting results, we performed two decomposition runs with the improved fitting routine of \gausspyplus\ (Sect.$\,\ref{sec:improved-fitting}$) with identical settings but different noise estimates corresponding to the maps of Fig.~\ref{fig:grs-01_noise}.
Panel~(b) in Fig.~\ref{fig:grs-01_noise_histogram} shows the difference between the number of fitted Gaussian components for the noise estimate using a fixed fraction of channels and the automated routine of \gausspyplus.
About $26\%$ of the spectra in the test field get fitted with a different number of Gaussian components and the total number of fitted components increases by $\sim 9\%$ for the fixed channel fraction approach.
Applying the flagging procedure of \gausspyplus\ with its default settings described in Sect.$\,\ref{sec:phase_1}$ to the two decompositions, we get that $43.8\%$ and $51.2\%$ of the fitted spectra would be selected for refitting if the automated noise estimate and fixed channel approach are used, respectively.
Compared to the \gausspyplus\ decomposition with the automated noise estimate, in the fixed channel approach the number of spectra flagged as having a number of components incompatible with their neighbours increase from $1.5\%$ to $6.1\%$ and the number of spectra having normalised residual values not matching a Gaussian distribution increases from $20.0\%$ to $25.1\%$.
Both of these increased numbers of flagged spectra are a good indication that the noise estimate using the fixed fraction of channels is yielding poorer decomposition results than the automated noise estimation routine incorporated in \gausspyplus.

\subsection{Comparison between the decomposition runs with \gausspy\ and \gausspyplus}
\label{sec:testfield-decomp}

In this section we present  decomposition runs of the GRS test field obtained with the original \gausspy\ algorithm and \gausspyplus.
The different \gausspyplus\ runs represent results after different stages of the algorithm (improved fitting routine, phase 1 and 2 of the spatially coherent refitting, referred to as Stage 1, 2, and 3, respectively) to better illustrate the changes and improvements obtained in each individual stage.

We decomposed $250$ randomly chosen spectra of the test field with the method outlined in Sect.$\,\ref{sec:training-set}$ to create the training set needed to infer optimal smoothing parameters for \gausspy.
\citet{Lindner2015} found that having two different smoothing parameters -- one parameter with a smaller value that accentuates the narrower peaks and another parameter with a higher value that is more suitable for broader peaks -- leads to huge improvements in the decomposition of HI spectra.
We also found that a two-phase decomposition approach using two different smoothing parameters $\alpha_{1}$ and $\alpha_{2}$ yields better decomposition results for the CO emission line spectra of the GRS survey.
For the same training set, the F$_{1}$ score (see Sect.$\,\ref{sec:gausspy}$) for the one-phase and two-phase decomposition approaches was $67.5\%$ and $74.7\%$, respectively.
We therefore used the smoothing parameters inferred from the two-phase decomposition of the training set, which yielded values of $\alpha_{1} = 2.89$ and $\alpha_{2} = 6.65$.
For the \gausspy\ decomposition we set $\text{SNR}_{1} = \text{SNR}_{2} = 3$.
We left all \gausspyplus\ parameters at their default settings, with the exception of setting $\Delta\mu_{\Max} = 4$ for Stage~3.

\begin{figure*}
  \centering
  \includegraphics[width=\textwidth]{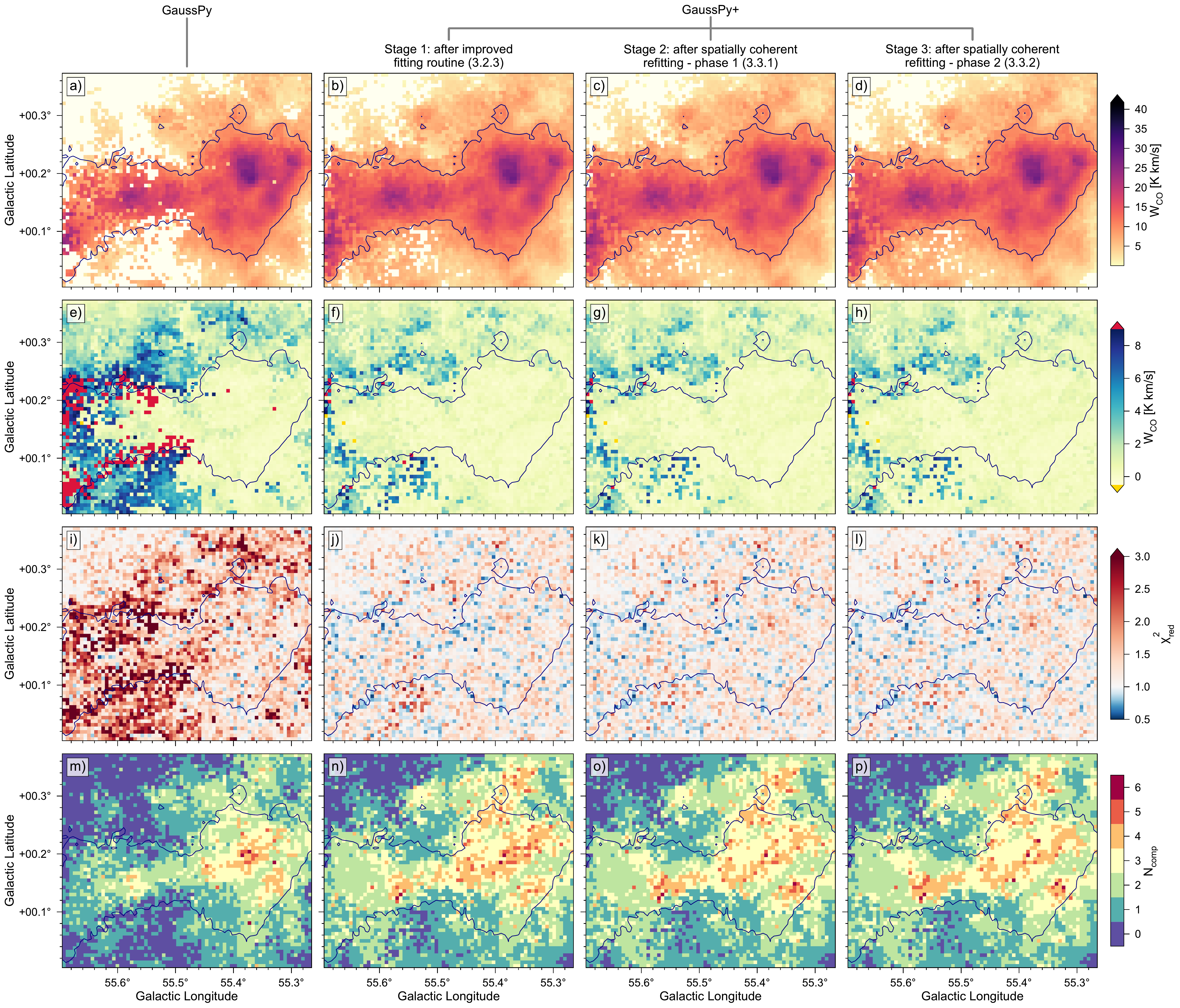}
  \caption{(from left to right:) Decomposition results for the original \gausspy\ algorithm and three stages of \gausspyplus\ (improved fitting routine, phase 1 and 2 of the spatially coherent refitting).
  The rows show (from top to bottom): zeroth moment maps of the decomposition results; residual maps obtained by comparing the zeroth moment maps of the decomposition with panel~(b) in Fig.~\ref{fig:grs-01_0mom_vs_sigma}; maps showing the $\chired$ values for the fit, with the goodness of fit calculation restricted to the channels estimated to contain signal (see Sect.$\,\ref{sec:signal-interval}$); and maps of the number of Gaussian fit components per spectrum.
  All panels are overplotted with the contour from panel~(b) in Fig.~\ref{fig:grs-01_0mom_vs_sigma}.
  The \gausspyplus\ decompositions show a clear trend towards more spatial coherence and an improvement in the quality of the fits for the regions with lower emission or higher noise levels.
  }
  \label{fig:test-field-comparison}
 \end{figure*}

Panels~(a)--(d) in Fig.~\ref{fig:test-field-comparison} show zeroth moment maps of the decomposition runs with the original \gausspy\ algorithm supplied with the improved noise estimation (panel~a) and \gausspyplus\ after the improved fitting stage (panel~b; Sect.$\,\ref{sec:improved-fitting}$), and after phase 1 (panel~c; Sect.$\,\ref{sec:phase_1}$) and phase 2 (panel~d; Sect.$\,\ref{sec:phase_2}$) of the spatially coherent refitting.
The zeroth moment maps were obtained by masking the same spectral channels as for the moment masked data in panel~(b) of Fig.~\ref{fig:grs-01_0mom_vs_sigma}.

Panels~(e)--(h) in Fig.~\ref{fig:test-field-comparison} show the corresponding zeroth moment maps of the residual.
In all three stages of \gausspyplus\ the performance in terms of the recovered flux is much better for the regions with lower S/N emission than the \gausspy\ decomposition, which was already noticeable in the case of synthetic spectra (Sect.$\,\ref{sec:synthetic-spectra}$, App.~\ref{app:test_decomposition}, and App.~\ref{app:test_recovery}).
For regions with high S/N \gausspy\ and all stages of \gausspyplus\ perform very well.\footnote{While the recovered flux is an essential criterion for the performance of the fit it may not give a good handle on the quality of the fits themselves.
For example the spectra might not be spatially coherent and might use many blended and broad components to fit the spectrum.}

The maps in panels~(i)--(l) of Fig.~\ref{fig:grs-01_0mom_vs_sigma} show the $\chired$ values for the fits, with the goodness of fit calculation restricted to the channels estimated to contain signal (see Sect.$\,\ref{sec:signal-interval}$).
We can see a clear improvement towards $\chired$ values closer to $1$ for the \gausspyplus\ decompositions compared to the \gausspy\ run.
In App.$\,$\ref{app:grs-reduced-chi-square} we demonstrate the importance of restricting the calculation of the $\chired$ values to regions in the spectrum that contain signal for the GRS test field.

The performance in recovered flux does not significantly change in the spatially coherent refitting phases of \gausspyplus, since the focus in these phases is shifted to reducing flagged features and making the fit results compatible with the neighbours instead of minimizing the residual.
Therefore the zeroth moment, residual and $\chired$ maps that show the quality of the fit results in terms of recovered flux do not change significantly between Stage~1 and Stage~3 of the \gausspyplus\ decomposition.

We can see more variation between panels~(m)--(p) in Fig.~\ref{fig:test-field-comparison}, which show maps of the number of fitted Gaussian components per spectrum for each decomposition run.
In the \gausspyplus\ decompositions the number of fitted components increases compared to the \gausspy\ run, which is due to the fitting of weaker emission lines in spectra containing increased noise levels (cf. panels~e--h) and the segmentation of very broad components into individual peaks.
Note also the progression towards more spatial coherence from panels~(m)--(p).

\begin{figure*}
    \begin{tabular}{cc}
      \includegraphics[width=\columnwidth]{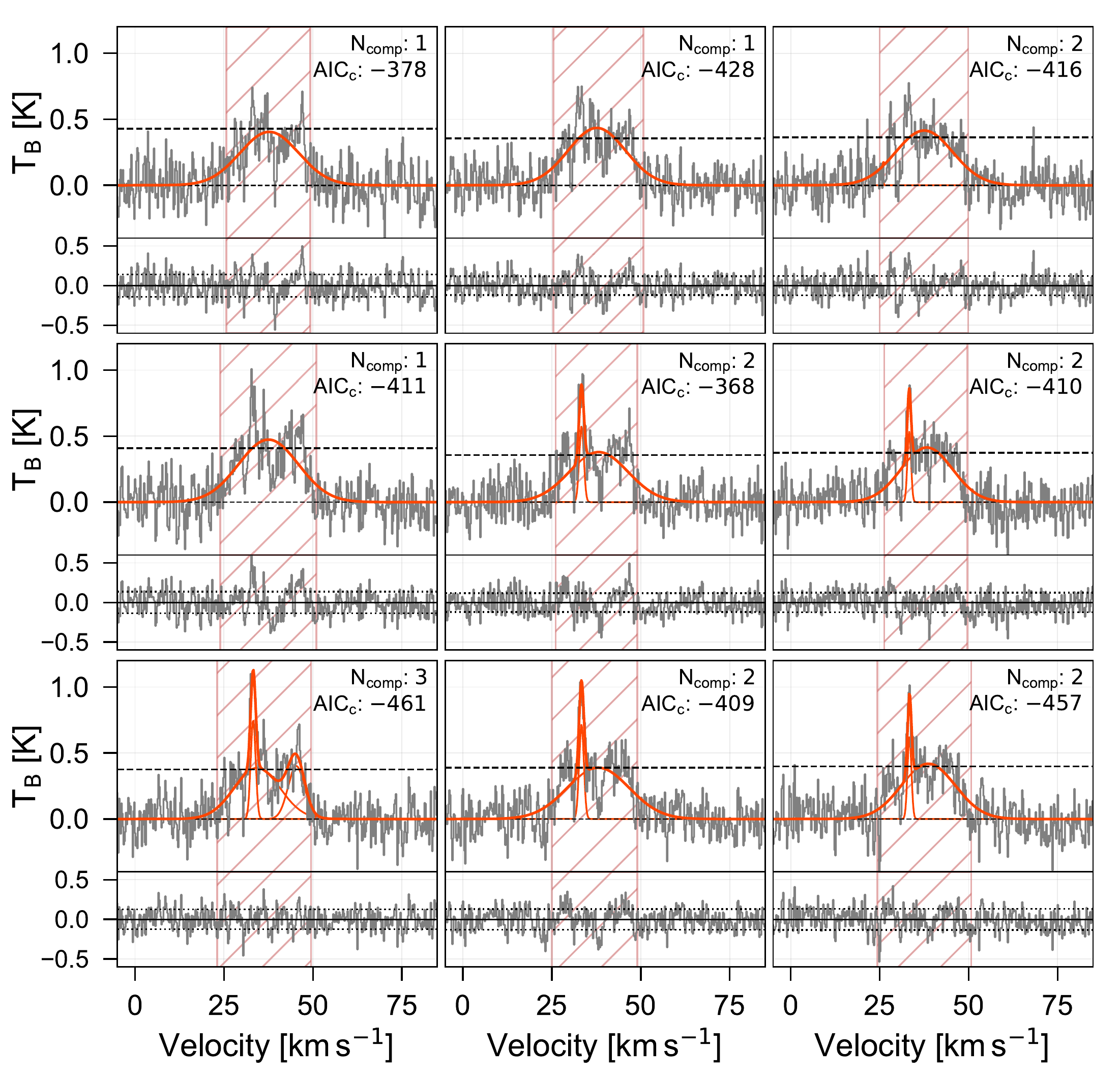} &   \includegraphics[width=\columnwidth]{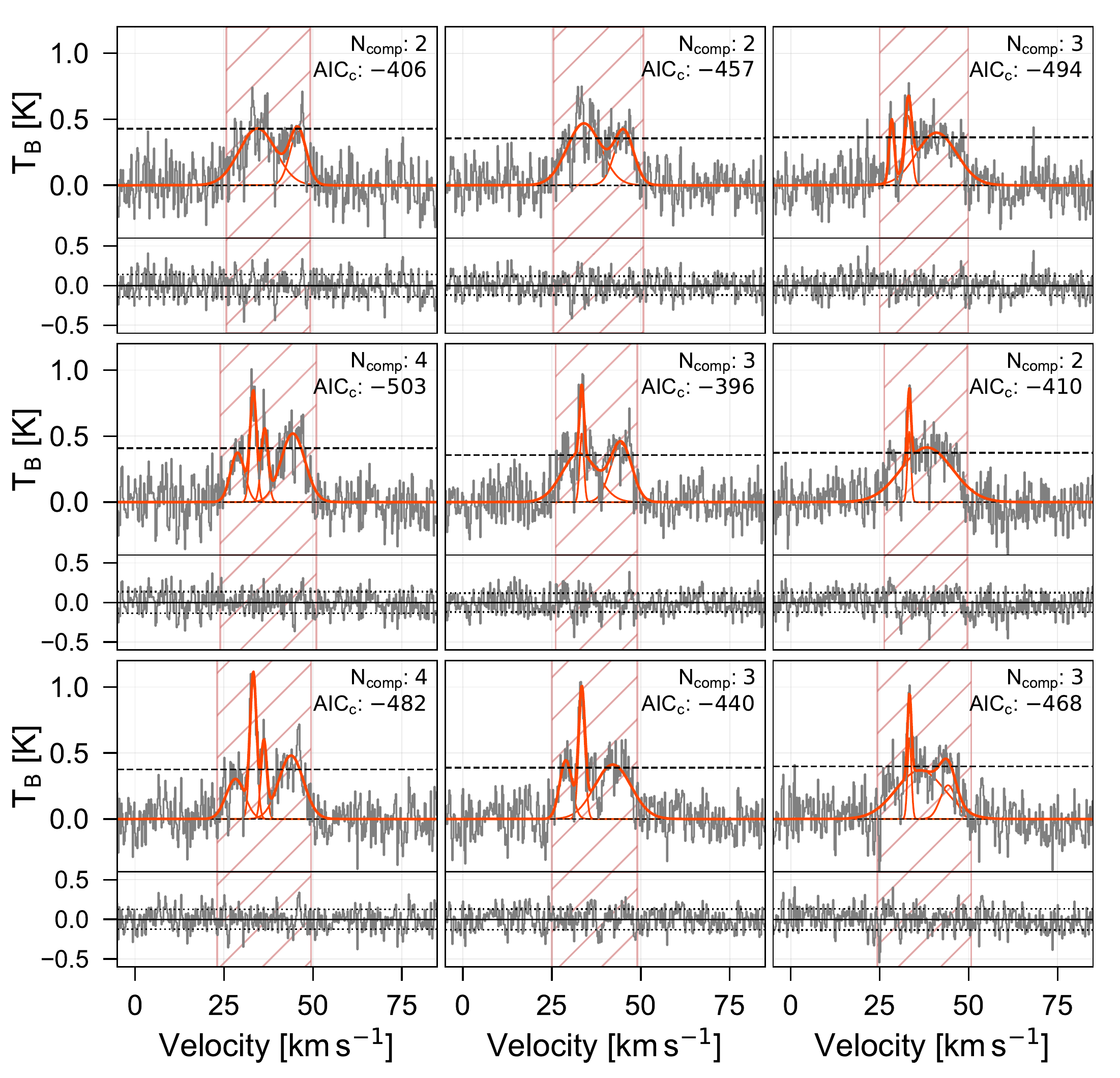} \\
    a) \gausspy & b) \gausspyplus\ (Stage 1) \\[6pt]
     \includegraphics[width=\columnwidth]{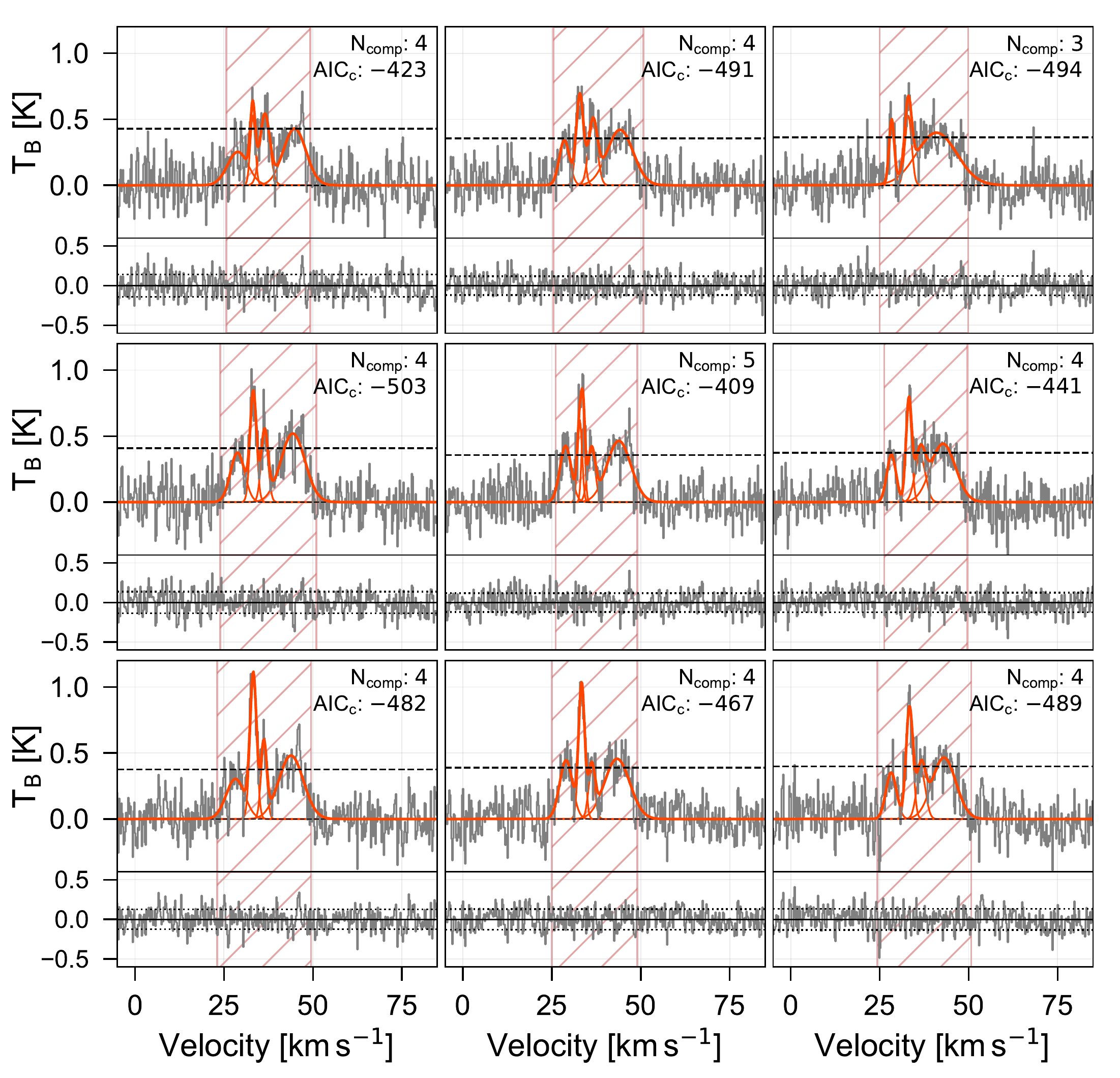} &   \includegraphics[width=\columnwidth]{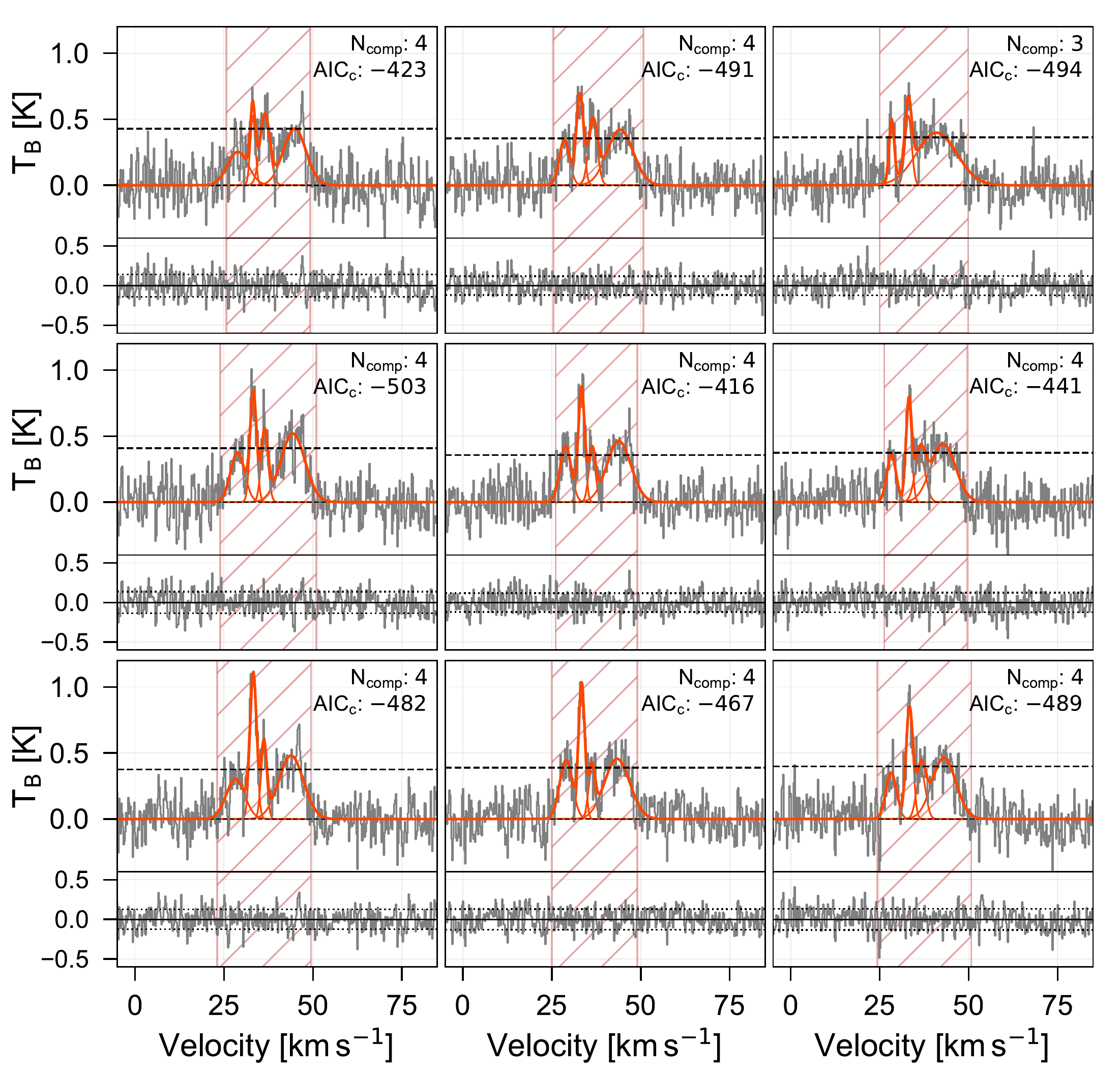} \\
    c) \gausspyplus\ (Stage 2) & d) \gausspyplus\ (Stage 3) \\[6pt]
    \end{tabular}
    \caption{Fitting results of nine neighbouring spectra in the GRS test field for the decomposition with \gausspy\ (a) and after Stage 1 -- 3 of \gausspyplus\ (b -- d, respectively). 
    Individual fit components and their combination are shown in thin and thick orange lines, respectively.
    Horizontal dashed black lines mark a S/N threshold of 3 and hatched areas indicate the identified signal intervals.
    The number of used fit components $\Ncomp$ and the resulting AIC$_{\text{c}}$ values are noted in the upper right corner of the main panel.
    The smaller subpanels show the residual with the horizontal dotted black lines marking values of $\pm \rms$.}
    \label{fig:example_spectra}
\end{figure*}

Figure$\,\ref{fig:example_spectra}$ further demonstrates this progression toward more spatial coherence by comparing the fitting results of \gausspy\ and Stage 1 -- 3 of \gausspyplus\ for nine neighbouring spectra from the GRS test field.
The signal peaks in these spectra show only moderate S/N ratios and \gausspy\ therefore tends to fit broad Gaussian components over most of the signal peaks.
Stage 1 of \gausspyplus\ already manages to improve upon these fitting results by separating the emission into more individual peaks; this improvement of the decomposition results can also be seen in the decreased residuals.
Stage 2 of \gausspyplus, which uses the information of already well-fit neighbouring spectra as input guesses for flagged spectra, can even further improve upon these results by creating more spatial coherence between the spectra.
Finally, Stage 3 of \gausspyplus, which tries to enforce spatial coherence between the centroid values of the fit components, improves the decomposition results once more, by getting rid of a fit component for the central spectrum that was inconsistent with the neighbouring fit solutions.

\begin{table}
    \caption{Comparison of parameters and flagged spectra for the decomposition runs with \gausspy\ and \gausspyplus.}
    \centering
    \small
    \renewcommand{\arraystretch}{1.2}
    \begin{tabular}{ccccc}
    \hline\hline
     & \gausspy\ & \gausspyplus\ & \gausspyplus\ & \gausspyplus\ \\
     &  & (Stage 1) & (Stage 2) & (Stage 3) \\
    \hline
$W_{\mathrm{CO,\,all}}$ & $73.0\%$ & $88.9\%$ & $89.6\%$ & $89.8\%$ \\
$W_{\mathrm{CO,\,contour}}$ & $84.0\%$ & $95.5\%$ & $95.6\%$ & $95.7\%$ \\
$\chi_{\mathrm{red,\,med}}^{2}$ & $1.436$ & $1.123$ & $1.121$ & $1.113$ \\
$\flag{tot}{}$ & $59.2\%$ & $43.8\%$ & $35.5\%$ & $38.0\%$ \\
$\flag{blended}{}$ & $5.8\%$ & $7.5\%$ & $2.9\%$ & $3.2\%$ \\
$\flag{neg.\,res.\,peak}{}$ & $2.6\%$ & $0.1\%$ & $0.0\%$ & $0.1\%$ \\
$\pazocal{F}_{\Theta}$ & $24.0\%$ & $22.7\%$ & $19.1\%$ & $21.6\%$ \\
$\pazocal{F}_{\Theta  > 50}$ & $10.6\%$ & $11.2\%$ & $9.6\%$ & $9.4\%$ \\
$\flag{residual}{}$ & $37.1\%$ & $20.0\%$ & $16.3\%$ & $16.5\%$ \\
$\pazocal{F}_{N_{\mathrm{comp}}}$ & $0.3\%$ & $1.5\%$ & $1.1\%$ & $1.0\%$ \\
    \hline
    \end{tabular}
    \label{tbl:stats-tf}
\end{table}

Table$\,\ref{tbl:stats-tf}$ compares parameters and the percentage of flagged spectra for the decomposition results for \gausspy\ and the three stages of \gausspyplus\ depicted in Fig.~\ref{fig:test-field-comparison}. 

The $W_{\mathrm{CO,\,all}}$ and $W_{\mathrm{CO,\,contour}}$ parameters give the fraction of recovered intensity values integrated along the spectral axis ($W_{\mathrm{CO}}$) for the whole test field and inside the contour of $5$~K~\kms, respectively.
The $W_{\mathrm{CO,\,all}}$ and $W_{\mathrm{CO,\,contour}}$ values were determined by comparing the moment maps of the decompositions (panels~a--d in Fig.~\ref{fig:test-field-comparison}) to the moment masked zeroth moment map of panel~(b) in Fig.~\ref{fig:grs-01_0mom_vs_sigma}.
As already noticeable in Fig.$\,\ref{fig:test-field-comparison}$, the performance of \gausspy\ and \gausspyplus\ is better for spectra containing high S/N emission peaks than for weaker emission lines.
With \gausspyplus\ we are able to recover about $90\%$ of the $W_{\mathrm{CO}}$ values contained in the entire test field and $\sim 95\%$ of the $W_{\mathrm{CO}}$ values contained inside the contour of $5$~K~\kms.
Compared to the \gausspyplus\ runs, the decomposition with the original \gausspy\ algorithm recovers about $12\%$ less flux inside the contour and $16\%$ less flux in the entire field.

The number of fitted Gaussian components $\Ncomp$ increase by about half for the \gausspyplus\ decompositions compared to the \gausspy\ run.
The median $\chired$ values ($\chi_{\mathrm{red,\,med}}^{2}$) of the \gausspyplus\ fitting results are also lower by $\sim 22\%$ than for the \gausspy\ results.

Table~\ref{tbl:stats-tf} also shows the fraction of spectra of the \gausspy\ and \gausspyplus\ results that would be flagged as not satisfying the quality criteria used in the first phase of the spatially coherent refitting (Sect.$\,\ref{sec:phase_2}$).
We use the default flagging criteria of \gausspyplus, which means that spectra get flagged if they have blended components ($\flag{blended}{}$), show negative residual features ($\flag{neg.\,res.\,peak}{}$), have broad components ($\pazocal{F}_{\Theta}$, determined with $f_{\Theta,\,\Max} = 2$), have residual data values whose distribution does not correspond to what is expected from Gaussian noise ($\flag{residual}{}$), or were fitted with a number of components that is not consistent with the number of components used in the fit solutions of neighbouring spectra ($\pazocal{F}_{N_{\mathrm{comp}}}$).
Note that $\pazocal{F}_{\Theta}$ indicates the fraction of spectra that contain broad components in relation to neighbouring components.
To better judge how many components with very large absolute FWHM values occur in the decompositions, we also list the fraction of spectra that contain components with FWHM values above $50$ spectral channels ($\pazocal{F}_{\Theta\,>\,50}$) that would imply very high velocity dispersion values of $\sim 4.3$~\kms.
The total flag value $\flag{tot}{}$ gives the percentage of spectra that were flagged by at least one of the individual flags.
For the \gausspy\ decomposition about $59\%$ of the spectra were flagged as not satisfying at least one of the flagging criteria, whereas this reduces to $\sim 35~\text{and}~38\%$ for stage 2 and 3 of \gausspyplus, respectively.
The fit results from stage 2 of \gausspyplus\ shows the lowest fraction of flagged spectra, which is not surprising given that this stage is designed for decreasing the number of flagged spectra. 
Stage 3 of \gausspyplus\ aims to increase the spatial coherence of the fit components, which is why the percentage of flagged spectra increase slightly again compared to the stage 2 fitting results. 
All three stages of \gausspyplus\ perform well in removing negative residual features and reducing fit results that lead to a residual whose distribution is inconsistent with Gaussian noise.
The percentage of spectra flagged with $\flag{blended}{}$, $\pazocal{F}_{\Theta > 50}$ and $\pazocal{F}_{\Ncomp}$ flags actually increases for Stage 1 of \gausspyplus\ compared to the results of \gausspy, which is likely just an effect of the increased number of fit components used in the \gausspyplus\ results. 
These flags are however reduced again in the spatially coherent refitting stages.

\begin{figure*}
  \centering
  \includegraphics[width=2\columnwidth]{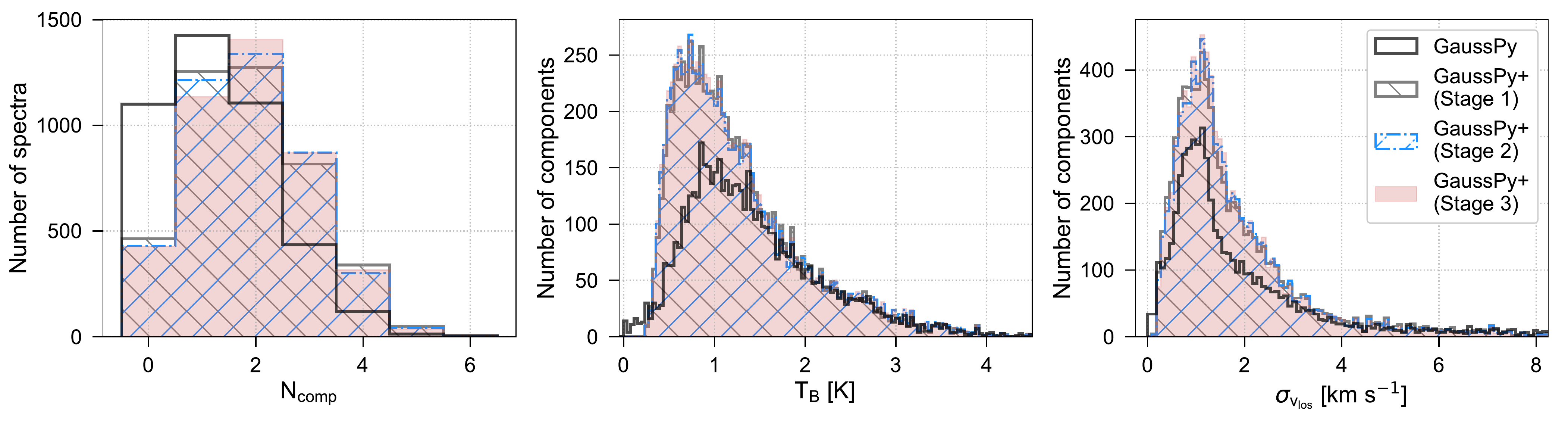}
  \caption{Distribution of fit parameters for the decomposition results of the GRS test field with \gausspy\ and the 3 stages of \gausspyplus. 
  (\textit{left}): Histogram of the number of fitted components per spectrum. 
  (\textit{middle}): Histogram of the amplitude values T$_{\mathrm{B}}$ of all Gaussian fit components.
  The bin size is 0.05~K.
  (\textit{right}): Histogram of the velocity dispersion values $\sigma_{\mathrm{v}_{\mathrm{los}}}$ of all Gaussian fit components.
  The bin size is 0.1~\kms.
  }
  \label{fig:test-field_ncomps_Tb_veldisp}
\end{figure*}

Finally, Fig.$\,\ref{fig:test-field_ncomps_Tb_veldisp}$ shows distributions of fit parameters for the decomposition results of \gausspy\ and the three stages of \gausspyplus. 
The left panel shows histograms of the number of fitted components per spectrum.
As was already demonstrated by panels m--p in Fig.$\,\ref{fig:test-field-comparison}$, \gausspyplus\ manages to fit more spectra than \gausspy, so that the total number of fitted components increases by about one third for the \gausspyplus\ stages.

The middle panel of Fig.$\,\ref{fig:test-field_ncomps_Tb_veldisp}$ shows histograms of the amplitude values of all fit components.
Comparing these distributions with the histogram of the estimated noise values shown in the left panel of Fig.$\,\ref{fig:grs-01_noise_histogram}$ reveals that \gausspyplus\ manages to fit many more components whose S/N value is only $\sim 3$ or lower.
The median S/N value of fit components decreases from 5.4 for the \gausspy\ decomposition to 4.3 for the \gausspyplus\ fit results.

The histograms of the velocity dispersion values for all fit components are given in the right panel of Fig.$\,\ref{fig:grs-01_noise_histogram}$.
The long tail towards increased $\sigma_{\mathrm{v}_{\mathrm{los}}}$ values is mostly due to fitted components with low S/N values; about half of the fit components with $\sigma_{\mathrm{v}_{\mathrm{los}}} > 4.3$~\kms in the \gausspyplus\ decomposition results of Stage 2 and 3 have S/N values $< 2$.


\section{Discussion}
\label{cha:discussion}
In this section, we list potential applications as well as limitations of \gausspyplus\ and give advice on settings to obtain optimal decomposition results.

\subsection{Applications and limitations of \gausspyplus}

The \gausspyplus\ algorithm should be applicable to any dataset that can be well described with Gaussian components; in particular it was designed to decompose large surveys of HI and CO isotopologues.
In case the line shape is better matched by a Voigt or Lorentzian profile (e.g. due to effects of pressure broadening) the decomposition with \gausspyplus\ will likely not give satisfactory results. 
The algorithm can also not fit the hyperfine structure of molecules such as NH$_{3}$ or N$_{2}$H+ directly.

Many of the individual routines implemented in \gausspyplus\, such as the noise estimation (Sect.$\,\ref{sec:noise-estimation}$), signal identification (Sect.$\,\ref{sec:signal-interval}$), and masking of noise artifacts (Sect.$\,\ref{sec:noise-spikes}$) can be used as standalone applications.
For example, the noise estimation can be used in combination with the signal identification to detect baseline shifts, unsubtracted continuum emission, or instrumental artifacts such as increased or amplified noise fluctuations.
Phase 1 of the spatially coherent refitting routine (Sect.$\,\ref{sec:phase_1}$) can also be used to just flag decomposition results without refitting them.

In its current version, \gausspyplus\ is not designed to deal with spectra that contain both emission and absorption lines.
If users would like to use \gausspyplus\ for the decomposition of emission lines that are expected to show strong self-absorption (such as the lowest rotational transitions of the $^{12}$CO molecule), we recommend to deselect the flagging of negative residual spikes, as in this case one would not want to fit a signal peak that has a dip in its center with two components.

The \gausspyplus\ algorithm will only perform well on spectra whose baseline is centered on a value of zero.
Incomplete continuum subtraction or baseline shifts of the spectrum will lead to wrong noise estimates, which in turn will give incorrect decomposition results, since core functionalities of \gausspyplus\ depend on the correctness of the estimated noise values.

The \gausspyplus\ algorithm can deal with large variations of the noise (see Sect.$\,\ref{sec:test-field}$).
However, since key steps of the algorithm are based on S/N thresholds, an inhomogeneous noise coverage or variation in the quality of the data will have an impact on the decomposition results.

In its current implementation \gausspyplus\ does not explicitly check for spatial coherence of the amplitude and FWHM values. 
In principle, these values should also become more coherent in the two phases of the spatially coherent refitting (Sect.$\,\ref{sec:spatial-refitting}$), where neighbouring fit solutions are used to improve the fit of a spectrum.
We focus on spatial coherence of the centroid positions, since it is a necessary requirement for correct amplitude and FWHM values. 
If Gaussian fit components are not placed correctly, their amplitude and FWHM values will by default be spatially inconsistent with neighouring fit solutions.
We also have to caution against constraining the FWHM parameters of Gaussian components with too restrictive limits based on fit solutions from neighbours.
In tests we performed, such a constraint could lead to Gaussian fit components with FWHM values close to the lower or upper limit of the constraint.
This effect caused artefacts in the distribution of all fitted FWHM values, but in case of smaller datasets this might not be easily noticeable.
We thus do not enforce limits for the width of the Gaussian fit components in any of the stages of \gausspyplus, apart from the requirement that the FWHM value has to be larger than the user defined $\Theta_{\Min}$ parameter, whose value defaults to the channel width of the dataset. 
This fitting without an upper limit and without a more constrained lower limit could allow fluctuations in the FWHM values between the Gaussian components of neighbouring spectra.

Our approach in phase 2 of the spatially coherent refitting will also favor structures with ellipsoid morphologies over possible ring-like structures (see Fig.$\,\ref{fig:schematic_phase_2}$).
Users thus should be cautious in using the spatially coherent refitting for centroid positions if the structures probed by the observations are not expected to be continuous over multiple neighbouring pixels or the data is not Nyquist sampled.

\subsection{Recommended settings for \gausspyplus}

We tested the default settings of \gausspyplus\ on synthetic spectra and line emission data from a $^{13}$CO survey and obtained very good decomposition results with them.
However, different datasets may require significantly different settings.
For example, in HI observations we would expect two distinct populations of narrow and very broad lineshapes corresponding to contributions from the cold and warm neutral medium respectively \citep[e.g.][]{Heiles2003}, which is not the case for observations of CO isotopologues. 
For the HI observations one would thus not flag and refit broad Gaussian components (Sect.$\,\ref{sec:broad-components}$), whereas this setting can lead to better decomposition results for the CO datasets.
Ultimately, it is the responsibility of the user to consider if the decomposition results of \gausspyplus\ are scientifically meaningful for the chosen application.

In our application of \gausspyplus\ on the GRS dataset we found it beneficial for the fitting to also retain weak components with amplitudes below a S/N threshold of 3. 
Since the decomposition of \gausspyplus\ performs a least squares minimization of the residual, the fit of higher peaks in a spectrum can be negatively affected if weak components get discarded or neglected.
We thus recommend to also accept components with $\snminfit < 3$ in the decomposition and only later on perform a cut based on their S/N values.

The \gausspyplus\ algorithm is designed to deal with spectra that contain only weak emission lines with S/N values around $3$ or even lower.
The quality check for the significance of a Gaussian component is specifically designed to help in such cases where \gausspyplus\ operates close to the noise. 
If the chosen settings for \gausspyplus\ produce too many false positives, users are advised to increase the chosen S/N limit and/or increase the value of the $\significance{\Min}$ threshold.
Conversely, in case the decomposition results of \gausspyplus\ are not including a significant fraction of signal peaks, users should try to decrease one or both of these parameter settings (see App.~\ref{app:snratio_vs_significance} for how changing both of these parameters affects the decomposition).

We designed \gausspyplus\ to be customizable to different datasets, which means that most of its parameters can be changed and finetuned by the user (see Table$\,$\ref{tbl:gausspyplus-keywords}).
However, the majority of parameters should yield good results for most datasets if left to their default settings.
To get first decomposition results only a small number of parameters (listed as \textit{essential parameters}) have to be specified by the user.
In case the decomposition does not yield good results we recommend to first change the \textit{essential parameters} before changing the parameters listed under \textit{more advanced settings} in Table$\,$\ref{tbl:gausspyplus-keywords}.


\section{Summary}

In this work, we present the \gausspyplus\ algorithm, a new fully automated Gaussian fitting package for the decomposition of emission line spectra.
The \gausspyplus\ algorithm is built upon \gausspy\ \citep{Lindner2015}, but significantly extends and improves upon its performance by the following added, fully automated functionality:

\begin{enumerate}
	\item
	Preparatory steps that can also be used as standalone applications.
	This includes methods to accurately estimate the noise (Sect.$\,\ref{sec:noise-estimation}$), identify signal peaks (Sect.$\,\ref{sec:signal-interval}$), and mask out noise artefacts (Sect.$\,\ref{sec:noise-spikes}$). 
	An additional routine (Sect.$\,\ref{sec:training-set}$) creates suitable training sets for the in-built machine learning process \gausspy\ uses to infer optimal parameter settings for the decomposition of a dataset.
	\item 
	Quality controls that are highly customizable to different datasets (Sect.$\,\ref{sec:checks}$).
	This includes a criterion that takes into account both the S/N values and the number of spectral channels of a signal feature or fitted Gaussian component (Sect.$\,\ref{sec:significance}$) and goodness of fit criteria to aid in the selection of the best fit solution for a spectrum (Sect.$\,\ref{sec:goodness-of-fit}$).
	Additional optional quality controls (Sect.$\,\ref{sec:quality-optional}$) allow the user to flag and refit undesired features in the decomposition such as blended Gaussian components, negative peaks in the residual, very broad Gaussian components, residual data points that are not normally distributed, or differences in the number of fitted components between neighbouring spectra.
	\item
	An improved fitting routine (Sect.$\,\ref{sec:improved-fitting}$) that is guided by the user-defined optional quality controls.
	\item 
	A spatially coherent refitting routine (Sect.$\,\ref{sec:spatial-refitting}$) that tries to refit spectra that do not pass the user-defined quality controls or spectra whose decompositions shows spatial incoherence with neighbouring fit solutions
\end{enumerate}

We thoroughly tested the performance of \gausspyplus\ on synthetic spectra designed to cover a wide range of spectral features expected in observations of emission lines of CO isotopologues. 
We found that it yields very good decomposition results that significantly outperform the original \gausspy\ algorithm in all tested cases (Sect.$\,\ref{sec:synthetic-spectra}$).
We also applied \gausspyplus\ to a test field from the Galactic Ring Survey (Sect.$\,\ref{sec:test-field}$) and showed that it can fit the data well resulting in considerable improvements in the decomposition compared to the original \gausspy\ algorithm. 

We conclude that the \gausspyplus\ algorithm is a powerful tool to analyze large Galactic plane surveys, such as GRS or SEDIGISM \citep{Schuller2017}. 
We will present and discuss its application on the entire GRS dataset in a forthcoming paper.

\begin{acknowledgements}
We would like to thank the anonymous referee for a very constructive, detailed and clear report that helped to significantly improve this work.
This project received funding from the European Union’s Horizon 2020 research and innovation program under grant agreement No 639459 (PROMISE). 
C.E.M acknowledges support from a National Science Foundation Astronomy and Astrophysics Postdoctoral Fellowship under Award No. AST-1801471.
This publication makes use of molecular line data from the Boston University-FCRAO Galactic Ring Survey (GRS). 
The GRS is a joint project of Boston University and Five College Radio Astronomy Observatory, funded by the National Science Foundation under grants AST-9800334, AST-0098562, \& AST-0100793.   
      \\\textbf{Code bibliography}:
      This research made use of \textsc{matplotlib} \citep{Hunter2007}, a suite of open-source python modules that provides a framework for creating scientific plots, \textsc{astropy}, a community-developed core Python package for Astronomy \citep{astropy}, and \textsc{aplpy}, an open-source plotting package for Python \citep{APLpy}
\end{acknowledgements}


\bibliographystyle{aa} 
\bibliography{bibliography} 

\begin{appendix} 


\section{Markov Chain}
\label{app:markov}

The basic principle or question behind this step is: given a certain peak in the spectrum, what is the probability that this peak was caused by random fluctuations of the noise?
This probability depends on the size of the spectrum, as the probability that random noise fluctuations cause a feature resembling a signal peak will increase with the number of spectral channels.
Note that the following probabilistic estimation does \textit{not} attempt to quantify the probability of a signal peak being a real feature, but tries to establish the probability of a peak being the result of random noise fluctuations.

This estimate proceeds as follows:

First, we convert the spectrum into a binary sequence by setting negative channels to a value of 0 and positive channels to a value of 1 (we treat channels that have an exact value of zero as positive channels).
Assuming that each channel can be treated independently from each other and is not correlated with its neighbouring channels, this binary sequence is analogous to a sequence of coin tosses, with the number of coin tosses equivalent to the number of spectral channels. 

This transformation thus allows us to work out the probability of a sequence of negative or positive channels being due to random noise fluctuations.
In case of pure white noise, the probability of a spectral channel having a positive or negative value is $\sfrac{1}{2}$.
To calculate the probability of a sequence of $n$ negative or positive spectral channels we use a one-step Markov chain with state space of $\lbrace 1,2,\cdots,n\rbrace$.
The $n\times n$ transition matrix $P_{i, j}$ that we use to determine the probability of a sequence of $n$ negative or positive consecutive channels has the following structure:

\begin{equation}
    P_{i,j} = 
       \begin{pmatrix}
          p_{i=1,j=1} & p_{i=1,j=2} & \cdots & p_{i=1,j=n} \\
          p_{i=2,j=1} & p_{i=2,j=2} & \cdots & p_{i=2,j=n} \\
          \vdots  & \vdots  & \ddots & \vdots \\
          p_{i=n,j=1} & p_{i=n,j=2} & \cdots & p_{i=n,j=n} \\
       \end{pmatrix}
\end{equation}

The rows $i$ give the possible states the system can be in (pre-transition states) and the column entries give the probability of transitioning to respective new states. 
That means that all of the elements in a row have to sum up to a probability of 1 ($\sum_{j=1}^{n} p_{i,j} = 1$).

The individual entries $p_{i,j}$ of the transition matrix have the following values:

\begin{equation}
    p_{i,j}=
        \begin{cases}
          \sfrac{1}{2}, & \text{for}\ i=1,2,\cdots,n-1 \text{~and}\ j=i+1 \\
          \sfrac{1}{2}, & \text{for}\ i=1,2,\cdots,n-1 \text{~and}\ j=1 \\
          1, & \text{for}\ i=n \text{~and}\ j=n \\
          0, & \text{otherwise}
        \end{cases}
\end{equation}

This allows us to determine the probability of finding a sequence of $n$ consecutive negative or positive channels in a spectrum with $N$ channels. 
We start in state 1 (the first spectral channel has either a positive or a negative value) and need to determine the probability of being in state $n$ (corresponding to a sequence of $n$ spectral channels with either positive or negative values) after $N-1$ Markov chain steps. 
\textit{This probability is given by the $p_{1,n}$ entry of the one-step transition matrix of the form $n\times n$ raised to the power of $N - 1$.}
We can thus compute the probability for any sequence of $n$ consecutive positive or negative channels in a spectrum with $\Nchan$ spectral channels with random values.

Let us illustrate this with the example of a Markov chain for $4$ consecutive negative or positive channels. 
In this case the Markov chain has a state space of $\lbrace 1, 2, 3, 4\rbrace$ and the transition matrix has the following form:

\begin{equation}
    P_{i,j} = 
       \begin{pmatrix}
          \sfrac{1}{2} & \sfrac{1}{2} & 0 & 0 \\
          \sfrac{1}{2} & 0 & \sfrac{1}{2} & 0 \\
          \sfrac{1}{2} & 0 & 0 & \sfrac{1}{2} \\
          0 & 0 & 0 & 1 \\
       \end{pmatrix}
       \label{eq:transition_matrix_example}
\end{equation}

In state 1 (which corresponds to row $i = 1$ of the transition matrix) we have a sequence of one positive or negative channel and we will always start with this state or revert to this state if the sign between neighbouring channels changes before we reached the full sequence of four consecutive channels. 
In state 2 (row $i=2$) and state 3 (row $i=3$) we have a sequence of two and three positive or negative channels, respectively.
State 4 (row $i=4$) is the absorbing final state, where we reached four consecutive positive or negative channels.
The individual column entries of each row then give the probabilities of moving to a new state.
In our example, the transition matrix element $p_{1,2}$ gives the probability of moving from state 1 to state 2 ($p_{1,2}=\sfrac{1}{2}$), and the element $p_{4,3}$ gives the probability of moving from state 4 to state 3 ($p_{4,3}=0$).

In our example we always start out with a spectral channel that has either a positive or negative value, so state 1 is just a sequence of 1 positive or negative channel. 
For state 1, there is a probability of $\sfrac{1}{2}$ that the system stays in state 1 (if the value of the next channel changes sign) or that it moves to state 2 (row $i=2$), in which we have two consecutive channels with the same sign.
For state 2 and state 3, there is again a probability of $\sfrac{1}{2}$ that the channel value changes sign and the system moves back to state 1, and a probability of $\sfrac{1}{2}$ that it moves to state 3 or the absorbing state 4, respectively. 

\begin{figure}
    \centering
    \includegraphics[width=\columnwidth]{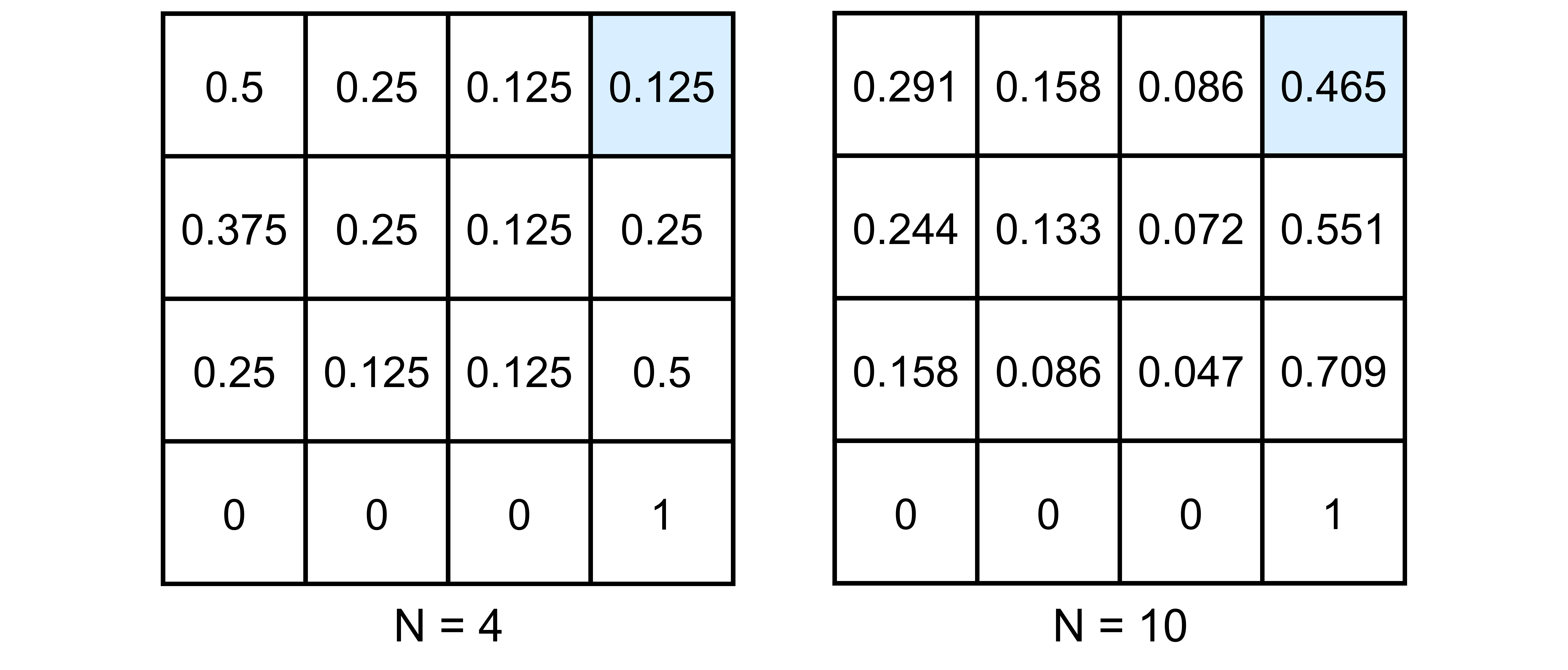}
    \caption{One-step Markov chain results for 4 consecutive negative or positive spectral channels in a sequence of 4 (\textit{left}) or 10 (\textit{right}) channels with random values.
    The value highlighted in blue gives the probability that 4 consecutive channels in the respective sequence are either positive or negative.
    }
    \label{fig:schematic_Markov_chain}
\end{figure}

Figure$\,$\ref{fig:schematic_Markov_chain} shows Markov chain results for 4 consecutive positive or negative channels in a sequence of 4 or 10 channels with random values (left and right panel, respectively).
These matrices were obtained by raising the one-step transition matrix of Eq.$\,$\ref{eq:transition_matrix_example} to the power of 3 and 9, respectively.
The last element in the first row of the matrices (highlighted in blue) gives the respective probabilities to get 4 consecutive positive or negative channels in random sequences of 4 or 10 channels.  

Given the random fluctuations of the noise, it becomes clear that the more spectral channels there are, the higher the probability of getting a sequence of $n$ channels with positive or negative value.
For example, the probability of having a sequence of ten consecutive positive or negative channels in a spectrum of 100 channels is $0.088$.
If the number of spectral channels doubles to 200, the probability of getting a sequence of ten consecutive positive or negative channels increases to $0.173$.

The noise estimation routine of \gausspyplus\ uses a user-defined probability threshold $\plimit$ (default value: $2\%$) to decide which features get masked out for the noise calculation in a spectrum with $\Nchan$ channels.
We use an iterative approach to calculate the minimum necessary number of consecutive positive or negative spectral channels $n$ for which $p_{1,n} < \plimit$.
We start by constructing a transition matrix for $n = 2$ and determine the $p_{1,n}$ value of $P^{\Nchan - 1}$.
If $p_{1,n} > \plimit$ we increase $n$ by one and repeat the calculation.
We stop these iterations once $p_{1,n} < \plimit$ and the final value of $n$ determines the minimum number of consecutive positive or negative channels a feature has to have to get masked out.

For example, for a spectrum with 700 spectral channels, features with more than 15 consecutive positive or negative spectral channels have a probability of less than $2\%$ to be caused by random noise fluctuations and will be thus masked out in the noise calculation routine. 


\section{Testing \gausspyplus\ on synthetic spectra}
\label{app:synthetic_spectra}

\subsection{Sample of synthetic spectra}
\label{app:test_sample}

\begin{figure}
    \centering
    \includegraphics[width=\columnwidth]{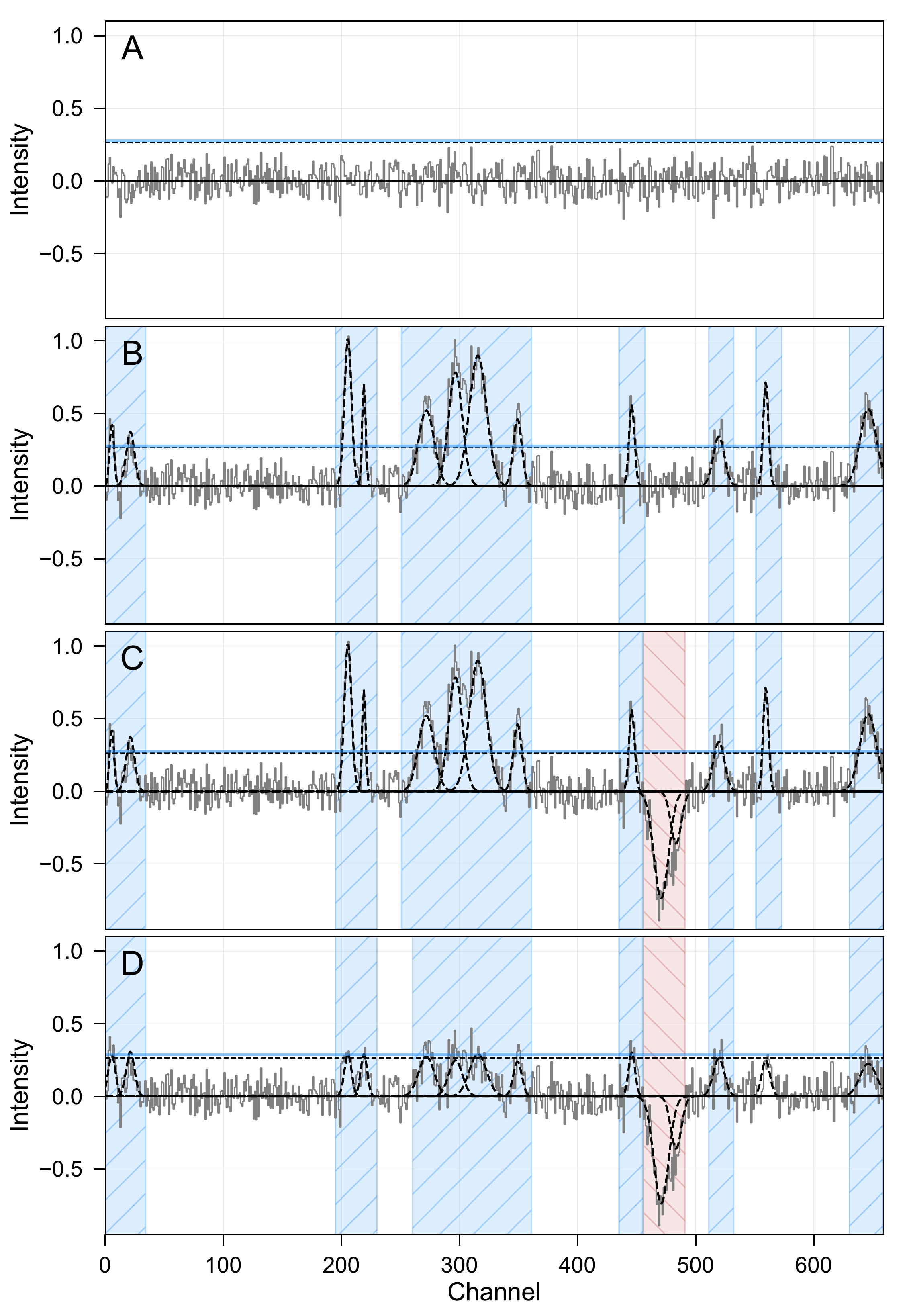}
    \caption{Example spectra from the four samples of synthetic spectra (A -- D) used to test the performance of \gausspyplus.
    Black dotted lines indicate individual Gaussian components of the signal and negative noise spikes.
    The horizontal dashed black lines show a S/N threshold of 3.
    Shaded areas indicate intervals that \gausspyplus\ classified as signal intervals (blue) and noise spikes(red). 
    Note that the noise is the same in all four panels.
    }
    \label{fig:synthetic_spectra_example}
\end{figure}

We created four different samples of $10,000$ synthetic spectra each, to mimic expected properties of spectra (see Fig.~\ref{fig:synthetic_spectra_example} for examples of each sample):
\begin{enumerate}[A:]
	\item
	White noise only.
	\item
	White noise and signal.
	For spectra in this sample up to $12$ Gaussian components ("signal") were added to the white noise of the spectra from sample~A.
	\item
	White noise, signal, and negative noise spikes.
	For spectra in this sample one or two negative Gaussian components ("noise spikes") were added to the spectra from sample~B to mimic instrumental artefacts.
	\item
	White noise, weak signal, and negative noise spikes.
	For spectra in this sample the positive Gaussian components from sample~C had their amplitudes reduced.
	The signal peaks can thus be hidden in the noise, which makes the decomposition very challenging.
\end{enumerate}

\begin{figure}
    \centering
    \includegraphics[width=\columnwidth]{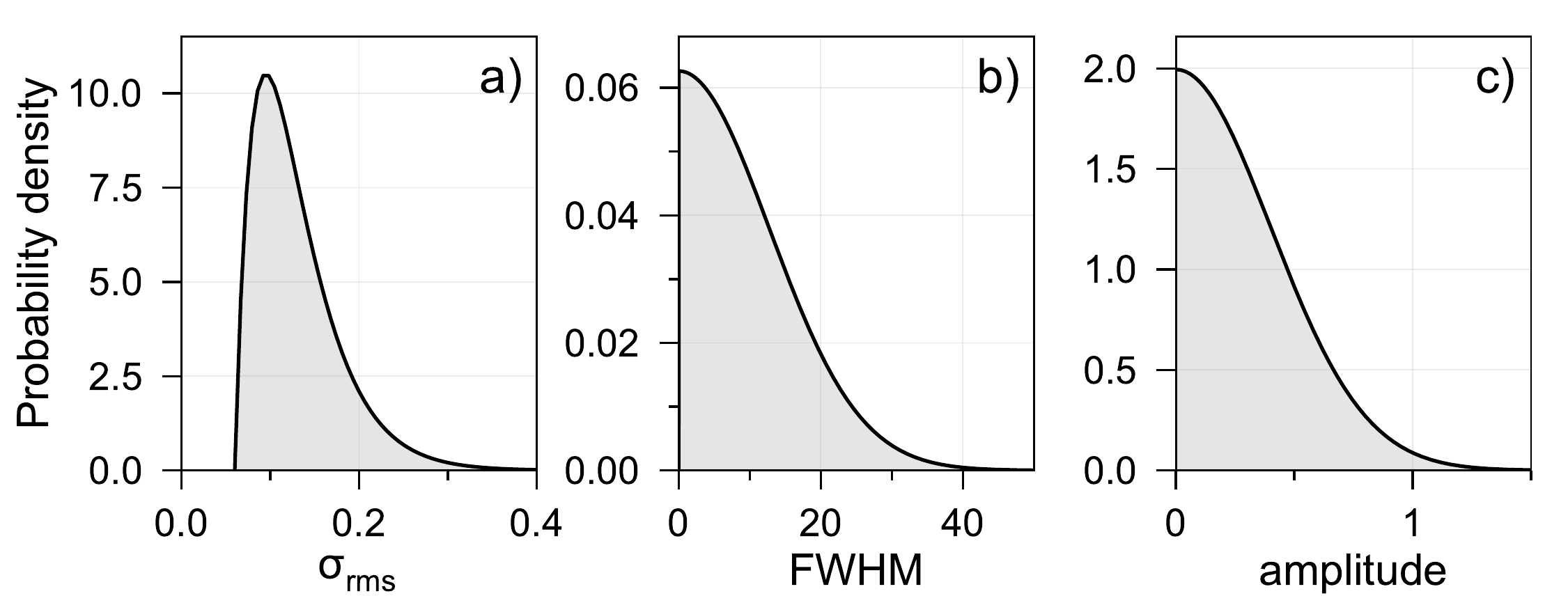}
    \caption{Probability distribution functions for $\rms$ (left), FWHM (middle), and amplitude values (right). 
    For the synthetic spectra, these distributions were randomly sampled to obtain the noise and Gaussian components of the signal.
    }
    \label{fig:app_synthetic_spectra_pdfs}
\end{figure}

The synthetic spectra were set up to closely mimic spectra from the GRS dataset with regards to the number of spectral channels (659), and expected noise and signal properties.
The $\rms$ value used to generate the white noise was randomly sampled from a Gamma distribution of the form 

\begin{equation}
	p\left(x\right) = x^{k-1}\frac{e^{-x/\theta}}{\theta^{k}\Gamma\left(k\right)},
\end{equation}

with $k=2$, and $\theta=0.35$.
To closely mimic the noise distribution of the GRS survey \citep[cf. Fig.~8 from][]{Jackson2006} we shifted the distribution by a value of $0.06$ and scaled it by a factor of $0.1$ (panel~a in Fig.~\ref{fig:app_synthetic_spectra_pdfs}).
The minimum $\rms$ value of our sample is $0.06$~K and we limited the maximum $\rms$ value to $0.4$~K.

The parameters of the Gaussian components of the signal were randomly sampled from distributions set up to resemble the signal peaks observed in the GRS dataset.
We sampled the FWHM values from a standard normal distribution scaled by a factor of $\sim 13$ (panel~b in Fig.~\ref{fig:app_synthetic_spectra_pdfs}).
We limited the FWHM to a maximum value of $50$ spectral channels.
We sampled the amplitude values from another standard normal distribution scaled by a factor of $0.4$ (panel~c in Fig.~\ref{fig:app_synthetic_spectra_pdfs}).
We limited the amplitude range to values of 
$\left[ 3.5\cdot\sigma_{\mathrm{rms}}, 2.5\right]$.
We sampled the mean values of the Gaussians from a uniform distribution over all 659 spectral channels.
For each spectrum, we required for every Gaussian signal component $i$ that: its $\significance{}$ value (Sect.$\,\ref{sec:significance}$) had to be $> 6$; its mean position $\mu_{i}$ had to be at a minimum distance of $\Theta_{j}$ to the mean position $\mu_{j}$ of the closest Gaussian signal component $j$, where $\Theta_{j}$ is the FWHM of the Gaussian component $j$; its FWHM value $\Theta_{i}$ had to be $< 20$~channels if its amplitude value $a_{i}$ was $> 1$.
The last condition was implemented to exclude components with both high amplitude values and broad linewidths.
This exclusion of the strongest components was only done to create a more challenging setup for the decomposition. 
Datasets with low to moderate spatial resolution such as GRS are likely to contain such features that can be caused by the broadening of lines due to the large spatial beamsize and large distances to the emitting physical objects.
However, these strong emission lines are fit well with \gausspy\ and \gausspyplus\ in case no strong blending with other lines is present, as is the case for our samples of synthetic spectra.

The parameters for the negative Gaussian components of the noise spikes were randomly sampled in mean position, amplitude, and FWHM from uniform distributions within the limits $\left[ 0, 659\right]$, $\left[ -4\cdot\sigma_{\mathrm{rms}}, -1.5\right]$, and $\left[ 1, 20\right]$, respectively.
We required that the noise spikes were placed at least a distance of $\Theta_{j}$ from the closest Gaussian signal component $j$.

The amplitude values of the Gaussian components for sample~D were sampled from a uniform distribution with the range $\left[ 2.5\cdot\sigma_{\mathrm{rms}}, 3.5\cdot\sigma_{\mathrm{rms}}\right]$.

\subsection{Performance of the automated noise estimation routine}
\label{app:test_noise}

Here we report the results of the automated noise estimation of \gausspyplus\ (Sect.$\,\ref{sec:noise-estimation}$) on the synthetic spectra from samples~A -- D discussed in the last section.

We used the default settings for the noise estimation routine ($\plimit=0.02$, $\Npad=5$), which means that sequences above $15$ consecutively positive or negative spectral channels get masked out for the noise estimation in addition to peaks that show high amplitude values. 

\begin{figure}
\centering
\includegraphics[width=\columnwidth]{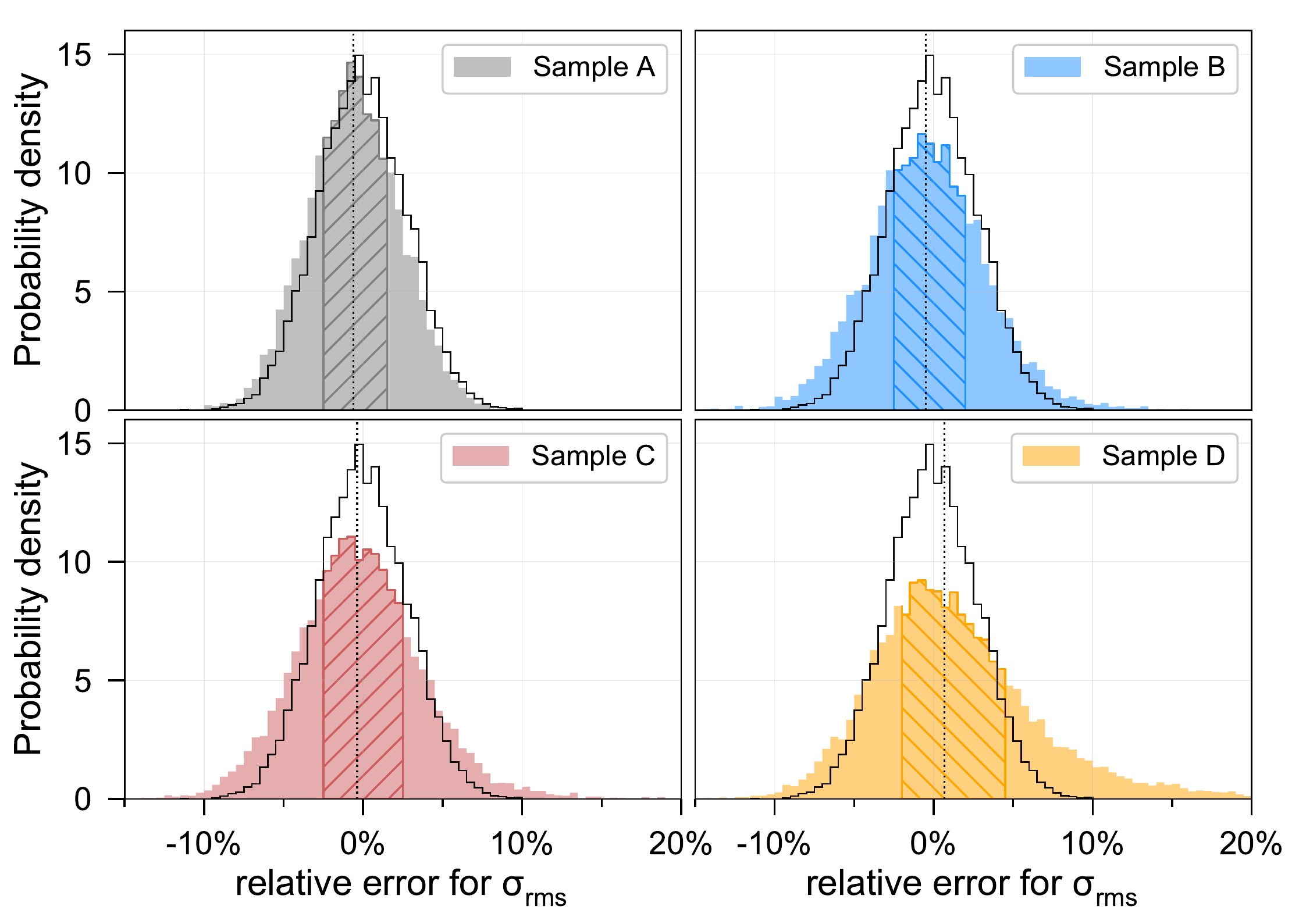}
\caption{Probability density distributions showing the results of our automated noise estimation for the sample of synthetic spectra containing: only white noise (upper left panel); white noise and signal (upper right panel); white noise, signal, and negative noise spikes (lower left panel); white noise, weak signal, and negative noise spikes (lower right panel).
The abscissa shows the determined root-mean-square noise value $\rms$ normalized by the true root-mean-square noise value $\rmstrue$ that was used to generate the white noise. 
Hatched areas and vertical dotted lines show the respective interquartile ranges and median value of the respective distributions.
The black solid line shows the distribution obtained by using all spectral channels from sample~A for the noise calculation.
See App.$\,$ for more details.
}
\label{fig:noise_test_pdf}
\end{figure}

Figure~\ref{fig:noise_test_pdf} shows probability density distributions of the relative errors of the $\rms$ values determined by \gausspyplus.
These relative errors were obtained by comparing the estimated $\rms$ values to the true noise values ($\rmstrue$) used to generate the white noise for all four samples (A -- D).
For comparison, we also show the probability distribution obtained if all channels in the spectra of sample~A are used for the calculation of the $\rms$ value (solid black line).
This distribution corresponds to the best we could do for the calculation of the $\rms$ value and its spread around the $\rmstrue$ value reflects inherent random effects of the noise that would be decreased if the number of spectral channels were increased.

For the majority of the synthetic spectra the noise estimation performed very well with the median of the distribution (dotted vertical line) being very close to the	$\rmstrue$ value and the interquartile ranges (hatched areas) within relative errors of $\pm 3\%$ and $\pm 4\%$ for samples~A -- C and sample~D, respectively.
Since the noise estimation always excludes the spectral channels with the highest negative and positive values (see Sect.$\,\ref{sec:noise-estimation}$), it tends to slightly underestimate the $\rms$ value for spectra containing only noise (sample~A).
For sample~B (white noise and signal), nearly all estimated $\rms$ values are within $\pm 10\%$ of $\rmstrue$.
For the spectra of sample~C the performance of the noise calculation is almost as good, which demonstrates that our method is robust to the presence of negative noise spikes or similar instrumental artefacts.
As expected, for sample~D (white noise, weak signal, noise spikes) we tend to overestimate the $\rms$ values.
However, given that a fraction of the signal peaks in these spectra is buried within the noise, the noise calculation still performs very well, with $\rms$ values within $\pm 10\%$ of $\rmstrue$ for about $93\%$ of the spectra. 

\subsection{Performance of the identification of signal intervals}
\label{app:test_signal_interval}

In this section we report on the results of the automated identification of signal intervals of \gausspyplus\ (Sect.$\,\ref{sec:signal-interval}$) on our samples of synthetic spectra (App.~\ref{app:test_sample}).

We used the default settings of \gausspyplus, with $\snmin=3$, $\significance{\Min}=5$, $\Nmin=100$, and $\Npad=5$.

For sample~A, whose spectra contain no signal, the signal identification had a false positive rate of $0.01\%$; that means out of a sample of $10\,000$ spectra with white noise there was only a single spectrum for which a signal interval was incorrectly identified.

\begin{figure}
    \centering
    \includegraphics[width=\columnwidth]{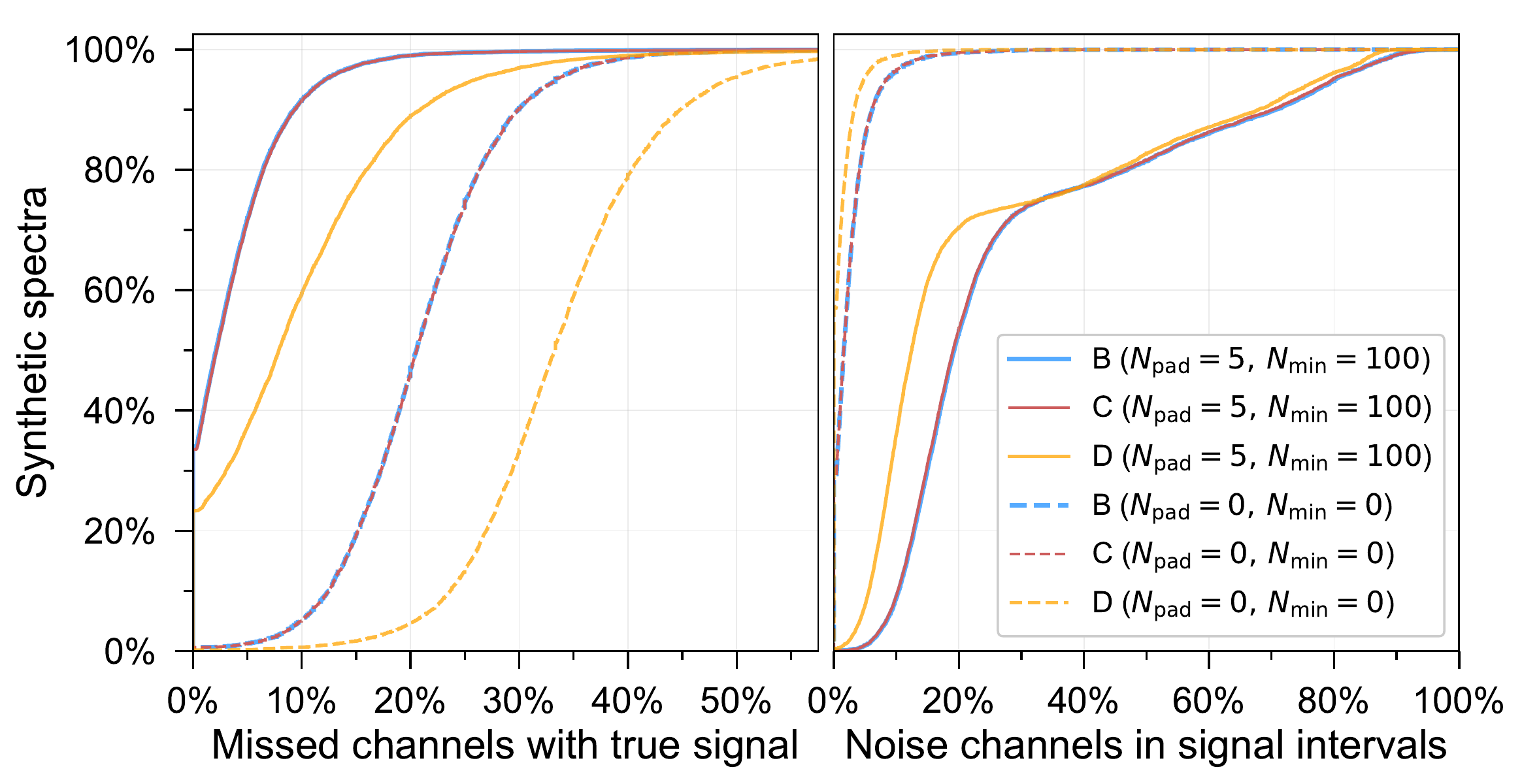}
    \caption{Results of the signal interval identification of \gausspyplus\ for our samples of synthetic spectra.
    (\textit{left}): Cumulative percentage of the synthetic spectra showing the fraction of unidentified spectral channels containing true signal.
    (\textit{right}): Cumulative percentage of the synthetic spectra showing the fraction of identified signal interval channels corresponding to noise.
    See App.$\,$\ref{app:test_signal_interval} for more details.
    }
    \label{fig:app_signal_interval_cumulative}
\end{figure}

The left panel in Fig.~\ref{fig:app_signal_interval_cumulative} shows the cumulative percentage of the synthetic spectra as a function of unidentified spectral channels that contain true signal.
We define the interval of channels containing true signal as all channels within $\mu_{i} \pm \Theta_{i}$ for a true Gaussian signal component $i$.
For $\sim 90\%$ of the spectra in sample~B and C, the fraction of unidentified spectral channels containing signal is $< 10\%$.
In case of weak signal (sample~D), the percentage of unidentified spectral channels with signal is still $< 20\%$ for $\sim 90\%$ of the spectra.
This performance is very good, given that many of the signal peaks in sample~D are by construction indistinguishable from noise features (with their amplitude values ranging from only $2.5\cdot\rms$ to $3.5\cdot\rms$).
The dashed lines indicate runs of the signal interval identification, for which we set the $\Npad$ and $\Nmin$ keywords to zero, meaning that there are no channels added on either side of the identified signal intervals.
The left panel in Fig.~\ref{fig:app_signal_interval_cumulative} demonstrates that we would miss a larger fraction of spectral channels containing true signal by setting $\Npad$ and $\Nmin$ to zero.

The right panel in Fig.~\ref{fig:app_signal_interval_cumulative} shows the cumulative percentage of the synthetic spectra as a function of the fraction of noise channels included in the identified signal intervals, again for the two runs in which we vary the $\Npad$ and $\Nmin$ values as for the left panel.
If $\Npad$ and $\Nmin$ are set to zero, only a very small fraction of noise channels is included in the estimated signal intervals.
As expected, this fraction increases if we extend the signal intervals on both sides by $\Npad = 5$ and require that the signal intervals contain a minimum number of channels per spectrum of $\Nmin = 100$.
However, this has no negative impact on the decomposition, since the signal intervals are only used in the goodness of fit calculations.
It would be more problematic if we set $\Npad$ and $\Nmin$ to zero, because in that case we would miss a higher fraction of real signal, which would not be considered in the goodness of fit estimates.

\subsection{Performance of the masking out of noise artifacts}
\label{app:test_noise_spike}

In this section we report on the performance of \gausspyplus\ in automatically masking noise spikes (Sect.$\,\ref{sec:noise-spikes}$) for the samples of synthetic spectra (App.~\ref{app:test_sample}).
We used the default settings for the $\snspike$ parameter that masks out all spectral features that contain negative values below $-5\times\rms$. 

Our routine managed to correctly identify $99.4\%$ and $98.8\%$ of all noise spikes with minimum values $< -5 \times \rmstrue$ in the synthetic spectra of samples~C and D, respectively. 
The small fraction of unidentified noise spikes with S/N ratios $< -5 \times \rmstrue$ was due to overestimates of the $\rms$ values.

The fraction of false positives---that means noise fluctuations that were incorrectly identified as noise spikes---was $0.02\%$ for both samples.

The performance of the masking of noise artifacts is also illustrated in Fig.$\,$\ref{fig:synthetic_spectra_example} and Fig.$\,$\ref{fig:spectra_comp_improved_gausspy}, where the shaded red areas indicate the spectral channels identified as noise spikes.

\subsection{Performance of the automated decomposition routine for the training set}
\label{app:test_trainingset}

As discussed in Sect.$\,\ref{sec:training-set}$, \gausspyplus\ can supply a training set for the determination of the best smoothing parameters for a dataset.
Here we discuss the performance results of the automated decomposition of spectra for the training set. 
We quantify the performance by comparing the resulting smoothing parameters $\alpha_{1}$ and $\alpha_{2}$ obtained from the decomposed training set with the smoothing parameters obtained for the same training set if the true known Gaussian parameters are supplied.
For the training sets, we randomly selected $250$ synthetic spectra from samples~B -- D (App.~\ref{app:test_sample}).
We then created two training sets for each sample by: i) decomposing the spectra via the method discussed in Sect.$\,\ref{sec:training-set}$; ii) supplying the true parameters for the Gaussian components of the synthetic spectra.

Table~\ref{tbl:test_trainingset} lists the result of the gradient descent technique applied by \gausspy\ to determine the best smoothing parameters for the training sets. 
The run in which the true values of the Gaussian components were supplied in the training set is indicated with "(true)".
For all runs the S/N ratio for the spectrum and its second derivative were set to $\text{SNR}_{1} = \text{SNR}_{2} = 3$.

\begin{table}
    \caption{Comparison of obtained smoothing parameter values $\alpha_{1}$ and $\alpha_{2}$ and the corresponding F$_{1}$ score for different training sets.}
    \centering
    \small
    \renewcommand{\arraystretch}{1.2}
    \begin{tabular}{cccc}
    \hline\hline
    Sample &$\alpha_{1}$ &$\alpha_{2}$ &F$_{1}$ score [$\%$]\\
    \hline
    B & $2.08$ & $4.91$ & $82.4$\\
    B (true) & $2.03$ & $4.91$ & $82.7$\\
    \hline
    C & $2.11$ & $4.89$ & $79.0$\\
    C (true) & $2.07$ & $4.87$ & $79.6$\\
    \hline
    D & $3.23$ & $4.98$ & $69.0$\\
    D (true) & $3.44$ & $5.09$ & $71.5$\\
    \hline
    \end{tabular}
    \label{tbl:test_trainingset}
\end{table}

For sample~B and C the runs for both training sets converge to essentially the same smoothing parameters $\alpha_{1}$ and $\alpha_{2}$.
For sample~D, the value for $\alpha_{1}$ inferred from the training set decomposed with our routine is slightly smaller than the parameter we get from the true values.
We tested the effect of this change by repeating the \gausspy\ decomposition for sample~D with the smoothing parameter values $\alpha_{1} = 3.44$ and $\alpha_{2} = 5.09$.
We then recomputed the percentage of correct identifications ($30.4\%$) and false positives ($6.9\%$) in the same way as for the values inferred from the decomposed training set given in Table$\,\ref{tbl:decomposition_correct_means}$ ($29.4\%$ and $6.5\%$ for the correct identifications and false positives, respectively).
This shows that the slight difference in the smoothing parameter inferred for sample~D has only a limited impact on the \gausspy\ decomposition results.

The comparison in Table$\,$\ref{tbl:test_trainingset} thus demonstrates that the automated method for creating training sets that is implemented in \gausspyplus\ works well, so that smoothing parameters close to the optimal value can be obtained from it.

\subsection{Performance of the Gaussian decomposition}
\label{app:test_decomposition}

Here we compare the performance of the decomposition of the original \gausspy\ algorithm and the improved fitting routine of \gausspyplus\ (Sect.$\,\ref{sec:improved-fitting}$) on our samples of synthetic spectra (App.~\ref{app:test_sample}).

\begin{figure*}
    \centering
    \includegraphics[width=2\columnwidth]{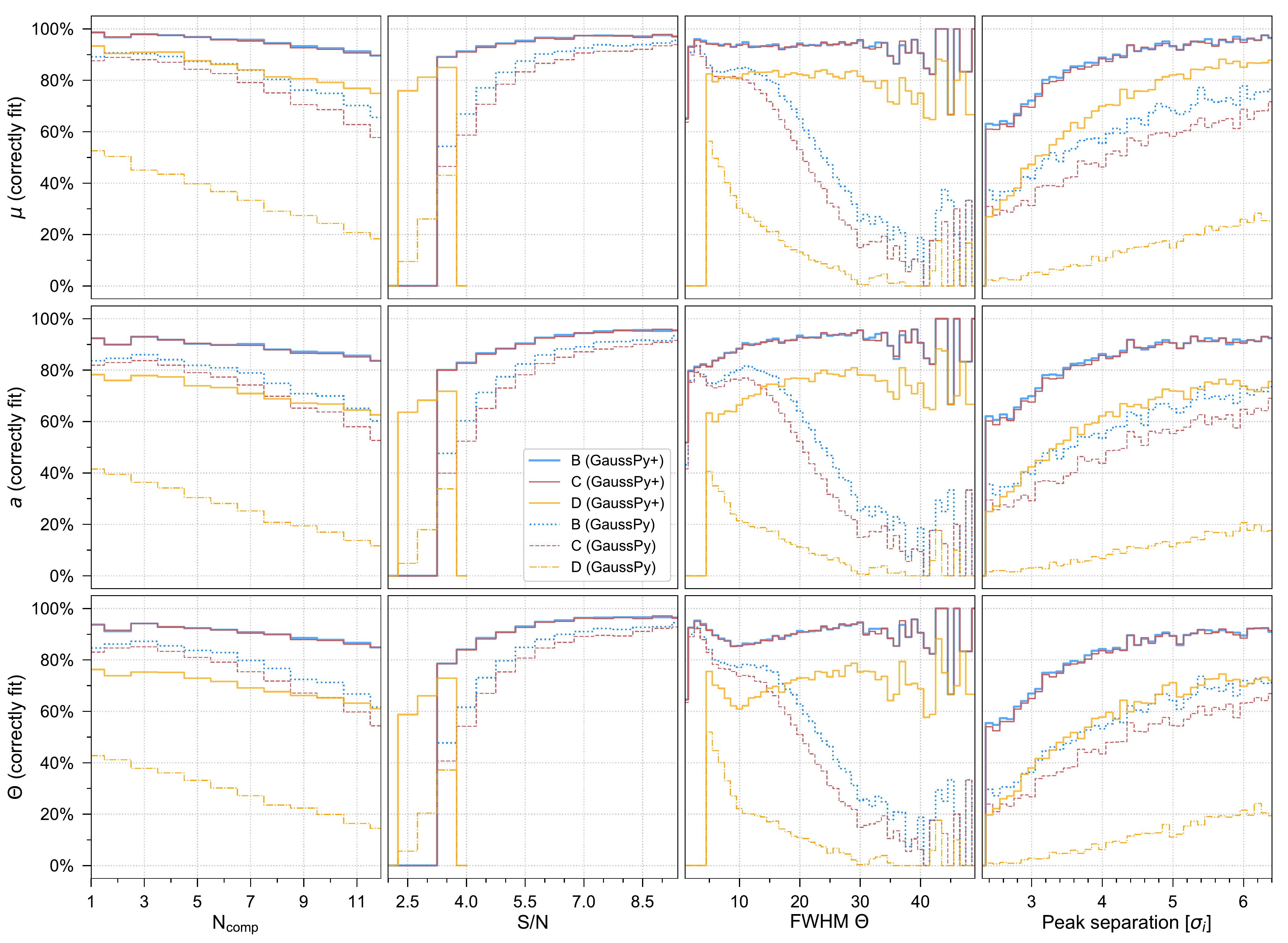}
    \caption{Performance of the \gausspy\ and \gausspyplus\ decomposition runs for samples of synthetic spectra.
    The ordinate in the upper, middle, and lower panels show the fraction of correctly fit Gaussian mean positions, amplitude values, and FWHM values, respectively, plotted against the number of true Gaussian components, the S/N ratio, and the true FWHM values in the left, center, and right panels, respectively.
    See App.$\,$\ref{app:test_decomposition} for more details.
    }
    \label{fig:app_correct_means+amps}
\end{figure*}

First we explore how the performance of the decomposition results of \gausspy\ and \gausspyplus\ for sample~B--D of the synthetic spectra (App.~\ref{app:test_sample}) varies with the number of components in the spectrum, the S/N ratio, and the width of the signal peaks.
We counted the mean position of fitted Gaussian components as correct if their values were within $\pm 2$ channels of the peak positions of the true underlying signal peak.
We counted amplitude and FWHM parameters as correctly fit if their values were within $\pm 20\%$ of the true value in addition to the requirement that the fitted mean position is within $\pm 2$ channels of the true position of the component.
Since for narrow signal peaks $20\%$ of the FWHM can amount to only a fraction of a channel we additionally count a FWHM parameter as correctly fit if its absolute error is within $\pm 2$ spectral channels of the correct FWHM value.

The left panels in Fig.~\ref{fig:app_correct_means+amps} show the percentage of correctly identified Gaussian fit parameters (mean position, amplitude and FWHM value from top to bottom, respectively) as a function of the number of components in the spectrum.
The \gausspyplus\ decomposition shows a very stable performance that is not much affected by a higher number of components or the existence of noise spikes.
Even in the case of signal peaks very close to the detection threshold (sample~D) it still yields a good performance.
In contrast, the ability of the original \gausspy\ algorithm to correctly decompose the components deteriorates by about $10-20\%$ for the synthetic spectra of sample~B and C, and up to $30\%$ for sample~D the more complex the spectra are.

The centre left panels in Fig.~\ref{fig:app_correct_means+amps} show the number of correctly determined Gaussian fit parameters as a function of the S/N~ratio. 
As expected, the performance results strongly depend on the S/N ratio.
However, compared to the results of the original \gausspy\ algorithm, the \gausspyplus\ decomposition gives a significantly better performance, especially in determining correct fit parameters for signal peaks with S/N values $\leq 3$, which can be heavily affected by the noise. 

The centre right panels in Fig.~\ref{fig:app_correct_means+amps} show the number of correctly determined Gaussian fit parameters as a function of the FWHM values of the true signal peaks.
In contrast to the \gausspy\ fit results, the performance of the \gausspyplus\ decomposition does not deteriorate with increasing width of the signal peaks, which means that both narrow and broad components are fit well.
The decomposition with the original \gausspy\ algorithm shows a much stronger dependence on the line width, and has difficulties in correctly decomposing broader components.

Finally, the right panels in Fig.~\ref{fig:app_correct_means+amps} show the percentage of correctly determined Gaussian fit parameters of the signal component $i$ as a function of peak separation to the closest neighbouring signal component $j$.
This peak separation is given as multiples of the standard deviation $\sigma_{i}$ of component $i$.
As expected, the performance of the decomposition with \gausspyplus\ decreases the closer two components are placed to each other as it gets exceedingly more difficult to correctly deblend them.
Nonetheless, the decomposition with \gausspyplus\ manages to fit about $\sim 60\%$ of even the most heavily blended components in sample~B and C correctly, which exceeds the performance of \gausspy\ by more than $20\%$.
For the challenging weak signal peaks of sample~D, the fraction of correctly decomposed components that were blended the most was lower ($\sim 20 \text{--} 30\%$).
However, the percentage of correct fits increases already significantly for moderate peak separations of $\sim 3 \text{--} 4 \times \sigma_{i}$ and reaches a stable high performance for even larger peak separations.
We test the performance of \gausspyplus\ for blended components in more detail in App.~\ref{app:test_recovery}.

Note also that the performance of \gausspyplus\ is unaffected by the presence of negative noise spikes, whereas the \gausspy\ decomposition results deteriorate in case such noise spikes are present.

We tried to choose fair criteria for the definition of when we count components in Fig.~\ref{fig:app_correct_means+amps} as correctly fit. 
Given that many of our signal peaks show only low to moderate S/N values, noise properties might already severely affect their lineshapes, so stricter criteria would not accept decomposition results that a human would likely classify as correctly fit.
Conversely, more relaxed criteria could allow too large absolute deviations from the correct parameter values. 
However, we repeated the analysis of Fig.~\ref{fig:app_correct_means+amps} for both stricter and more relaxed criteria and we do recover the same general trends: performance results that exceed the decomposition of \gausspy\ and are almost unaffected by the number of components in the spectrum, the FWHM value or the presence of noise spikes, and increase with higher S/N values or larger peak separations. 

\begin{figure}
    \centering
    \includegraphics[width=\columnwidth]{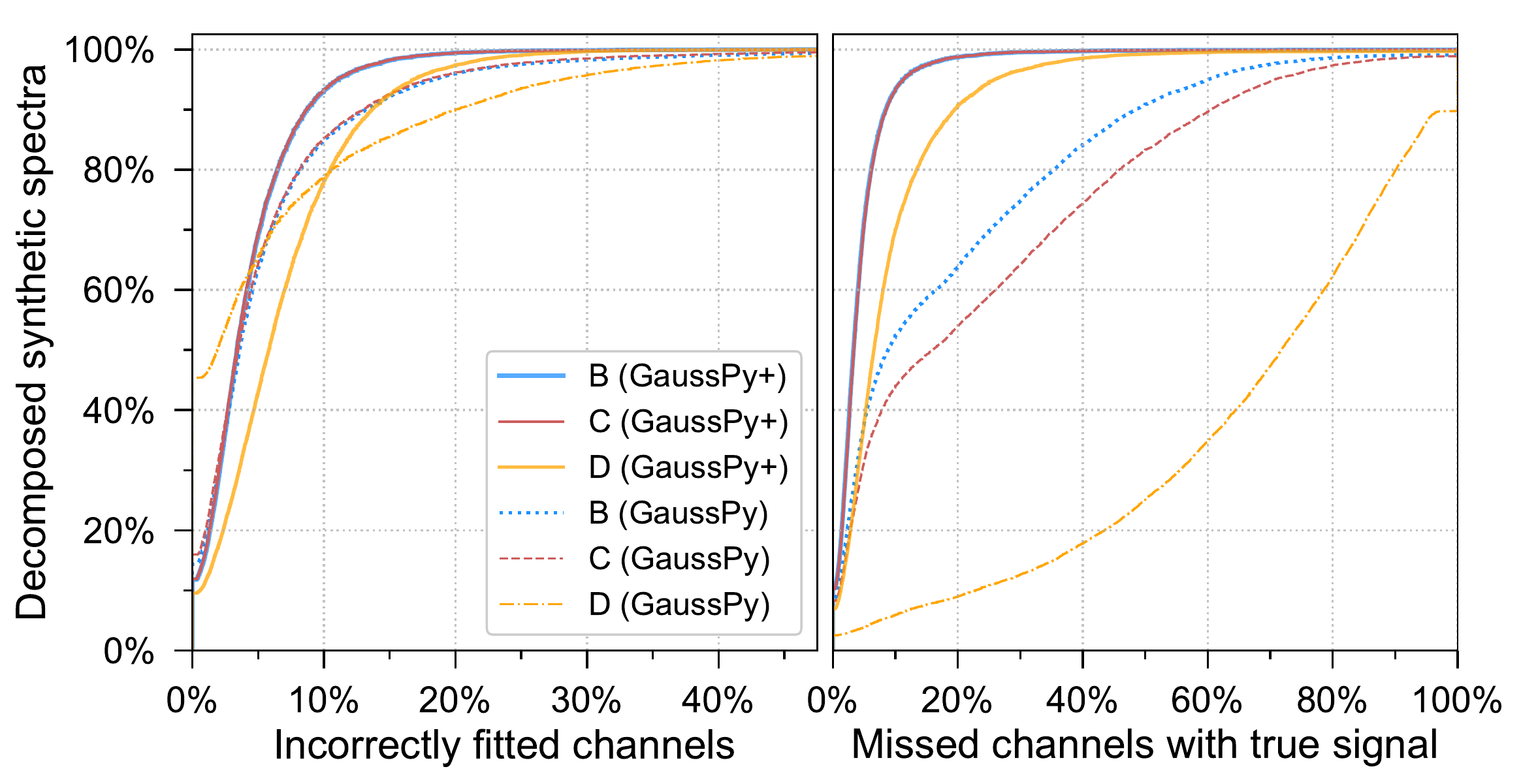}
    \caption{Comparison of the decomposition results obtained with \gausspy\ and \gausspyplus\ for our samples of synthetic spectra.
    (\textit{left}): Cumulative percentage of decomposed spectra showing the fraction of spectral channels that were incorrectly fit.
    (\textit{right}): Cumulative percentage of decomposed spectra showing the fraction of spectral channels containing true signal that were not fit.
    See App.$\,$\ref{app:test_decomposition} for more details.
    }
    \label{fig:app_decomp_interval_cumulative}
\end{figure}

Next, we compare the number of fitted spectral channels with the channels containing true signal for the \gausspy\ and \gausspyplus\ decompositions.
We define the interval of fitted channels or channels containing true signal as all channels within $\mu_{i} \pm \Theta_{i}$ for a fitted or true Gaussian component $i$.
The left panel in Fig.~\ref{fig:app_decomp_interval_cumulative} shows the cumulative percentage of decomposed synthetic spectra as a function of the percentage of incorrectly fitted spectral channels.
Both \gausspy\ and \gausspyplus\ show a very good performance with a low fraction of false positives.
The improved results of \gausspyplus\ are due to its ability to more correctly identify individual signal peaks where \gausspy\ fits a single Gaussian components over multiple peaks.

The right panel of Fig.~\ref{fig:app_decomp_interval_cumulative} shows the cumulative percentage of decomposed synthetic spectra as a function of spectral channels containing true signal that were not fit by Gaussian components.
For all three samples of synthetic spectra \gausspyplus\ significantly improves the decomposition results of \gausspy\ by fitting more components at their correct positions.
The improvement is especially striking in case of the spectra containing only weak signal of sample~D.

Figure~\ref{fig:app_decomp_interval_cumulative} thus illustrates that \gausspyplus\ manages to fit significantly more channels containing true signal than \gausspy; moreover, it does so without fitting too many noise features.

\begin{figure*}
    \centering
    \includegraphics[width=2\columnwidth]{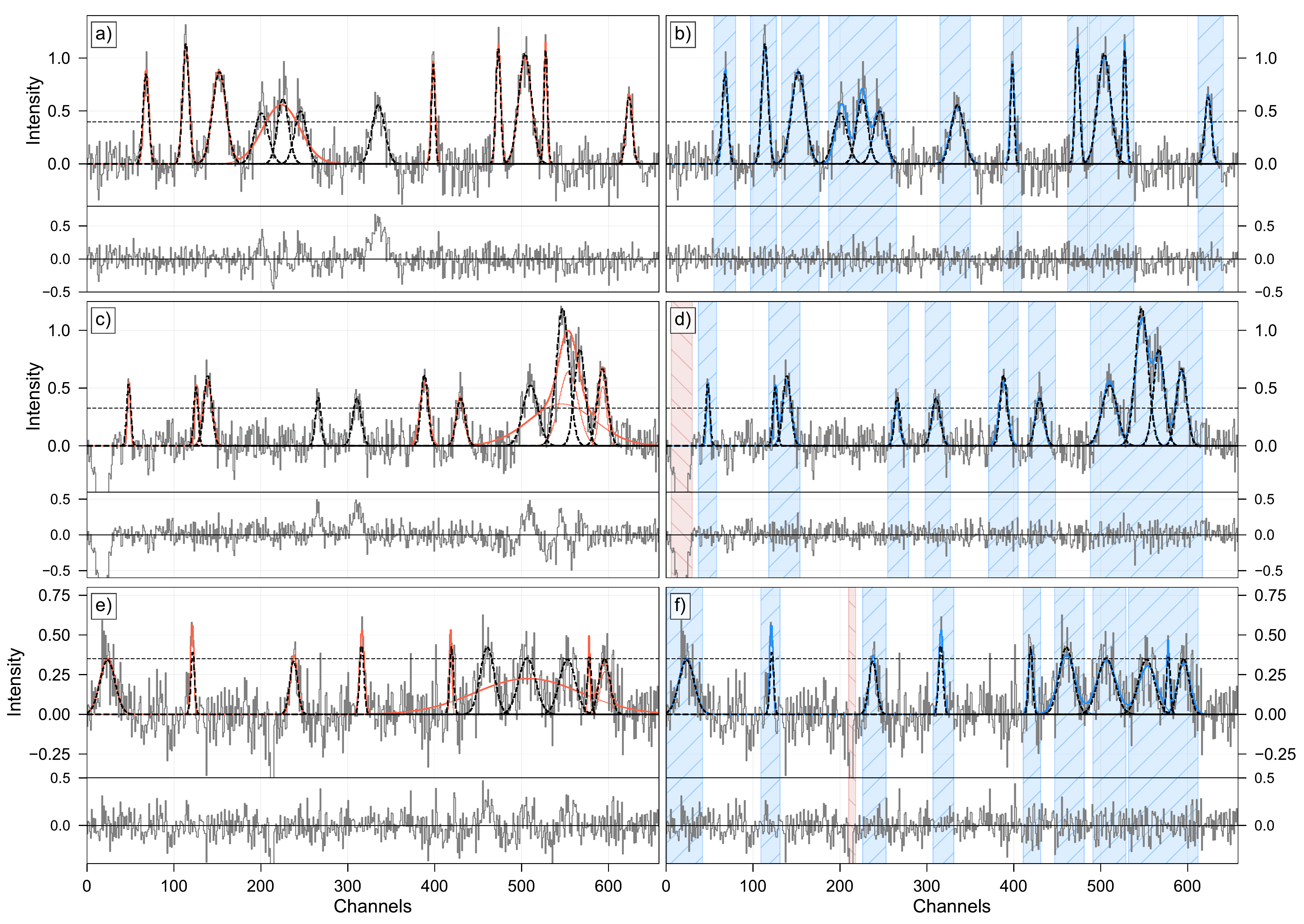}
    \caption{Example spectra illustrating the better performance of the improved fitting routine of \gausspyplus\ (Sect.$\,\ref{sec:improved-fitting}$) compared to the original \gausspy\ algorithm.
    The upper (a, b), middle (c, d), and lower (e, f) panels show synthetic spectra from samples~B, C, and D, respectively.
    The panels on the left (a, c, e) show the decomposition results obtained with the original \gausspy\ algorithm and panels to the right (b, d, f) show the corresponding decomposition results from our improved fitting routine \gausspyplus.
    The correct individual Gaussian components are indicated in dotted black lines; individual Gaussian components and their combined intensity from the decomposition run with \gausspy\ and \gausspyplus\ are indicated in dotted and solid red and dotted and solid blue lines, respectively.  
    The smaller panels below the spectrum show the corresponding residual.
    Dashed black lines indicate a S/N ratio of $3$.
    Blue and red shaded areas show the automatically identified signal and noise spike intervals, respectively.
    }
    \label{fig:spectra_comp_improved_gausspy}
\end{figure*}

The improved performance of \gausspyplus\ is further illustrated in Fig.~\ref{fig:spectra_comp_improved_gausspy}, which contrasts decompositions of the original \gausspy\ algorithm (left panels) with decompositions obtained with our improved fitting routine (right panels) for synthetic spectra from sample~B--D. 
Figure~\ref{fig:spectra_comp_improved_gausspy} shows that the original \gausspy\ algorithm sometimes has problems in decomposing mildly blended signal peaks and signal peaks at the edge of the spectrum, whereas \gausspyplus\ has no problems in fitting those components correctly.
Note also that \gausspyplus\ does a good job of identifying signal peaks and noise artefacts.

\subsection{Recovery of identical components with different S/N ratios and degrees of blendedness}
\label{app:test_recovery}

Here we quantify how well the improved fitting algorithm (Sect.$\,\ref{sec:improved-fitting}$) of \gausspyplus\ is able to recover blended components.
For this, we create a sample of $11\,340$ synthetic spectra  that contain two identical Gaussian signal peaks.
The parameters of the signal peaks could have the values: $[3, 3.5, ..., 7]$ for the S/N ratio; $[5, 10, ..., 30]$ spectral channels for the FWHM; and $[1, 1.2, ..., 5]\cdot\sigma_{i}$ for the separation of the mean positions of the signal peaks.
We created ten spectra of each possible parameter combination and added different noise sampled from a $\rms$ value of 0.13 to each spectrum\footnote{The number of spectral channels (659) and the $\rms$ value were again chosen to closely mimic properties of the GRS dataset.}.

We constructed a training set by randomly selecting $500$ spectra of different parameter combinations and inferred smoothing parameters $\alpha_{1}$ and $\alpha_{2}$ by supplying the true values of the signal peaks. 
Since our aim here is to establish the performance of our decomposition given ideal settings, we supplied the true paramater values as solutions instead of decomposing the training set with the method described in Sect.$\,\ref{sec:training-set}$.
From this training set we inferred smoothing parameters values of $\alpha_{1} = 2.16$ and $\alpha_{2} = 6.19$ that led to an F$_{1}$ score of $76.8\%$.
We then performed decompositions with the original \gausspy\ algorithm and the improved fitting routine of \gausspyplus\ (Sect.$\,\ref{sec:improved-fitting}$), leaving all the settings at their default values.

\begin{figure}
    \centering
    \includegraphics[width=\columnwidth]{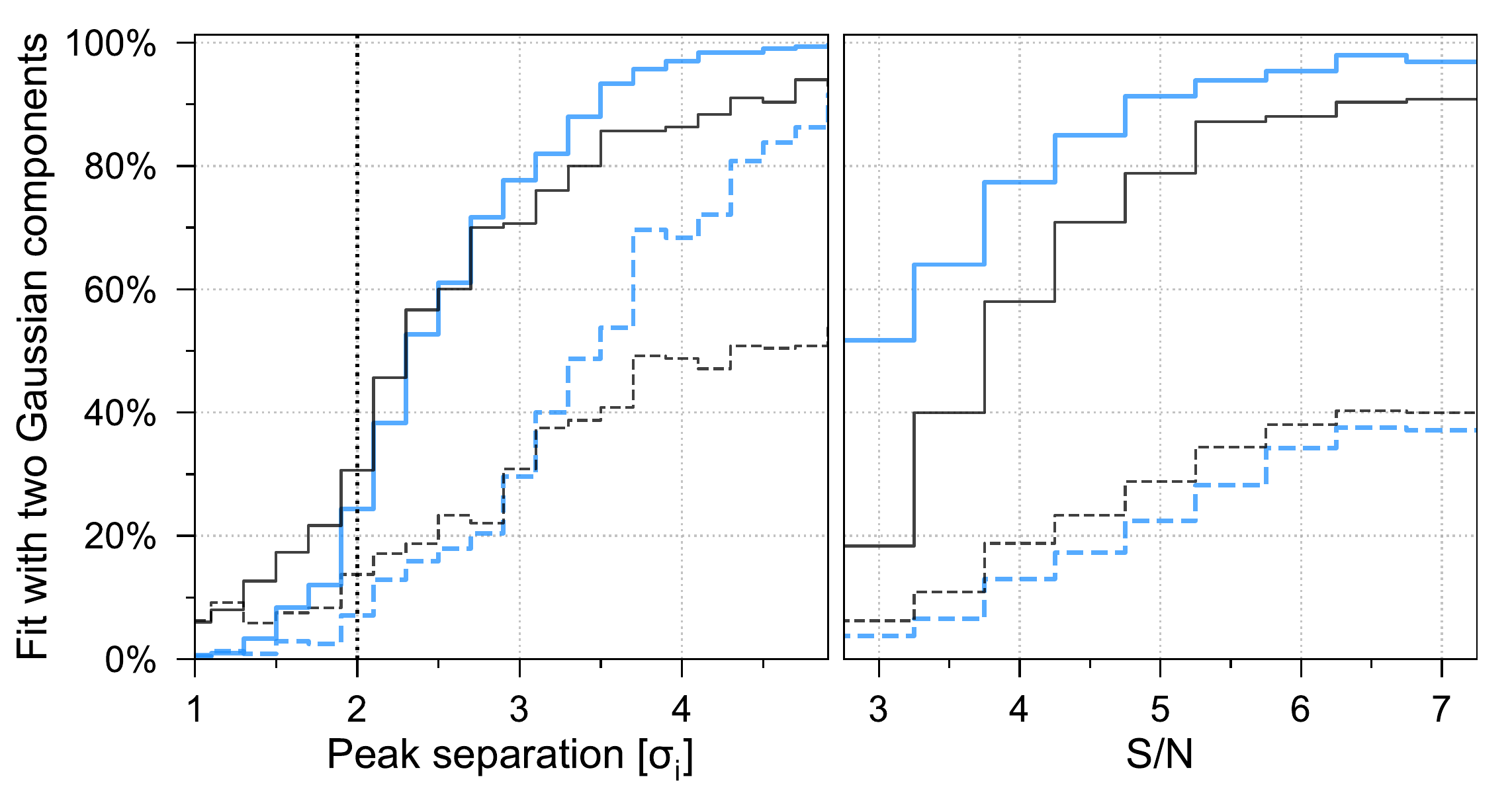}
    \caption{Decomposition results of a sample of synthetic spectra with two identical Gaussian components, whose S/N ratio, peak separation and FWHM parameter were varied.
    Blue and black lines indicate the results obtained for the runs with \gausspyplus\ and \gausspy, respectively.
    (\textit{left}:) Percentage of fitted spectra with two Gaussian components as a function of peak separation for S/N ratios $< 5$ (dashed lines) and $\geq 5$ (solid lines).
    The dotted vertical line indicates the separation threshold for two identical Gaussian components without noise. 
    (\textit{right}:) Percentage of fitted spectra with two Gaussian components as a function of their S/N ratio for peak separations of $< 3\cdot\sigma_{i}$ (dashed lines) and $\geq 3\cdot\sigma_{i}$ (solid lines).
    }
    \label{fig:app_test_simple_spectra}
\end{figure}

Figure~\ref{fig:app_test_simple_spectra} shows the performance results of the two decomposition runs.
The left panel shows the percentage of fits using two Gaussian components as a function of peak separation, split into a sample with low to moderate S/N ratios ($< 5$, dashed lines) and high S/N ratios ($\geq 5$, solid lines).
The vertical dotted line indicates the separation threshold for two identical Gaussian components in case of no noise (see also Sect.$\,\ref{sec:blended-comps}$).
In case of low to moderate S/N ratios it becomes very difficult to differentiate two similar Gaussian components if their peak positions are separated by less than about $3.5$ times their standard deviation.
For higher S/N ratios identical signal peaks can be located closer together until they essentially become indistinguishable from a single component; for the decomposition with \gausspyplus\ the signal peaks have to be located closer than $\sim 2.5 \cdot \sigma_{i}$ until the majority of signal peaks will be fit with two components.
Note that \gausspyplus\ by design fits preferentially a single instead of two components if the peaks are only separated closely, as in such cases a fit with a single Gaussian component will already be a good match to the combined signal peaks and the simplest fit solution is preferred without additional information (e.g. from neighbouring fit solutions) to inform the fit.
For larger peak separation \gausspyplus\ exceeds the performance of \gausspy, especially in the case of low to moderate S/N values.

The right panel of Fig.~\ref{fig:app_test_simple_spectra} shows the percentage of decomposition results using two Gaussian fit components, split into two samples with small ($< 3 \cdot\sigma_{i}$, dashed lines) and large ($\geq 3 \cdot\sigma_{i}$, solid lines) peak separations.
For small peak separations, \gausspy\ and \gausspyplus\ will preferentially fit the signal peaks with a single component, even if the S/N ratio is high. 
For larger peak separations the two-component fit solution is dominant and the percentage of spectra fit with two components increases significantly for high S/N ratios.

Note that since the decomposition was performed without any additional knowledge about the signal peaks (as could be imposed by neighbouring spectra in spatially coherent decompositions), it can become very challenging to correctly fit signal peaks with low S/N ratios, as random fluctuations of the noise can significantly change their shape. 
Moreover, the two identical signal peaks we placed in the spectra will combine to a symmetric peak that might be equally well fit by a single or two components if they are heavily blended.
Spectral features of two blended components of different shape will cause an asymmetry that can make it easier to decompose them correctly.

\section{Performance details for \gausspyplus}

\subsection{Performance and execution time for the decomposition of the training set}
\label{app:time_training_set}

We compared the decomposition results and runtime of the \textsc{SLSQPLSQFitter} fitting routine used to create training sets for \gausspy\ (Sect.$\,\ref{sec:training-set}$) with the runtime of the improved fitting routine of \gausspyplus\ (Sect.$\,\ref{sec:improved-fitting}$).
We used both fitting techniques to decompose sample~B of our synthetic spectra (App.$\,$\ref{app:test_sample}).
For both algorithms we distributed the decomposition over 50 CPUs.

In terms of performance the decomposition with the \textsc{SLSQPLSQFitter} could correctly identify $95.4\%$ of the signal components and had a false positive fraction of $1.5\%$.
Both of these values exceed the corresponding numbers for the results of \gausspyplus\ ($93.7\%$ and $1.6\%$, respectively, cf. Table$\,\ref{tbl:decomposition_correct_means}$), which confirms that our routine for creating training sets produces high quality decompositions. 

\begin{table}
    \caption{Comparison of the execution times for sample~B of the synthetic spectra.}
    \centering
    \small
    \renewcommand{\arraystretch}{1.2}
    \begin{tabular}{ccc}
    \hline\hline
     & t$_{\mathrm{real}}$ [min] & t$_{\mathrm{CPU}}$ [min]\\
    \hline
    \textsc{SLSQPLSQFitter} & 43.06 & 1868.87\\
    \gausspyplus\ & 2.59 & 110.01\\
    \hline
    \end{tabular}
    \label{tbl:time-comp_training_routine}
\end{table}

Table$\,$\ref{tbl:time-comp_training_routine} lists the results of the execution times: t$_{\mathrm{real}}$ is the elapsed wall clock time from start to finish of the execution of the decomposition and t$_{\mathrm{CPU}}$ is the total amount of spent CPU time.
These results show that the \textsc{SLSQPLSQFitter} fitting routine is about an order of magnitude slower than \gausspyplus, which is why we recommend to use the former routine only for the decomposition of spectra for the training set. 

\subsection{Execution time for the GRS test field}
\label{app:execution_time_grs}

In this section we discuss the execution time of the \gausspyplus\ algorithm for the decomposition of the GRS test field using the default settings of \gausspyplus\ and distributing the computation over 50 CPUs.

\begin{table}
\caption{Comparison of the execution times for the \gausspyplus\ decomposition of the GRS test field.}
    \centering
    \small
    \renewcommand{\arraystretch}{1.2}
\begin{tabular}{ccccc}
\hline\hline \multirow{ 2}{*}{Method} & t$_{\mathrm{real}}$ & f$_{\mathrm{real}}$ & t$_{\mathrm{CPU}}$ & f$_{\mathrm{CPU}}$ \\
 & [min] & [\%] & [min] & [\%] \\
\hline
Training set creation & 0.39 & 4.5 & 16.33 & 21.3 \\
Training & 5.61 & 64.2 & 20.86 & 27.2 \\
Preparation & 0.08 & 0.9 & 0.28 & 0.4 \\
Decomposition Stage 1 & 0.32 & 3.6 & 9.55 & 12.4 \\
Decomposition Stage 2 & 0.44 & 5.0 & 8.03 & 10.5 \\
Decomposition Stage 3 & 1.91 & 21.8 & 16.69 & 21.8 \\
\hline
\end{tabular}
\label{tbl:time-comp_grs}
\end{table}

Table~\ref{tbl:time-comp_grs} shows an overview of the execution time for all stages of \gausspyplus\ in terms of wall clock time t$_{\mathrm{real}}$ and total CPU time t$_{\mathrm{CPU}}$ as well as their respective relative percentages f$_{\mathrm{real}}$ and f$_{\mathrm{CPU}}$.
The entire \gausspyplus\ decomposition for the GRS test field needed t$_{\mathrm{real}} = 8.74$~min and t$_{\mathrm{CPU}} = 76.74$~min.

Since the total size of the GRS test field ($4200$ spectra) is relatively small, the creation of the training set and training with \gausspy\ amounted to a significant contribution to f$_{\mathrm{real}}$ and f$_{\mathrm{CPU}}$, which would be reduced for larger datasets, where the decomposition steps will need a larger fraction of the total time.
We also report the individual times for the execution of the three decomposition stages of \gausspyplus: the improved fitting routine (Sect.$\,\ref{sec:improved-fitting}$; Stage~1), phase 1 of the spatially coherent refitting (Sect.$\,\ref{sec:phase_1}$, Stage~2), and phase 2 of the spatially coherent refitting (Sect.$\,\ref{sec:phase_2}$, Stage~3).
Execution times for the spatially coherent refitting stages will typically depend on how many criteria are used in the flagging of spectra in Stage~2 and the minimum weight threshold $\pazocal{W}_{\Min}$ the user selects in Stage~3 (cf. Fig~\ref{fig:app_iterations}).

\subsection{Effect of varying minimum S/N ratio and significance}
\label{app:snratio_vs_significance}

Here we test how changing the values of the minimum S/N ratio $\snmin$ and significance parameter $\significance{fit}$ affects the decomposition results for our samples of synthetic spectra (App.$\,$\ref{app:test_sample}).
We use $\snmin$ and $\significance{fit}$ values of $[2.5, 3, 3.5, 4]$ and $[4, 5, 6, 7]$ respectively, and perform a decomposition with \gausspyplus\ for every combination of those values (16 in total).
For the spectra of sample~A that contain only white noise we found that with significance values $\significance{fit} \geq 5$ no noise features were fit. 
For a significance value of $\significance{fit} = 4$ and $\snmin$ values of 2.5, 3, 3.5, and 4, \gausspyplus\ incorrectly fitted 38, 31, 21, and 8 noise features, respectively.

\begin{figure}
    \centering
    \includegraphics[width=\columnwidth]{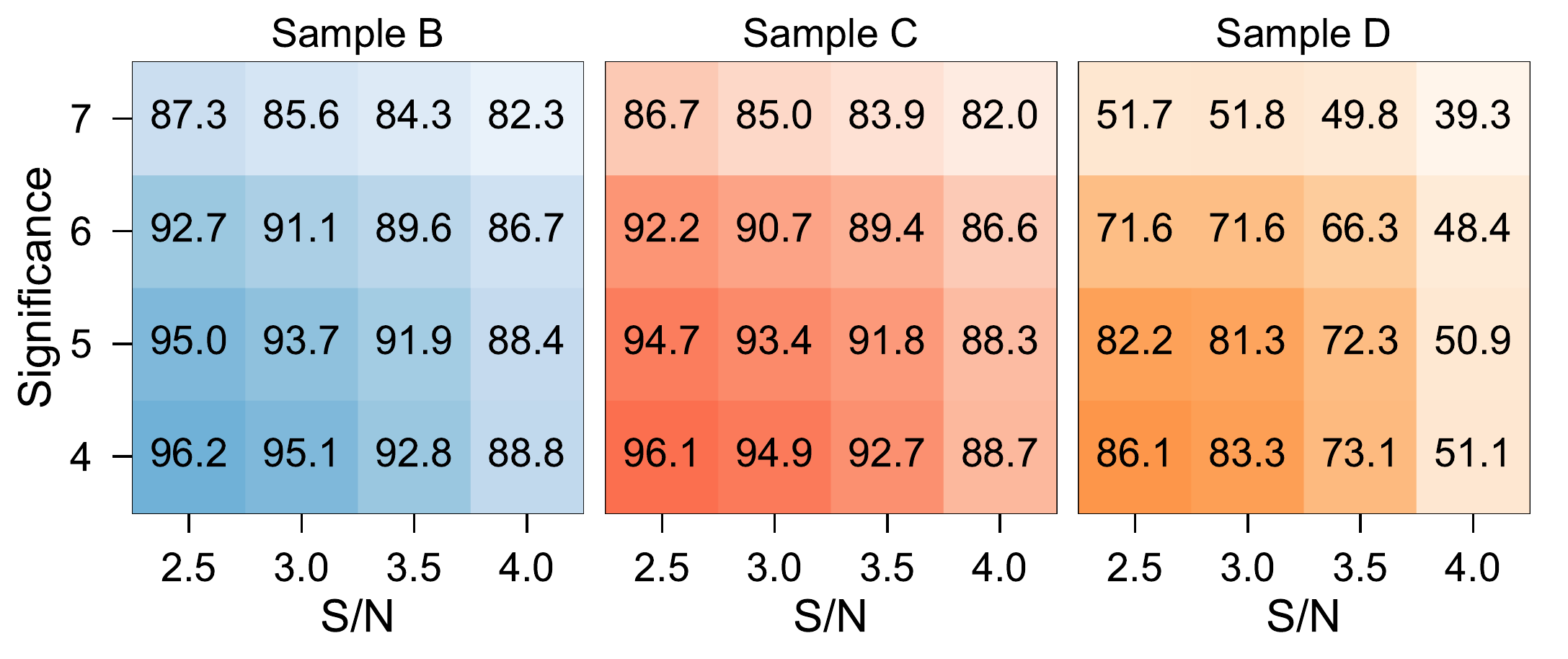}
    \caption{Percentage of correctly identified mean positions of Gaussian components in the decomposition of samples~B--D (left to right) with varying values for the minimum S/N ratio and significance parameters.
    }
    \label{fig:app_snr_significance_tp}
\end{figure}

\begin{figure}
    \centering
    \includegraphics[width=\columnwidth]{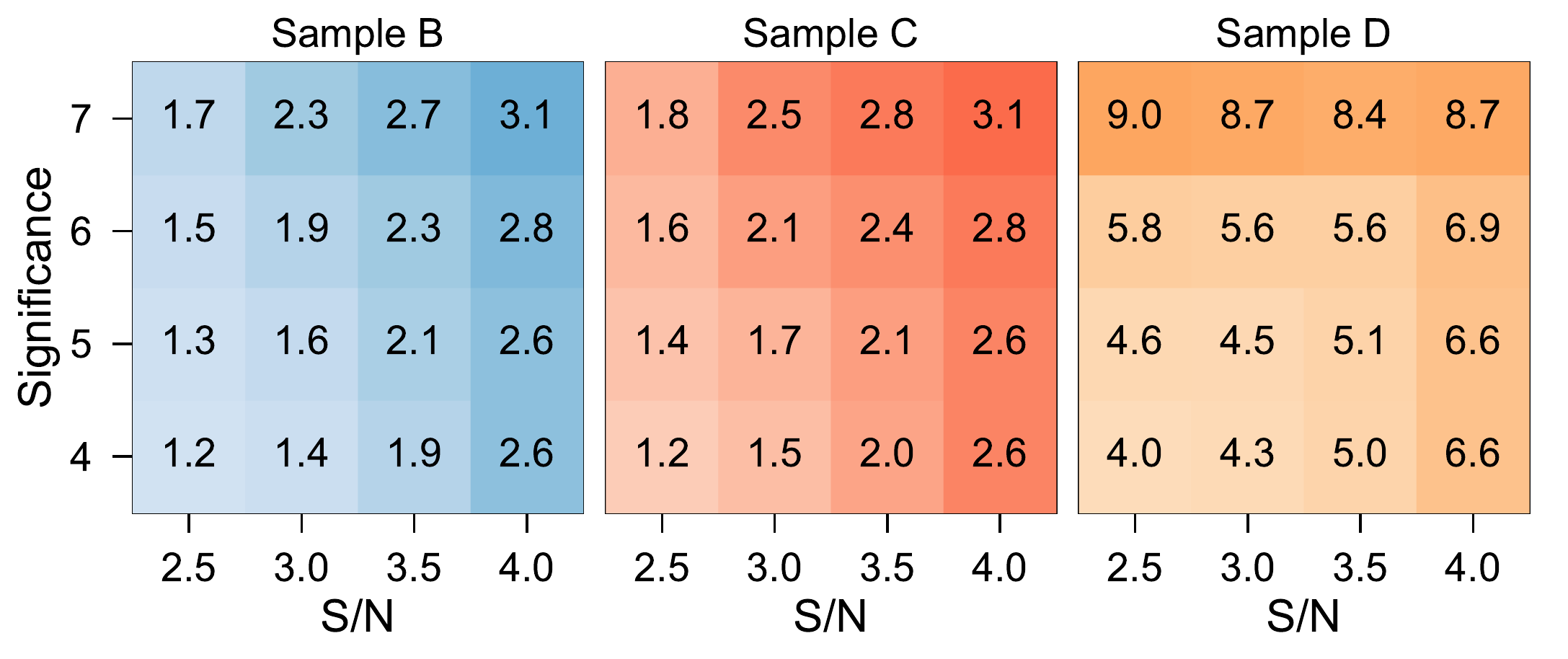}
    \caption{Percentage of incorrectly identified mean positions of Gaussian components in the decomposition of samples~B--D (left to right) with varying values for the minimum S/N ratio and significance parameters.
    }
    \label{fig:app_snr_significance_fp}
\end{figure}

For samples~B--D we calculated the percentage of correctly and incorrectly fitted mean position values of Gaussian components for each decomposition run, which are shown in Fig.$\,$\ref{fig:app_snr_significance_tp} and Fig.$\,$\ref{fig:app_snr_significance_fp}, respectively.
We count the mean position of a Gaussian component as correctly detected if it is within $\pm 2$ channels of the true value. 
If the mean position value of a fitted component was more than $4$ channels away from the true mean positions of all signal components in the spectrum we counted it as an incorrect identification. 
Note that the decomposition with $\significance{fit} = 5$ and $\snmin = 3$ corresponds to the \gausspyplus\ run at its default settings we presented in Sect.$\,\ref{sec:synthetic-spectra}$.
Figure~\ref{fig:app_snr_significance_tp} and Fig.$\,$\ref{fig:app_snr_significance_fp} demonstrate the interdependence between the $\significance{fit}$ and $\snmin$ parameters.
In general, increasing one of these parameters has adverse effects on the percentage of correct and incorrect detections of Gaussian components in the synthetic spectra of samples~B--D.
However, this adverse effect can be offset by decreasing the value for the other parameter.
The results from Fig.$\,$\ref{fig:app_snr_significance_fp} show that setting the $\significance{fit}$ and $\snmin$ parameters to higher values can lead to a big increase in incorrectly identified fit components. 
This increase is due to the large fraction of signal components with low S/N ratios in our synthetic spectra. 
If we set the $\snmin$ or $\significance{fit}$ values to higher values, those components are either prevented from being fit or are incorrectly fit with one broad component instead of multiple narrower ones. 

Figure~\ref{fig:app_snr_significance_tp} and Fig.$\,$\ref{fig:app_snr_significance_fp} also demonstrate that we could have improved the decomposition results reported in Sect.$\,\ref{sec:synthetic-spectra}$ by choosing a lower minimum S/N ratio of $\snmin = 2.5$.
In principle we could have even further improved upon that result by also decreasing the required significance value to $\significance{fit} = 4$, but the results from the decomposition of sample~A demonstrate that this setting would already allow the fit of noise features.

Ultimately, the choice for the values of the $\snmin$ and $\significance{fit}$ parameters needs to be guided by the dataset.
For the synthetic spectra we used perfect Gaussian noise properties, which will likely not be the case for real observational data.
Thus users might want to set higher values for the $\significance{fit}$ parameters to exclude the fitting of noise features, even though it might result in a reduction of fitted weaker signal peaks. 
Note also that we constructed the synthetic spectra of samples~B--D to contain a large fraction of signal peaks with amplitudes close to or even below a S/N ratio of 3 to test how well the \gausspyplus\ decomposition with default settings works for weak signal peaks. 
Decompositions of datasets for which users expect signal peaks with high S/N ratios will thus likely benefit from an increase of the values for the $\snmin$ and $\significance{fit}$ parameters.

\subsection{Performance of in-built and optional quality control procedures}
\label{app:performance_quality_control}

Here we discuss the performance of the in-built and optional quality control procedures described in Sect.$\,\ref{sec:fwhm}\text{--}\ref{sec:channel-range}$ and Sect.$\,\ref{sec:negative-residual}\text{--}\ref{sec:blended-comps}$.

For $\sim34\%$ of the spectra of the GRS test field at least one of the in-built quality control procedures was used to remove one or more components in the decomposition with the improved fitting routine of \gausspyplus.
For sample A -- D of the synthetic spectra the percentage of spectra for which components were removed due to failing the in-built quality controls was $\sim3\%, 20\%, 20\%\text{ and }22\%$, respectively.
The comparatively larger fraction of spectra with rejected fit components in the decomposition of the GRS test field was mostly due to the presence of low-intensity signal peaks that did not satisfy the criterion for the amplitude value and imperfect noise properties, which led to the fitting of noise peaks that did not satisfy the requirement for the significance value.

\begin{table}
\caption{Number of fit components removed by the in-built quality control procedures for the decomposition of the GRS test field and the synthetic spectra.}
    \centering
    \small
    \renewcommand{\arraystretch}{1.2}
\begin{tabular}{ccccc}
\hline\hline  & $\Theta$ & $a$ & $\significance{fit}$ & $\mu$ \\
 & (3.2.1.1)\tablefootmark{a} & (3.2.1.2)\tablefootmark{a} & (3.2.1.3)\tablefootmark{a} & (3.2.1.4)\tablefootmark{a} \\
\hline
GRS test field & 136 & 705 & 1127 & 16 \\
Sample A & 17 & 7 & 263 & 0\\
Sample B & 673 & 669 & 836 & 506\\
Sample C & 690 & 632 & 829 & 492\\
Sample D & 447 & 585 & 1975 & 434\\
\hline
\end{tabular}
\label{tbl:performance_inbuilt_control}
\tablefoot{ 
    \footnotesize 
    \tablefoottext{a}{See corresponding section for a description of the parameter.\\}}
\end{table}

Table~\ref{tbl:performance_inbuilt_control} gives the exact number of fit components that were removed due to the in-built quality controls using the default settings of \gausspyplus. 
In general, the significance criterion was most often used and thus is the strictest criterion, followed by the requirement of a minimum S/N value for the fitted amplitude and a minimum value for the fitted FWHM.
Since the synthetic spectra were set up to also contain emission in the outermost channels, the criterion checking whether the fitted mean position was within the channel range was also used frequently to correct fit results for these spectra.
Note that the sequence of how the in-built quality controls are used matters as e.g. a component that already failed the requirements for the amplitude value will not be subjected to the significance criterion anymore (cf. Fig.$\,\ref{fig:flowchart_in-built_quality_check}$). 
Thus, had we checked the significance criterion first, it would have been responsible for removing even more components.

Also note that while we report here only on the performance of the in-built quality criteria for the improved fitting routine, these criteria are also used in all refit attempts in the spatially coherent refitting phases.

\begin{table}
\caption{Number of new best fit solutions obtained by utilizing the optional quality control procedures for the decomposition of the GRS test field and the synthetic spectra.}
    \centering
    \small
    \renewcommand{\arraystretch}{1.2}
\begin{tabular}{cccc}
\hline\hline  & neg. res. peak & broad & blended \\
 & (3.2.2.1)\tablefootmark{a} & (3.2.2.2)\tablefootmark{a} & (3.2.2.3)\tablefootmark{a} \\
\hline
GRS test field & 25 & 542 & 14\\
Sample A & 0 & 0 & 0\\
Sample B & 133 & 353 & 40\\
Sample C & 137 & 354 & 49\\
Sample D & 2 & 683 & 1\\
\hline
\end{tabular}
\label{tbl:performance_optional_control}
\tablefoot{ 
    \footnotesize 
    \tablefoottext{a}{See corresponding section for a description of the parameter.\\}}
\end{table}

Table~\ref{tbl:performance_optional_control} lists the number of successful refits using the optional quality control procedures for refitting negative residual features, broad and blended components for the GRS test field and the four samples of synthetic spectra.
The refitting of broad fit components into multiple narrower individual components was the criterion that led to most successful refits, followed by the refitting of components that caused negative residual features. 
This mostly reflects the generally low S/N values of signal peaks in the spectra, for which \gausspy\ often fits a single broad Gaussian component over multiple individual signal peaks (cf. Fig.~\ref{fig:spectra_comp_improved_gausspy}).
The refitting of features labelled as blended did not yield that many successful refits.
The signal peaks in the synthetic spectra were constructed in such a way as to not show heavily blended components, which explains the decreased use of this criterion.
For the GRS test field, deviations of emission lines from a Gaussian line shape could have caused the fit of multiple blended components, which resulted in decreased residuals or AICc values that could not be matched with the fit of a single component. 

\subsection{Refit iterations of the spatially coherent refitting phases}
\label{sec:refit_iterations}

In this section we discuss the performance of the two phases of spatially coherent refitting (Sect.$\,\ref{sec:phase_1}$ and $\ref{sec:phase_2}$).

\begin{figure}
    \centering
    \includegraphics[width=\columnwidth]{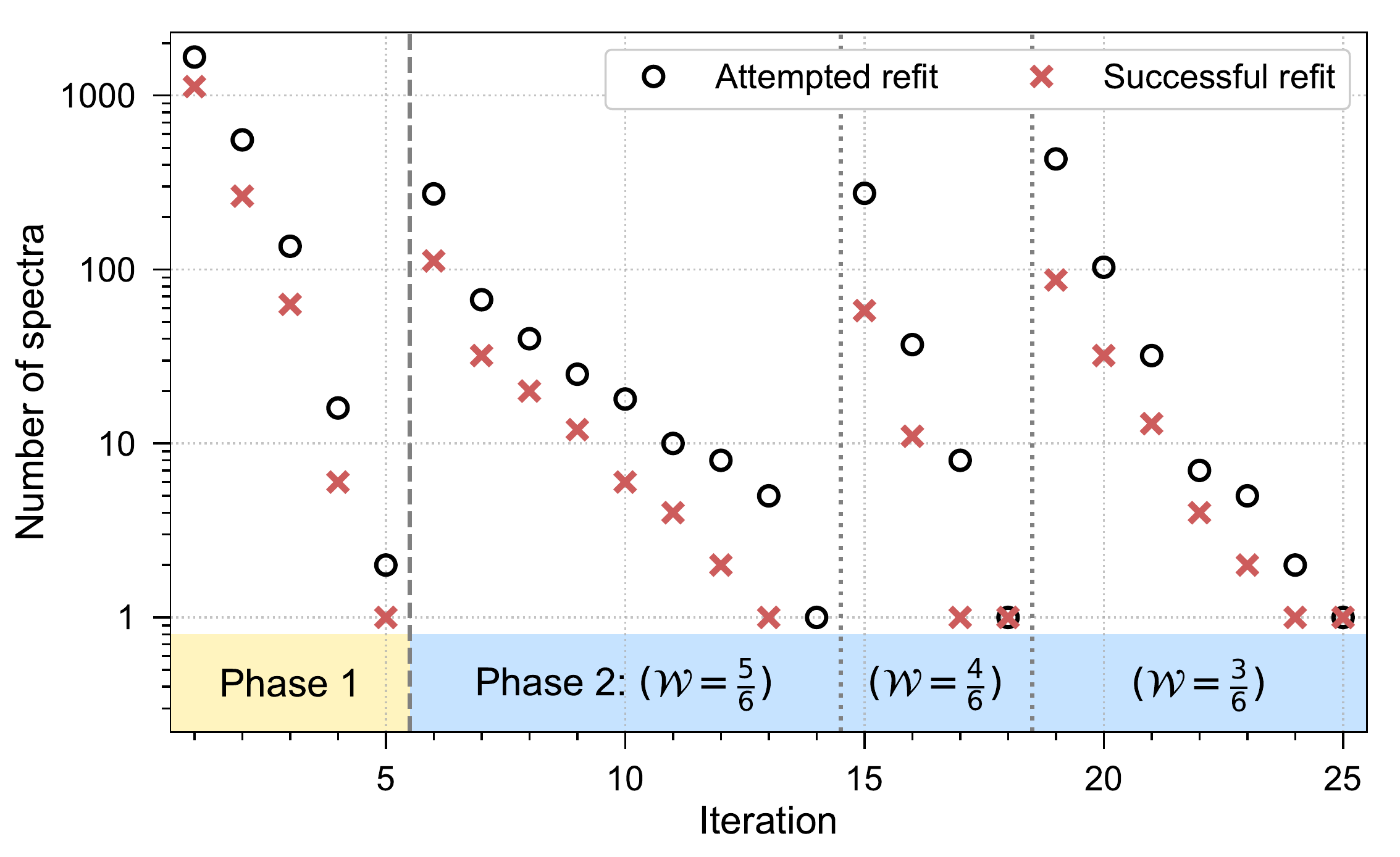}
    \caption{Number of refit attempts and successful refits of spectra of the GRS test field for each iteration in the two phases of the spatially coherent refitting.
    See App.~\ref{sec:refit_iterations} for more details.
    }
    \label{fig:app_iterations}
\end{figure}

For the GRS test field, the two phases of spatially coherent refitting of \gausspyplus\ needed 25 iterations in total to converge to a final fit solution.
Figure~\ref{fig:app_iterations} shows the number of attempted and successful refits for all iterations.
Most of the attempted and successful refits occur in phase 1, which needed 5 iterations. 
Since in a new iteration we will only refit spectra if they had not been flagged in the previous iteration or at least one of the fit solutions of its neighbours got updated, fewer spectra will be refit in each progressing iteration, which is demonstrated by the steep decrease of refit attempts in Figure~\ref{fig:app_iterations}. 
For example, in the first iteration of phase 1, 1839 out of the 4200 spectra were flagged and selected for refitting. 
The \gausspyplus\ algorithm tried to refit 1664 of these flagged spectra\footnote{For the remaining 175 flagged spectra no unflagged neighbouring fit solutions were available.} with new fit solutions derived from neighbouring spectra, $\sim 68\%$ of which got a new best fit solution.
In the second iteration, we only tried to refit 556 flagged spectra, of which $\sim 47\%$ got a new best fit solution.

Figure~\ref{fig:app_iterations} further shows the performance of phase 2 of the spatially coherent refitting, which proceeded in three stages, because in the default settings of \gausspyplus\ the minimum required weight threshold $\pazocal{W}$ is reset to a lower value two times.
The runtime of phase~2 can therefore be decreased by setting a higher minimum weight threshold (e.g., $\pazocal{W}_{\Min} = \sfrac{4}{6}$), which should already lead to good spatial coherence between the neighbouring fit solutions.

Note that the total number of refitting iterations for the spatially coherent refitting phases depends on the size of the data cube as well as on how many flags are set in phase 1 and to which weight threshold we go down to in phase 2. 

In terms of total added and subtracted number of components for the decomposition of the GRS test field, phase 1 removed 226 components and added 295, whereas phase 2 subtracted 84 components and added 191 components.
About $13\%$ of the added components in phase 2 led to fit solutions being flagged as blended.

\section{Normality tests}
\label{app:normality-tests}

As discussed in Sect.$\,\ref{sec:goodness-of-fit}$, as a goodness of fit check we subject the normalised residual to two normality tests to decide whether the data points of the residual are normally distributed and thus consistent with Gaussian noise.

We found that a combination of the two-sided Kolmogorov-Smirnov (K-S) test \citep{Kolmogorov1933, Smirnov1939} and the normality test based on D’Agostino and Pearson \citep[D-P; ][]{Dagostino1971, Dagostino1973} yielded the most reliable means to detect unfitted signal peaks in the residual.

We tested the performance of each normality test for mock residuals that we created by adding a single Gaussian component to white noise.
We used six different combinations of the S/N and significance values for the Gaussian components. 
We also varied the number of spectral channels between $100 \text{--} 1000$ in steps of $100$.
We produced $1000$ spectra for each possible combination of Gaussian signal component and number of spectral channels for a total of $60\,000$ spectra. 
We then applied the normality tests to each of these mock residuals to check which test could most reliably identify the leftover signal component by rejecting the null hypothesis of normally distributed residual values.

\begin{figure}
    \centering
    \includegraphics[width=\columnwidth]{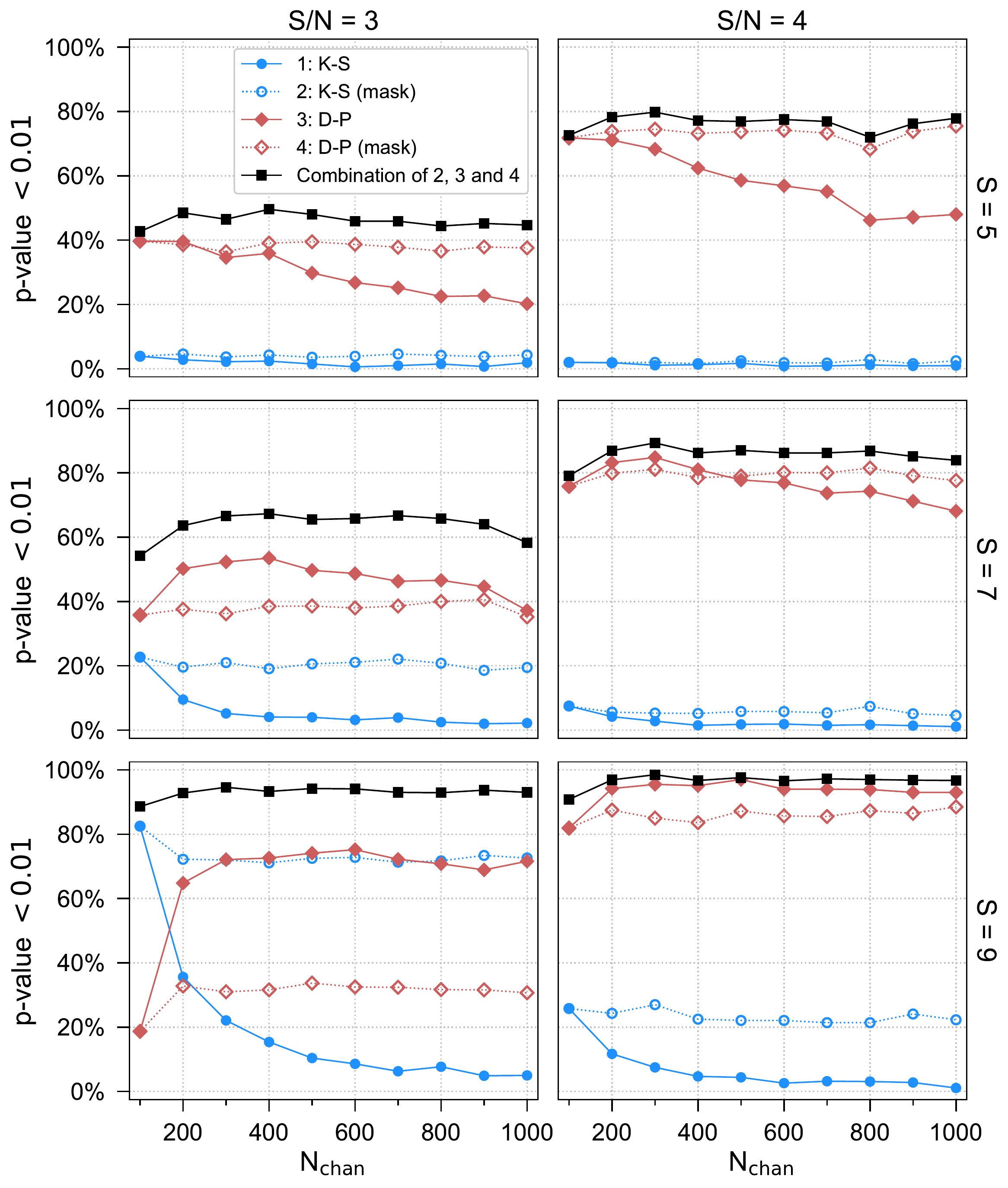}
    \caption{Comparison of the performance of different normality tests for mock residuals as a function of the number of spectral channels.
    The residuals contain a single Gaussian signal component with a signal-to-noise ratio of 3 or 4 (left and right panels, respectively) and significance values of 5, 7, and 9 (upper, middle, and lower panels, respectively).
    See App.~\ref{app:normality-tests} for more details.
    }
    \label{fig:app_normaltest}
\end{figure}

Figure~\ref{fig:app_normaltest} shows the performance of the normality tests for the different combinations.
On the ordinate we plot the percentage of spectra for which the normality tests yielded p-values below the default threshold in \gausspyplus\ of $1\%$, indicating that the residual data points are not normally distributed.
The results of the K-S and D-P test are shown in blue and red, respectively.
Moreover, we applied both normality tests on the whole residual and only the residual data points within the identified signal ranges, which is indicated by the filled and unfilled symbols, respectively. 
The black line shows the performance of the null hypothesis testing included in \gausspyplus, which combines the results of the D-P test applied to the full residual and the results of the D-P and K-S tests applied to only the residual data points within the identified signal intervals. 
For this combination, we use the smallest $p$-value resulting from these three normality tests.
Figure~\ref{fig:app_normaltest} demonstrates that this combination results in an increased ability to detect leftover signal peaks in the residual for both narrow and broad components (with low and high significance values, respectively).
We are able to identify the majority of signal peaks in the residual with $\significance{} \geq 7$ \text{or} $\text{S/N} = 4$ and the identification fraction reaches nearly $100\%$ for the strongest tested components ($\significance{} = 9$, $\text{S/N} = 4$)
Moreover, this improved performance is independent of the number of spectral channels. 
In comparison, the individual results of the K-S and D-P tests show a decreased performance and even a complementary behaviour for broader Gaussian signals with lower S/N values ($\significance{} = 9$, $\text{S/N} = 3$) and low number of spectral channels ($< 300$).

To check the fraction of false positives identified by the normality tests, we checked their performance also for Gaussian noise only, for which we removed the signal component from all residuals used in Figure~\ref{fig:app_normaltest}. 
We evaluate the spectra again in groups of $1000$ spectra and report the median, minimum and maximum false positive rate for all groups as the fraction of spectra for which the hypothesis tests yielded a $p$-value $< 1\%$ and thus would not pass our criterion for normally distributed residuals.

\begin{table}
\caption{Percentage of false positives identified by the normality tests.}
\label{tbl:fp_normality_test}
\centering
    \small
    \renewcommand{\arraystretch}{1.2}
\begin{tabular}{cccc}
    \hline\hline
	Test & Combination & K-S & D-P \\
	\hline
Median & $2.2\%$ & $5.5\%$ & $7.6\%$ \\
Minimum & $1.4\%$ & $0.3\%$ & $1.1\%$ \\
Maximum & $3.3\%$ & $11.5\%$ & $13.9\%$ \\
\hline
\end{tabular}
\end{table}

Table~\ref{tbl:fp_normality_test} lists the false positive rates.
The combination of the normality tests as implemented in \gausspyplus\ leads to the best performance over different channel ranges, as evidenced by the reduced median false positive rate compared to the individual normality tests.
The K-S and D-P tests produce higher false positive rates with increasing numbers of spectral channels, whereas the combination of the two tests performed best for the highest number of spectral channels we probed. 

\section{$\chired$ calculations for the GRS test field}
\label{app:grs-reduced-chi-square}

A problem in determining the $\chired$ value (Sect.$\,\ref{sec:goodness-of-fit}$) is that it depends on the number of channels in the spectrum.
If the spectrum consists of many channels that contain only noise, low $\chi^{2}$ values and $\chired$ values close to $1$ follow even if the performance of the fit is not satisfactory in the part of the spectrum where there is signal.

To avoid this problem we identify the regions likely to contain signal already in the noise estimation step (see Sect.$\,\ref{sec:signal-interval}$) and use only these regions for the $\chired$ calculations.
We also mask out negative noise spike features that tend to produce high $\chired$ values even for spectra whose signal features were well fit (see Sect.$\,\ref{sec:noise-spikes}$).

\begin{figure}
    \centering
    \includegraphics[width=\columnwidth]{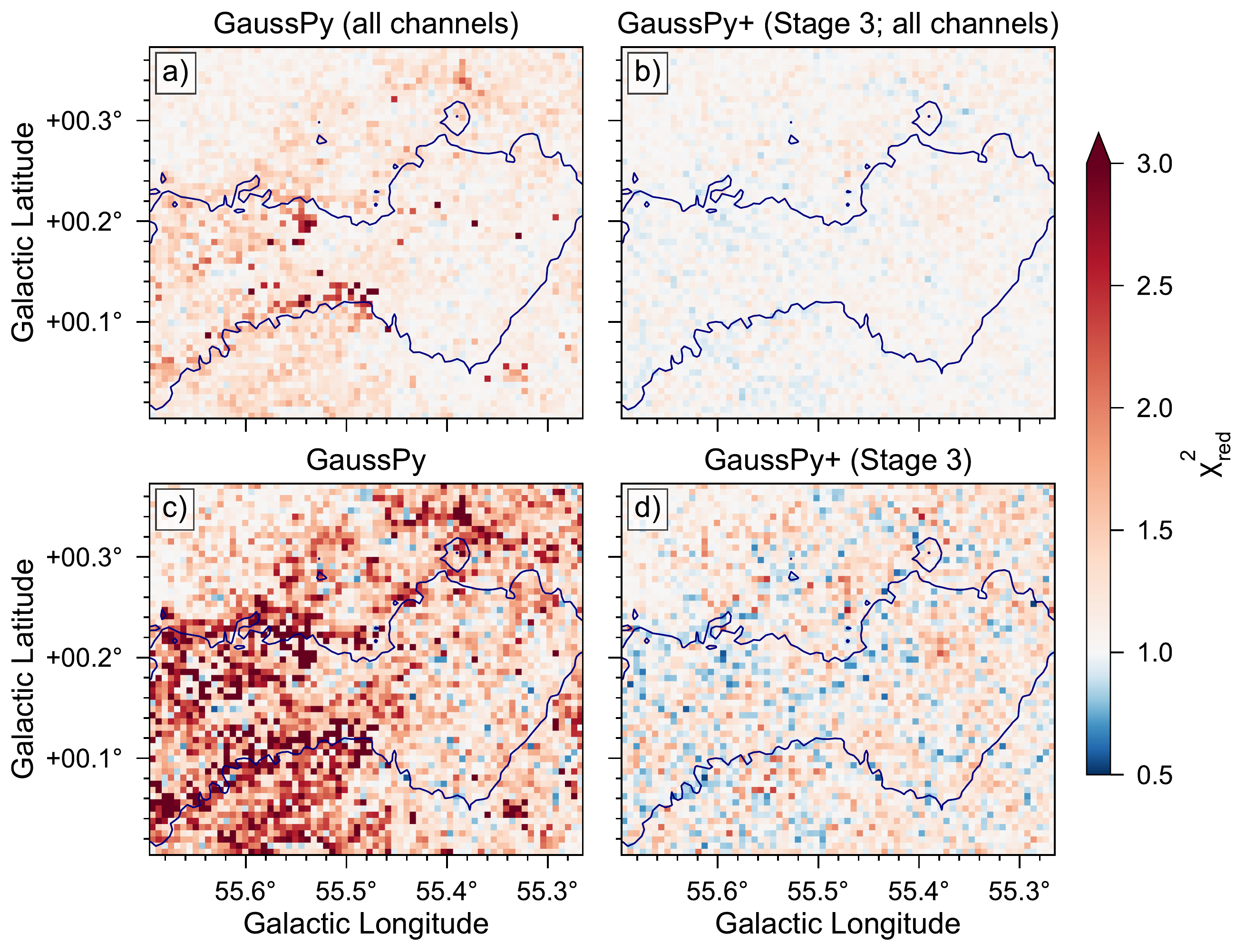}
    \caption{Maps showing the $\chired$ values for the \gausspy\ (left panels) and stage~3 of the \gausspyplus\ (right panels) decomposition results, calculated by using either all available spectral channels (upper panels) or restricted to the spectral channels estimated to contain signal.
    All panels are overplotted with the contour from panel~(b) in Fig.~\ref{fig:grs-01_0mom_vs_sigma}.
    Panels~(c) and (d) are identical to panels~(i) and (l) in Fig.~\ref{fig:test-field-comparison}.
    }
    \label{fig:grs-01_rchi2_all_channels}
\end{figure}

\begin{figure}
    \centering
    \includegraphics[width=\columnwidth]{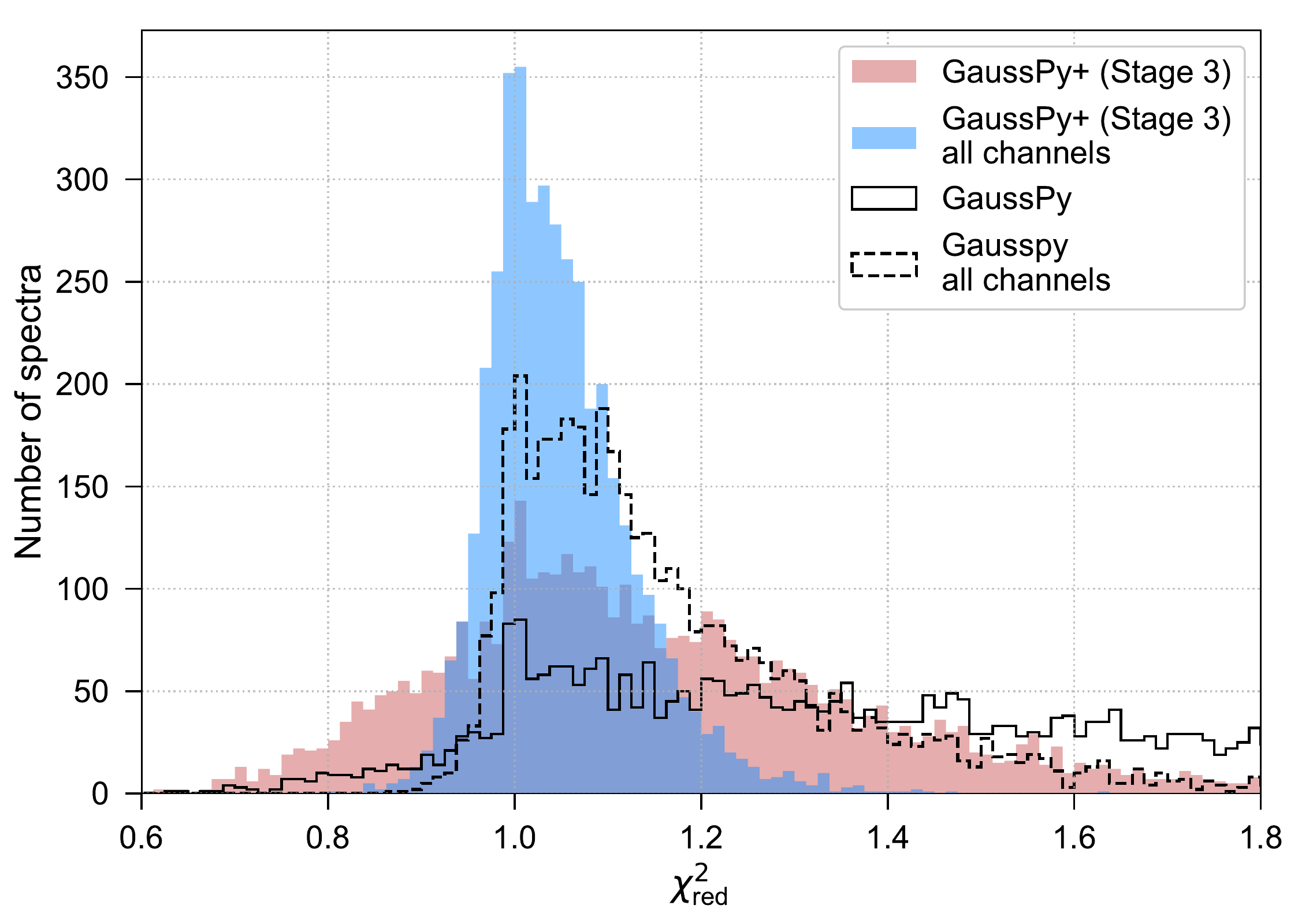}
    \caption{Comparison of the distribution of the $\chired$ values for the decomposition results of \gausspy\ and stage 3 of \gausspyplus\ restricted to spectral channels estimated to contain signal and calculated over the whole spectral range.
    }
    \label{fig:grs-01_rchi2_hist_ncomps_rchi2}
\end{figure}

To illustrate the importance of restricting the $\chired$ calculation to intervals containing signal we recomputed the goodness of fit calculations for the decomposition results of the GRS test field (Sect.$\,\ref{sec:testfield-decomp}$) obtained with \gausspy\ and after stage~3 of \gausspyplus\ by using all available spectral channels.
Panels~(a) and (b) in Fig.$\,$\ref{fig:grs-01_rchi2_all_channels} show the recomputed $\chired$ values using all 424 spectral channels.
For comparison, we also show the maps of $\chired$ values again that was obtained by restricting the goodness of fit calculations to spectral channels estimated to contain signal (panels~c and d, which are identical to panels~m and p in Fig.$\,\ref{fig:test-field-comparison}$).
Fig.~\ref{fig:grs-01_rchi2_hist_ncomps_rchi2} gives the corresponding histograms.
Both figures clearly illustrate how the goodness of fit values are artificially reduced if most of the spectral channels included in the calculation contain only noise. 
Using all available spectral channels for the goodness of fit calculations thus makes it more challenging to use the $\chired$ values to decide which fit results were not successful.

\section{Symbols, \gausspyplus\ keywords and default values}

\begin{table}
    \caption{Symbols used throughout the text.}
    \centering
    \small
    \renewcommand{\arraystretch}{1.2}
    \begin{tabularx}{\columnwidth}{cX}
    \hline
    Symbol & Description \\
    \hline
$\mu_{i}$ & offset or mean position of Gaussian fit component $i$ \\
$a_{i}$ & amplitude of Gaussian fit component $i$ \\
$\sigma_{i}$ & standard deviation of Gaussian fit component $i$ \\
$\Theta_{i}$ & FWHM value of Gaussian fit component $i$ \\
$\chired$ & reduced chi-squared; chi-squared per degree of freedom \\
AICc & corrected Akaike information criterion \\
F$_{1}$ score & measure of the accuracy for the decomposition of the training set (Sect.$\,\ref{sec:gausspy}$) \\
$\rms$ & root-mean-square noise of the spectrum \\
$\rmsTa$ & root-mean-square noise of the spectrum given in antenna temperature values $T_{A}^{*}$ \\
$\Nchan$ & number of channels in a spectrum \\
$\Ncomp$ & number of fitted Gaussian components in a spectrum \\
$\significance{data}$ & significance estimates for signal peaks (Sect.$\,\ref{sec:significance}$) \\
$\significance{fit}$ & significance estimates for fitted Gaussian components (Sect.$\,\ref{sec:significance}$) \\
$\pazocal{F}_{tot}$ & sum of the $\flag{blended}{}$, $\flag{neg.\,res.\,peak}{}$, $\pazocal{F}_{\Theta}$, $\flag{residual}{}$, and $\pazocal{F}_{N_{\mathrm{comp}}}$ flags \\
$\pazocal{W}$ & weight threshold \\
    \hline
    \end{tabularx}
    \label{tbl:symbols}
\end{table}

Table~\ref{tbl:symbols} gives the description of symbols used throughout the text.

Table~\ref{tbl:gausspyplus-keywords} gives an overview of the parameter settings of \gausspyplus, listing their corresponding default values and symbols used throughout the text.
To get first decomposition results users only need to supply values for the parameters listed under \textit{essential parameters}.
In case the decomposition does not yield good results we recommend to first use different values for the \textit{essential parameters}.
If this should not improve the results users can vary the parameters listed under \textit{more advanced settings}.

\begin{table*}
    \caption{\gausspyplus\ keywords mentioned throughout the text.}
    \centering
    \small
    \renewcommand{\arraystretch}{1.2}
    \begin{tabularx}{2\columnwidth}{cXcc}
    \hline
    Symbol & Description & \gausspyplus\ keyword & Default\\
\hline\\[-1.0em]\multicolumn{4}{c}{\textbf{essential parameters}} \vspace{1pt}\\       
$\alpha_{1}$ & first smoothing parameter used in \gausspy\ decomposition (Sect.$\,\ref{sec:gausspy}$) & \texttt{alpha1} & \texttt{None} \\
$\alpha_{2}$ & second \gausspy\ smoothing parameter; only used in two-phase decomposition (Sect.$\,\ref{sec:gausspy}$) & \texttt{alpha2} & \texttt{None} \\
$\snmin$ & minimum S/N ratio for signal peaks in the data (Sect.$\,\ref{sec:snratio}$) & \texttt{snr} & $3$ \\
$\significance{\Min}$ & Minimum significance value for signal peaks and fitted Gaussian components (Sect.$\,\ref{sec:significance}$) & \texttt{significance} & $5$ \\
\hline\\[-1.0em]\multicolumn{4}{c}{\textbf{more advanced settings}}\vspace{1pt}\\       
$\Delta\mu_{\Max}$ & maximum difference in offset positions of Gaussian components for grouping (Sect.$\,\ref{sec:phase_1}$) & \texttt{mean\_separation} & $2$\tablefootmark{*} \\
$\Delta\Theta_{\Max}$ & maximum difference in FWHM values of Gaussian components for grouping (Sect.$\,\ref{sec:phase_1}$) & \texttt{fwhm\_separation} & $4$\tablefootmark{*} \\
$\Theta_{\Min}$ & minimum value for the FWHM of fitted Gaussian components & \texttt{min\_fwhm} & $1$\tablefootmark{*} \\
$\Theta_{\Max}$ & maximum value for the FWHM of fitted Gaussian components & \texttt{max\_fwhm} & \texttt{None}\tablefootmark{*} \\
$f_{a}$ & factor by which the maximum data value is multiplied to get a maximum limit for the fitted amplitude $a_{i}$ & \texttt{max\_amp\_factor} & $1.1$ \\
$f_{\Theta}$ & factor by which the FWHM value of a fit component has to exceed all other (neighbouring) fit components to get flagged (Sect.$\,\ref{sec:broad-components}$) & \texttt{fwhm\_factor} & $2$ \\
$f_{\mathrm{sep}}$ & factor to determine the minimum required separation between two fit components before they are counted as blended (Sect.$\,\ref{sec:blended-comps}$) & \texttt{separation\_factor} & $1 / \sqrt{2\, \text{ln}\,2}$ \\
$f_{w}$ & factor that determines the weight given to neighbouring spectra located at a distance of 1 and 2 pixels (Sect.$\,\ref{sec:phase_2}$) & \texttt{weight\_factor} & $2$ \\
$\flag{neg.\,res.\,peak}{}$ & flag criterion for negative residual features (Sect.$\,\ref{sec:negative-residual}$) & \texttt{flag\_neg\_res\_peak} & \texttt{True} \\
$\pazocal{F}_{\Theta}$ & flag criterion for broad fit components (Sect.$\,\ref{sec:broad-components}$) & \texttt{flag\_broad} & \texttt{True} \\
$\flag{blended}{}$ & flag criterion for blended fit components (Sect.$\,\ref{sec:blended-comps}$) & \texttt{flag\_blended} & \texttt{True} \\
$\flag{residual}{}$ & flag criterion for fit results whose normalised residual values do not pass the Kolmogorov-Smirnov test for normality (Sect.$\,\ref{sec:normaltest}$) & \texttt{flag\_residual} & \texttt{True} \\
$\pazocal{F}_{N_{\mathrm{comp}}}$ & flag criterion for fit results whose number of components are not compatible with neighbouring fits (Sect.$\,\ref{sec:ncomps}$) & \texttt{flag\_ncomps} & \texttt{True} \\
$\Npad$ & number of spectral channels added to the left and right of signal intervals (Sect.$\,\ref{sec:signal-interval}$) & \texttt{pad\_channels} & $5$\tablefootmark{*} \\
$\Nmin$ & minimum number of spectral channels the signal intervals in a spectrum must have (Sect.$\,\ref{sec:signal-interval}$) & \texttt{min\_channels} & $100$\tablefootmark{*} \\
$\Delta N_{\Max}$ & maximum allowed difference in $\Ncomp$ between fit solution and weighted median number of components determined from all immediate neighbours (Sect.$\,\ref{sec:ncomps}$) & \texttt{max\_diff\_comps} & $1$ \\
$\Delta N_{\mathrm{jump}}$ & maximum allowed difference in $\Ncomp$ between individual neighbouring spectra (Sect.$\,\ref{sec:ncomps}$) & \texttt{max\_jump\_comps} & $2$ \\
$N_{\mathrm{jump}}$ & maximum number of allowed $N_{\mathrm{jump}}$ occurrences for a single spectrum (Sect.$\,\ref{sec:ncomps}$) & \texttt{n\_max\_jump\_comps} & $1$ \\
$\plimit$ & probability threshold for features of consecutive positive or negative channels to be counted as more likely to be a noise feature (Sect.$\,\ref{sec:noise-estimation}$, App.$\,$\ref{app:markov}) & \texttt{p\_limit} & $0.02$ \\
$p$-value  & $p$-value for the null hypothesis that the residual resembles a normal distribution (Sect.$\,\ref{sec:phase_1}$) & \texttt{min\_pvalue} & $0.01$ \\
$\snminfit$  & minimum S/N ratio ($=a_{i}/\sigma_{\mathrm{rms}}$) for fitted Gaussian components (Sect.$\,\ref{sec:snratio}$) & \texttt{snr\_fit} & \texttt{None} \\
$\snminneg$ & minimum S/N ratio for negative peaks in the spectrum (Sect.$\,\ref{sec:negative-residual}$) & \texttt{snr\_negative} & \texttt{None} \\
$\snspike$ & S/N threshold for noise spikes (Sect.$\,\ref{sec:noise-spikes}$) & \texttt{snr\_noise\_spike} & $5$ \\
$\text{SNR}_{1}$ & S/N threshold used by \gausspy\ for the original spectrum & \texttt{snr\_thresh} & \texttt{None} \\
$\text{SNR}_{2}$ & S/N threshold used by \gausspy\ for the second derivative of the smoothed spectrum & \texttt{snr2\_thresh} & \texttt{None} \\
$\order$ & minimum number of spectral channels a peak has to contain on either side (Sect.$\,\ref{sec:training-set}$) & \texttt{order} & $6$\tablefootmark{*} \\
$\pazocal{W}_{\Min}$ & minimum weight threshold before phase 2 of the spatially coherent refitting routine is terminated (Sect.$\,\ref{sec:phase_2}$) & \texttt{min\_weight} & $0.5$ \\
    \hline
    \end{tabularx}
    \label{tbl:gausspyplus-keywords}
    \tablefoot{ 
    \footnotesize 
    \tablefoottext{*}{Have to be specified in channel units.}}
\end{table*}

\end{appendix} 

\end{document}